\documentclass[12pt,a4paper,final]{iopart}

\usepackage{iopams}  
\usepackage{graphicx}
\usepackage[breaklinks=true,colorlinks=true,linkcolor=blue,urlcolor=blue,citecolor=blue]{hyperref}
\usepackage{amsfonts}

\expandafter\let\csname equation*\endcsname\relax
\expandafter\let\csname endequation*\endcsname\relax
\usepackage{amsmath}

\usepackage{amssymb}
\usepackage{bm}
\usepackage{dcolumn}
\usepackage{graphicx}
\usepackage{graphics}
\usepackage[latin1]{inputenc}
\usepackage{latexsym}
\usepackage{rotating}
\usepackage{hyperref}
\usepackage[all]{hypcap} 
\usepackage{xspace} 
\usepackage[usenames,dvipsnames]{color}
\usepackage{mathrsfs}
\usepackage{subfigure}

\usepackage{ulem}
\usepackage{outlines}
\usepackage{enumitem}
\setenumerate[1]{label=\Roman*.}
\setenumerate[2]{label=\Alph*.}
\setenumerate[3]{label=\roman*.}
\setenumerate[4]{label=\alph*.}

\definecolor {darkgreen}{rgb}{0.2,0.7,0.2}

\newcommand\be{\begin{equation}}
\newcommand\ba{\begin{eqnarray}}
\newcommand\ee{\end{equation}}
\newcommand\ea{\end{eqnarray}}

\newcommand\bw{\begin{widetext}}
\newcommand\ew{\end{widetext}}

\newcommand{\nn}{\nonumber}

\begin{document}
\title[Hereditary Effects to Third Post-Newtonian Order]{Hereditary Effects in Eccentric Compact Binary Inspirals to Third Post-Newtonian Order}

\author{Nicholas Loutrel}
\address{eXtreme Gravity Institute, Department of Physics, Montana State University, Bozeman, MT 59717, USA.}
\ead{nicholas.loutrel@montana.edu}

\author{Nicol\'as Yunes}
\address{eXtreme Gravity Institute, Department of Physics, Montana State University, Bozeman, MT 59717, USA.}
\ead{nicolas.yunes@montana.edu}

\date{\today}

\begin{abstract} 

While there has been much success in understanding the orbital dynamics and gravitational wave emission of eccentric compact binaries in the post-Newtonian formalism, some problems still remain. 
The largest of these concerns hereditary effects: non-linear phenomena related to the scattering off of the background curved spacetime (tails) and to the generation of gravitational waves by gravitational waves (memory). 
Currently, these hereditary effects are only known numerically for arbitrary eccentricity through infinite sums of Bessel functions, with closed-form, analytic results only available in the small eccentricity limit. 
We here calculate, for the first time, closed-form, analytic expressions for all hereditary effects to third post-Newtonian order in binaries with \emph{arbitrary} eccentricity.
For the tails, we first asymptotically expand all Bessel functions in high eccentricity and find a superasymptotic series for each enhancement factor, accurate to better than $10^{-3}$ relative to post-Newtonian numerical calculations at all eccentricities.  
We further improve the small-eccentricity behavior of the superasymptotic series by generating hyperasymptotic expressions for each enhancement factor, typically accurate to better than $10^{-8}$ at all eccentricities. 
%
%
%
For the memory, we discuss its computation within the context of an osculating approximation of the binary's orbit and the difficulties that arise. 
Our closed-form analytic expressions for the hereditary fluxes allow us to numerically compute orbital phases that are identical to those found using an infinite sum of Bessel functions to double numerical precision.
\end{abstract}



\maketitle

\section{Introduction}
\label{intro}

There has never been a more exciting era for gravitational wave (GW) astrophysics.~The advanced Laser Interferometer Gravitational Wave Observatory (aLIGO)~\cite{Harry:2010zz} has made the first detection of GWs~\cite{GW150914} and presented us with a new window through which to observe the universe. These detectors will be joined in the coming years by the advanced Virgo detector in Italy in 2016~\cite{TheVirgo:2014hva}, the KAGRA detector in Japan in 2018~\cite{Uchiyama:2004vr}, and the LIGO India detector in 2019~\cite{Unnikrishnan:2013qwa}. These instruments are poised to make continued detections of GWs and answer fundamental questions about the nature of the gravitational interaction~\cite{Yunes:2016jcc}.

One of the predominant type of events expected to be seen by these ground-based interferometers are the late inspiral and coalescence of binary systems comprised of compact objects, specifically neutron stars (NSs) and black holes (BHs). Over the past several decades, the following primary picture of such binaries has emerged: by the time the orbit has shrunk enough that the GWs emitted are in the sensitivity band of detectors, the binary will necessarily have extremely small orbital eccentricity ($e \ll 0.1$). This is expected because the quadrupolar nature of GWs makes them extremely efficient at circularizing binaries that form with large orbital separations, such as the Hulse-Taylor pulsar~\cite{Hulse:1974eb}.

In more recent years, however, this view has been challenged. Studies of dense stellar environments, such as globular clusters and galactic nuclei, have shown that more exotic formation channels could lead to a population of binaries with non-negligible eccentricity ($e > 0.1$) that emit GWs in the sensitivity band of ground-based detectors~\cite{2009MNRAS.395.2127O, lee2010,2003ApJ...598..419W,Kushnir:2013hpa,2013PhRvL.111f1106S,Antognini:2013lpa,2013ApJ...773..187N,2014ApJ...781...45A,Antonini:2015zsa}. Close encounters between two compact objects on unbound orbits could lead to dynamical captures, either through GW emission or dynamical friction, which could have eccentricities close to unity~\cite{2009MNRAS.395.2127O, lee2010}. Similarly, close encounters between multiple objects could form hierarchical triple systems which, through Kozai-Lidov oscillations and other three-body resonances~\cite{2003ApJ...598..419W,Kushnir:2013hpa,2013PhRvL.111f1106S,Antognini:2013lpa,2013ApJ...773..187N,2014ApJ...781...45A,2013ApJ...773..187N,Antonini:2015zsa}, could drive the inner binary to high eccentricity. Although the event rate for such highly eccentric signals is expected to be small (maybe 1-2 events per year), these rate estimates are quite uncertain; a single detection would provide unique and highly desirable information about a different class of binary systems and about theoretical physics in the extreme gravity regime~\cite{Loutrel:2014vja}.

What is the best way to detect such events? The most efficient method is matched filtering, provided one knows accurately the shape of the expected signal. In this search strategy, a set of theoretical models or \emph{templates} are used to fit the data and extract the parameters of the model. To construct such templates, precise knowledge of the dynamics of the system generating the GWs is required. Inaccuracies in the templates can potentially throw off detection or systematically bias the recovered parameters. Currently, most matched-filtering searches use non-eccentric template families, which could miss non-eccentric signals all together if the orbital eccentricity exceeds $0.05$~\cite{Huerta:2013qb, Huerta:2014eca}. 

One potential solution is to construct eccentric template families. Numerical templates, calculated for example through the numerical solution to the post-Newtonian (PN) equations of motion\footnote{In this framework, the field equations are solved through a weak-field and small-velocity expansion. A term of $n$-PN order is of ${\cal{O}}\left(v^{2n}\right)$ relative to its leading-order expression, where $v$ is the characteristic velocity of the system.}, are computationally expensive and can be affected by numerical error in long evolutions. Analytic templates are, in principle, more computationally efficient and provide deeper insight into the physical processes that control GW emission, but their construction is mathematically difficult. Post-circular templates, valid in the small eccentricity limit, were developed in~\cite{PhysRevD.80.084001} and improved in~\cite{Tessmer:2010sh, Tessmer:2010ii, Huerta:2014eca, Moore:2016qxz, Tanay:2016zog}, but by construction they are only valid for small eccentricity binaries. GWs from highly elliptical binaries are more a set of intense \emph{bursts} centered around pericenter passage than a continuous signal, rending their analytic modeling very difficult (though there has been some very recent progress~\cite{Tanay:2016zog, Forseth:2015oua}.)
 
A promising, though sub-optimal, method of detecting GWs from highly elliptic binaries that relies less on accurate templates is power stacking~\cite{Tai:2014bfa}. If a set of $N$ bursts is found within a detector's data stream, the power can be added within each burst, with the amplification\footnote{The scaling with $N$ is a result of the stacking of power rather than amplitude as done in matched filtering.} of the signal-to-noise ratio (SNR) scaling as $N^{1/4}$ (assuming all bursts have the same SNR.) Adding up the bursts in a sequence is crucial because any one burst will likely be very weak (SNRs of $1$--$2$), unless the elliptical binary is absurdly close to Earth. Power stacking, however, only works if one has \emph{prior knowledge} of where the bursts will occur in time-frequency space. To that end, a time-frequency track for highly elliptical systems, known as the burst model, was developed in~\cite{Loutrel:2014vja}.

Although power stacking is far more robust to systematic mismodeling error than matched filtering, it is still not immune~\cite{Tai:2014bfa}, in particular when considering the accuracy of GW fluxes. The latter provide a radiation-reaction force in the PN formalism that forces the binary to inspiral, through the balance of the binary energy lost by the system to the energy carried away by GWs from the system. In template based searches, inaccurate GW fluxes introduce errors in the Fourier transform of the waveform through the stationary phase approximation used in the post-circular extensions~\cite{PhysRevD.80.084001} of the so-called TaylorF2 waveform family~\cite{Huerta:2014eca}. In power stacking searches, similar inaccuracies introduce an error in the evolution of the orbital elements as the orbit osculates due to the emission of bursts, affecting the time-frequency tracks of the burst model~\cite{Loutrel:2014vja}. An accurate and complete treatment of the GW fluxes is clearly essential to extend any PN template family, the post-circular waveform family and the burst model to higher PN order.

Within the PN framework, the fluxes can be broken down into two distinct sets -- the \textit{instantaneous} terms and the \textit{hereditary} terms~\cite{Blanchet:2013haa} -- that represent physically distinct processes. Consider a compact binary that inspirals due to GW emission and an observer at spatial infinity. As the binary inspirals, the GWs propagate along the curved spacetime generated by the binary and they are observed at spatial infinity some retarded time later. To leading PN order, the GWs are, however, treated as though they are propagating on flat spacetime. The instantaneous flux describes the direct linear emission of GWs from the source at the instant corresponding to the current retarded time. These terms depend on the time variation of the binary's multipole moments and enter the fluxes at integer PN orders.

Hereditary terms, on the other hand, are a result of the nonlinear nature of General Relativity, and they are labeled ``hereditary'' because they depend on integrals over the entire past lifetime of the source. Hereditary terms can be further broken down into two distinct subsets: the \textit{tail} terms and the \textit{memory} terms.  The tail terms~\cite{Blanchet379, Blanchet383, Blanchet:1988, Blanchet:1992br, Blanchet:1993ec, Blanchet:1995fr, Rieth:1997mk, Arun:2007rg} describe how the time varying radiative multipole moments interact with the curved spacetime of the binary, which to leading PN order is characterized by the ADM mass of the system. This nonlinear interaction causes the waves to scatter as they propagate, and since they are odd under time reversal, they enter the fluxes at half-integer PN orders. Tail terms enter first at 1.5PN order relative to the leading PN order instantaneous flux through a quadratic monopole-quadrupole interaction. Similar tail terms also result from higher-order multipoles; at 3PN order, the so-called ``tail squared'', (tail)$^2$, and tail-of-tails, tail(tails), terms enter the fluxes~\cite{Blanchet:1997ji, Blanchet:1997jj} through cubic monopole-quadrupole interactions that describe double scattering (i.e.~the GWs scatter twice off of the curvature of spacetime as they propagate to spatial infinity).

The second type of hereditary fluxes are the memory terms~\cite{Christodoulou:1991, Wiseman:1991, Thorne:1992, Blanchet:1992br, Blanchet:1997ji, Arun:2004ff, Favata:2008yd, Favata:2011qi}. The GWs emitted by the binary are not simply waves that propagate along a background spacetime, but they are themselves a source of spacetime curvature. As a result, the GWs generate their own GWs as they propagate to spatial infinity. The memory is also responsible for permanently changing length scales, for example, of a ring of test particles or the arms of an interferometer (see e.g.~\cite{Lasky:2016knh,Garfinkle:2016nhe} for a recent discussion of the detectability of this effect in aLIGO.) The memory enters the fluxes beginning at 2.5PN order relative to the leading PN order instantaneous flux through the integral of derivatives of the quadrupole moment squared.

If we desire to create a model that is as accurate a representation of Nature as possible, we must include the nonlinear tail and memory terms in the GW fluxes. As it currently stands, the tail terms are only known numerically for arbitrary eccentricity through infinite sums of Bessel functions~\cite{Arun:2007rg}. Closed-form, analytic results are available for small eccentricities as Taylor series. Other closed-form, analytic results approximately valid at all eccentricities are available through Pad\'{e} approximants~\cite{Tanay:2016zog} or through the "factoring" of high-eccentricity terms~\cite{Forseth:2015oua}. The latter two methods fundamentally rely on re-summing a Taylor series about small eccentricity, and thus their accuracy can be problematic unless a sufficiently high number of terms in the small-eccentricity expansion are kept. 

\subsection{Executive Summary}

The goal of this paper is to find closed-form, analytic expressions for the hereditary flux terms (both the tail and memory terms) that are valid for arbitrarily eccentric compact binaries and to 3PN order relative to the leading PN order instantaneous flux. As we shall see, this will require the construction of superasymptotic and hyperasymptotic series for the tail fluxes, as well as the development of an osculating approximation for the evaluation of memory integrals. 

The calculation of the tail terms begins with a Fourier decomposition in multipole moments, as detailed in~\cite{Arun:2007rg}. To compute the energy and angular momentum fluxes to 3PN order, one requires the Fourier decomposition of the 1PN mass quadrupole moment, which was first discussed in~\cite{Arun:2007rg} and is presented here in detail for the first time. With the Fourier decomposition at hand, one can then define the enhancement factors for each of the PN terms in the fluxes. These enhancement factors depend on sums of the Fourier coefficients of the multipole moments, which can be expressed in terms of the Bessel function $J_{p}(p e_{t})$ and its derivative with respect to the argument, with $p$ the Fourier index and $e_{t}$ the ``temporal'' eccentricity.

We re-sum the Bessel series expressions for all of the tail enhancement factors to 3PN order by employing the uniform asymptotic expansion of the Bessel function:
\begin{align}
J_{p}(p e_{t}) \sim K_{1/3}\left(\frac{2}{3} \zeta^{3/2} p\right) + {\cal{O}}\left(\frac{1}{p}\right)
\\
J_{p}'(p e_{t}) \sim K_{2/3}\left(\frac{2}{3} \zeta^{3/2} p\right) + {\cal{O}}\left(\frac{1}{p}\right)
\end{align}
where $K$ is the modified Bessel function of the second kind and $\zeta$ is a known function of $e_{t}$. This asymptotic expansion is valid when $p \rightarrow \infty$, and in this limit, we can replace the sums in the enhancement factors with integrals over the Fourier index. These steps are fundamentally valid in the high eccentricity limit ($e_{t} \sim 1$), so we expand the result of the integral about $\epsilon = 1 - e_{t}^{2} \ll 1$. Through comparison with numerical PN results for the enhancement factors, we show that the resulting series in $\epsilon$ are asymptotic and that they can be truncated at very low order to achieve an optimally-truncated, \emph{superasymptotic series}~\cite{Boyd}, i.e.~one that achieves the minimum possible error relative to an ``exact'' numerical answer. We also show that these optimally-truncated superasymptotic series can have their accuracy further improved by matching them to small eccentricity expansion of the enhancement factors, generating a \emph{hyperasymptotic series}~\cite{Boyd} with typical relative errors of $\lesssim 10^{-8}$ compared to numerical PN results at \emph{all} eccentricities; the only exceptions are two 2.5PN order enhancement factors that are accurate to $\lesssim 10^{-4}$.

The calculation of the memory terms require the development of an osculating approximation. The memory depends on a set of integrals that, in turn, depend on how the orbital elements of the binary evolve due to the emission of gravitational radiation. To evaluate these integrals analytically, we specialize to systems that form with orbital eccentricities close to unity, such that the binary evolution effectively becomes a discrete set of osculating Keplerian ellipses. To lowest order in this approximation, the orbital elements of the binary jump instantaneously from one orbit to the next. Applying the orbit average to this lowest order osculating approximation results in the memory terms vanishing, due to their periodic nature  over the current orbit. We discuss how to rectify this by extending the osculating approximation to higher order in a multiple scale expansion.

Are these closed-form, analytic expressions for the hereditary fluxes at 3PN order sufficiently accurate for template-based and power-stacking searches? The answer to this question depends sensitively on the SNR of the signal, and thus, its closeness to Earth and the noise level of the detector. Therefore, without a careful Monte Carlo, data analysis exploration it is difficult to answer this question quantitatively. Qualitatively, however, we have explored this question by numerically evolving the PN equations of motion with the hereditary fluxes prescribed either through our closed-form expressions or through infinite Bessel sums. We found that the difference between these two orbital phases (calculated from 10Hz to the last stable orbit) is within double precision. 

The remainder of this paper presents the details of the results discussed above. Section~\ref{review} reviews the basics of the radiation reaction problem in GR. Following this, we analytically calculate the Fourier decomposition of the 1PN mass quadrupole in Section~\ref{Fourier} and we define all of the enhancement factors through 3PN order in Section~\ref{enhance}. Section~\ref{Bessel} describes the details of the uniform asymptotic expansion of the Bessel functions and how to apply it to re-sum the enhancement factors, with the results given in Section~\ref{super}. In Section~\ref{memory}, we evaluate the memory terms in the fluxes to 3PN order in Section and compute the harmonic decomposition in~\ref{harmonic}. Finally, we develop the osculating approximation and discuss the results after orbit averaging in Section~\ref{osculate}. The main results of this paper are closed-form, analytic expressions for the tail terms [Eqs.~\eqref{eq:Ptails}-\eqref{eq:Qtails},~\eqref{P-tail}-\eqref{G-tail},~\eqref{varphi-super}-\eqref{vargamma-super}, \eqref{theta-super}-\eqref{tildetheta-super}, \eqref{alphasuper}-\eqref{tildechisuper}, \eqref{phi-hyper}-\eqref{tildephi-hyper}, \eqref{psi-hyper}-\eqref{chi-hyper}] 
to 3PN order. For the remainder of this paper, we use geometric units where $G = 1 = c$.

\section{Radiation Reaction in General Relativity}
\label{review}
We begin by providing a review of the mathematical structure of the GW fluxes and the machinery that is necessary to evaluate them, including a description of the quasi-Keplerian (QK) parameterization and orbit-averaging.

\subsection{Basics}
 
Consider a binary of compact objects inspiralling due to the emission of gravitational radiation. As the GWs propagate to spatial infinity, they extract orbital energy, $E$, and angular momentum, $L_{i}$, from the binary. The loss of orbital energy and angular momentum are given by the balance laws~\cite{blanchet-review},
\begin{align}
\label{balance}
\frac{dE}{dt} &= - {\cal{P}}\,,
\quad
\frac{dL_{i}}{dt} = - {\cal{G}}^{i} = - \frac{1}{2} \epsilon_{ijk} {\cal{G}}^{jk} = - {\cal{G}} \; \hat{L}^{i}\,,
\end{align}
where $\epsilon_{ijk}$ is the three-dimensional Levi-Civita symbol, while ${\cal{P}}$ and ${\cal{G}}^{jk}$ are the energy and angular momentum flux due to the emission of gravitational radiation. Typically, these quantities have two components. The first are the fluxes of energy and angular momentum that reach spatial infinity, specifically ${\cal{P}}_{\infty}$ and ${\cal{G}}^{jk}_{\infty}$. The second occurs in binary systems where at least one component is a BH, specifically the energy and angular momentum fluxes through the BH's horizon, ${\cal{P}}_{H}$ and ${\cal{G}}^{jk}_{H}$~\cite{Mino:1997bx}. However, the fluxes through the horizon are typically small compared to the fluxes to spatial infinity, and we thus neglect them here. 
 
The energy and angular momentum fluxes to spatial infinity can be calculated using the short-wave approximation~\cite{Isaacson:1968ra, Isaacson:1968gw, PW},
\begin{align}
\label{sw-P}
{\cal{P}}_{\infty} &= \frac{1}{32 \pi} \oint_{R \rightarrow \infty} R^{2} \left(\partial_{t_{R}}h^{jk}_{\rm TT} \partial_{t_{R}}h_{jk}^{\rm TT}\right) d\Omega\,,
\\
\label{sw-G}
{\cal{G}}^{jk}_{\infty} &= \frac{1}{16 \pi} \oint_{R \rightarrow \infty} R^{2} \left[2 h^{p[j}_{\rm TT} \partial_{t_{R}}h^{k] {\rm TT}}_{p} - \partial_{t_{R}} h^{pq}_{\rm TT} x^{[j}\partial^{k]} h_{pq}^{\rm TT}\right] d\Omega\,,
\end{align}
where $h_{jk}$ is a metric perturbation (away from Minkowski spacetime) that describes GWs, $t_{R} = t - R$ is the retarded time of the binary, and $\partial_{t_{R}}$ denotes a partial derivative with respect to $t_{R}$.
The TT symbol indicates that $h_{jk}$ has been projected into the transverse traceless subspace, using 
\begin{align}
P_{ijkl} &= P_{i (k} P_{l) j} - \frac{1}{2} P_{ij} P_{kl}\,,
\\
P_{ij} &= \delta_{ij} - N_{i} N_{j}
\end{align}
where $N_{i} = [{\rm sin}\theta \; {\rm cos}\phi, {\rm sin}\theta \; {\rm sin}\phi, {\rm cos}\theta]$ denotes the location of the source in the sky.

In order to calculate what these fluxes are for the binaries under consideration, we first need to compute $h_{jk}$. Using the post-Minkowskian (PM) formulation~\cite{Thorne:1980rm, Blanchet379, blanchet-review, PW} to solve the relaxed Einstein field equations, $h_{jk}$ is given to to all orders by
\begin{align}
\label{hjk}
h_{jk} &= \frac{4}{R} \sum_{l = 2}^{\infty} \frac{1}{l!} \left[N_{L-2} U_{jk}^{L-2}(\tau) - \frac{2 l}{l+1} N^{a}_{L-2} \epsilon_{ab (j} V_{k)}^{bL-2}(\tau)\right]
\end{align}
where $U_{L}$ and $V_{L}$ are the mass and current-type radiative multipole moments~\cite{Blanchet379, Thorne:1980rm, PW}. The subscript $L$ is shorthand notation for the multi-index $L = i_{1} i_{2} ... i_{l}$. Formally, the summation on $l$ goes to infinity, but in practice it is truncated at some PM order. Since the PM and PN formalisms are interconnected, the PM order of truncation is typically taken to be the same as the PN order. 

The radiative multipole moments $U_{L}$ and $V_{L}$, which are functions of retarded time, are related to the source multipole moments by matching the wave zone and the near zone metric perturbations in an overlap or buffer region. For the purposes of this paper, the most important multipoles are the mass quadrupole and octopole, and the current quadrupole. As an example, the radiative mass quadrupole to 3PN order can be broken down as follows~\cite{Blanchet:1995fg, 2002PhRvD..65f4005B, Blanchet:2001ax, Blanchet:2004ek, Faye:2012we}
\begin{align}
\label{eq:mass-quad-break-down}
U_{jk} &= U_{jk}^{\rm inst} + U_{jk}^{\rm tails} + U_{jk}^{\rm mem}\,,
\end{align}
where the instantaneous radiative mass quadrupole is related to the source multipoles by
\begin{align}
U_{jk}^{\rm inst}(t_{R}) &= \ddot{I}_{jk} + \frac{1}{7} \overset{(5)}{I}_{a<j} I_{k>}^{a} - \frac{5}{7} \overset{(4)}{I}_{a<j} \dot{I}_{k>}^{a} - \frac{2}{7} \overset{(3)}{I}_{a<j} \ddot{I}_{k>}^{a} + \frac{1}{3} \epsilon_{ab<j} \overset{\!\!\!\!\!\!\!(4)}{I_{k>}^{a}} J^{b} 
\nn \\
&+ 4 \frac{d^2}{dt^{2}} \left[\ddot{W} I_{jk} - \dot{W} \dot{I}_{jk}\right]\,.
\end{align}
In this equation, 
\begin{equation}
\label{mass-quad}
I_{jk} = \mu x_{<j} x_{k>} + {\cal{O}}({\rm 1PN})
\end{equation}
is the source mass quadrupole, $J_{b}$ is the source current dipole, $W$ is the gauge monopole moment defined in~\cite{2002PhRvD..65f4005B} and the $<>$ brackets stand for the symmetric trace-free (STF) projection. Similarly, the radiative mass octopole and current quadrupole can be written in terms of the source mass octopole and the source current quadrupole
\begin{align}
\label{mass-oct}
I_{jkl} = \mu \sqrt{1 - 4 \eta} \; x_{<j} x_{k} x_{l>} + {\cal{O}}({\rm 1PN})\,,
\\
\label{curr-quad}
J_{jk} = \mu \sqrt{1 - 4\eta} \; \epsilon_{ab<j} x_{k>} x^{a} v^{b} + {\cal{O}}({\rm 1PN})\,,
\end{align}
respectively. In Eq.~\eqref{eq:mass-quad-break-down}, the tail and memory radiative mass quadrupoles are related to the source multipoles through
\begin{align}
\allowdisplaybreaks[4]
\label{U-tail}
U_{jk}^{\rm tails}(t_{R}) &= 2 M \int_{0}^{\infty} d \tau \overset{(4)}{I}_{jk}(t_{R} - \tau) \left[{\rm ln}\left(\frac{\tau}{2 r_{0}}\right) + \frac{11}{12}\right]
\nn \\
& + 2 M^{2} \int_{0}^{\infty} d \tau \overset{(5)}{I}_{jk}(t_{R} - \tau) \left[ {\rm ln}^{2} \left(\frac{\tau}{2 r_{0}}\right) + \frac{57}{70} {\rm ln}\left(\frac{\tau}{2 r_{0}}\right) + \frac{124627}{44100}\right]\,,
\\
\label{U-mem}
U_{jk}^{\rm mem}(t_{R}) &= - \frac{2}{7} \int_{0}^{\infty} d\tau \overset{(3)}{I}_{a<j}(t_{R} - \tau) \overset{\!\!(3)}{I^{a}}_{k>}(t_{R} - \tau)\,.
\end{align}
where $r_{0}$ is a regularization factor that drops out after performing the integration and computing an observable. The contribution proportional to the ADM mass $M$ in Eq.~\eqref{U-tail} is the 1.5PN order tail term, while the term proportional to $M^{2}$ is the 3PN order tail-of-tail term. Notice that although the instantaneous contribution depends on the source multipoles evaluated only at retarded time $t_{R}$, the tail and memory contributions depend on the integrals of the source mass quadrupole over the lifetime of the source, and thus, they are referred to as \emph{hereditary} terms. 
 
Summarizing, the energy and angular momentum fluxes at spatial infinity can be computed by simply performing the TT projection of $h_{jk}$ in Eq.~\eqref{hjk}, with the radiative multipole moments expressed in terms of the source multipole moment, and inserting the result into Eqs.~\eqref{sw-P} and~\eqref{sw-G}~\cite{Arun:2007sg, Arun:2007rg, Arun:2009mc, Blanchet:1995fg, 2002PhRvD..65f4005B, Blanchet:2001ax, Blanchet:2004ek}. Since the radiative multipoles can be split into instantaneous, tail, and memory contributions, we similarly split the fluxes as
\begin{align}
\label{eq:PNdecomp-E}
{\cal{P}}_{\infty} &= {\cal{P}}_{\infty}^{\text{inst}} + {\cal{P}}_{\infty}^{\text{tails}} + {\cal{P}}_{\infty}^{\text{mem}}\,,
\\
\label{eq:PNdecomp-L}
{\cal{G}}_{\infty} &= {\cal{G}}_{\infty}^{{\text{inst}}} + {\cal{G}}_{\infty}^{{\text{tails}}} + {\cal{G}}_{\infty}^{{\text{mem}}}\,,
\end{align}
and consider each of them separately. The instantaneous fluxes have been computed previously for arbitrary eccentricity and to 3PN order in~\cite{Arun:2007sg, Arun:2009mc}, so in this paper we focus on evaluating the tail and memory fluxes for arbitrarily eccentric binaries.

\subsection{Quasi-Keplerian Parametrization}
\label{Kep}
The picture from the previous section is clear: the GWs emitted by a binary system depend on the time varying source multipole moments, and thus, if we desire to compute ${\cal{P}}_{\infty}$ and ${\cal{G}}_{\infty}$, we need to first calculate the orbital trajectories of the binary components. We recall that to obtain the fluxes to 3PN order, we only need the source moments to leading PN or \emph{Newtonian} order, with the exception of the source mass quadrupole, which must be calculated to 1PN order. For this reason, we detail below the calculation of the orbital trajectories of the binary components also to 1PN order, although this can easily be extended to higher PN order if needed. 

\begin{figure}[ht]
\centering
\includegraphics[clip=true,scale=0.34]{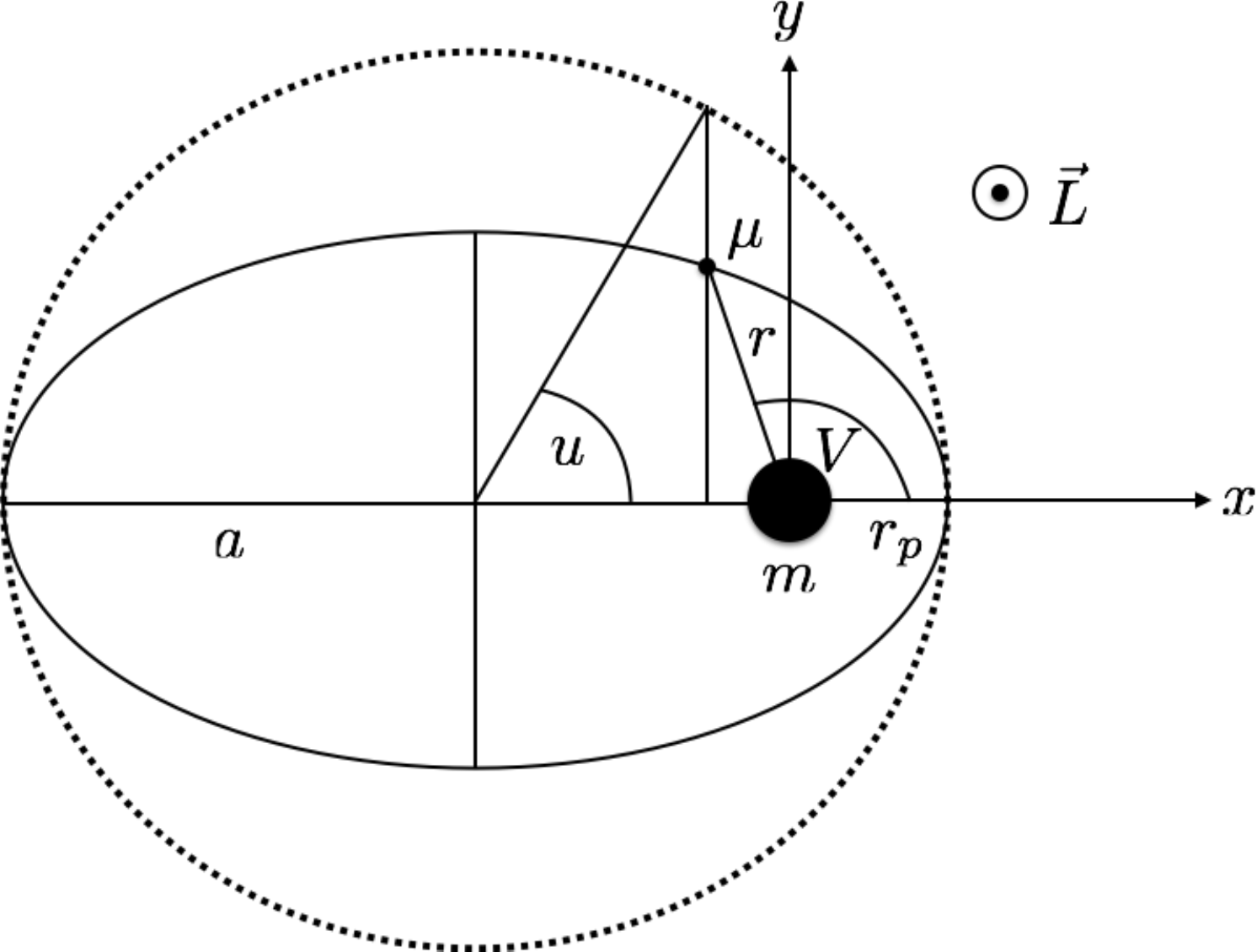}
\caption{\label{orbit} Diagram of Keplerian elliptical orbits in an effective one body frame. A point mass $\mu$ orbits around a central mass $m$ located at the focus of the ellipse. The semi-major axis $a$, pericenter distance $r_{p}$, and orbital angular momentum $\vec{L}$ are constants of the orbit when radiation reaction is neglected. The orbital radius $r$ is the line connecting $m$ and $\mu$, while the true anomaly $V$ is the angle from pericenter, which is located along the x-axis, to $r$. The eccentric anomaly $u$ is the angle measured from the x-axis to a line whose end points are the center of the ellipse and a point on a circle that is concentric to the ellipse and of radius $a$, where the end point on the circle is determine by a line parallel to the y-axis that passes through $\mu$.}
\end{figure}

We begin by describing our binary system in more detail. Let the two objects in the binary have masses $m_{1}$ and $m_{2}$. For simplicity, we work in an effective one body frame where a body of mass $m = m_{1} + m_{2}$ sits stationary at the focus of an ellipse, and a smaller mass $\mu = m_{1} m_{2} / m$ moves around the ellipse. We take the system to be orbiting in the xy-plane, so that the orbital angular momentum is aligned with the z-axis. We take the longitude of pericenter to be zero, so that pericenter is aligned with the positive x-axis. The radius of the orbit $r$ is the length of the line connecting $\mu$ and $m$, with the true anomaly $V$ being the angle between the positive x-axis and $r$. Figure~\ref{orbit} details this, as well as the definition of the eccentric anomaly $u$, which is similar to the true anomaly only defined in an elliptical coordinate system.

The QK parametrization~\cite{zbMATH03938612, zbMATH04001537, Damour:1988mr, 0264-9381-12-4-009, Schaefer1993196, Memmesheimer:2004cv} provides a solution to the PN equations of motion that govern the orbital dynamics of the binary system by treating the orbits as Keplerian ellipses in a suitably chosen precessing reference frame. There are three equations that govern the solution to the equations of motion in the QK parametrization. The first is the radial equation:
\begin{equation}
\label{r-eqn}
r = a \left[1 - e_{r} {\rm cos}(u)\right]\,,
\end{equation}
where $r$ is the radius of the orbit, $a$ is the semi-major axis of the Keplerian ellipse, $e_{r}$ is the radial eccentricity, and $u$ is the eccentric anomaly (see Fig.~\ref{orbit}). The second is the famous Kepler equation, which governs the time evolution of the eccentric anomaly,
\begin{equation}
\label{t-eqn}
\ell = u - e_{t} {\rm sin}(u) + {\cal{O}}(\text{2PN})\,,
\end{equation}
where $e_{t}$ is the temporal eccentricity, $\ell = n (t - t_{p})$ is the mean anomaly, $n = 2 \pi / P$ the mean motion where $P$ is the orbital period, and $t_{p}$ the time of pericenter passage. Finally, the azimuthal equation governs the evolution of the orbital phase $\phi$,
\begin{equation}
\label{phi-eqn}
\phi - \phi_{p} = K V(u) + {\cal{O}}(\text{2PN})\,,
\end{equation}
where $\phi_{p}$ is the longitude of pericenter, $K$ is the periastron advance due to relativistic precession, and $V(u)$ is the true anomaly given by~\cite{Damour:2004bz, Arun:2007rg}
\begin{align}
\label{V-eqn}
V(u) &= u + 2 \; \rm{arctan}\left[\frac{\beta_{\phi} {\rm sin}(u)}{1 - \beta_{\phi} {\rm cos}(u)}\right]\,,
\end{align}
with
\begin{align}
\label{beta}
\beta_{\phi} &= \frac{1 - \sqrt{1 - e_{\phi}^{2}}}{e_{\phi}}\,,
\end{align}
and $e_{\phi}$ the azimuthal eccentricity. At Newtonian order, all three of the eccentricities in the QK parametrization reduce to the Newtonian orbital eccentricity and $K = 1$. Although the radial equation [Eq.~\eqref{r-eqn}] is valid to all PN orders, Eqs.~\eqref{t-eqn} and~\eqref{phi-eqn} are only valid up to 1PN order. At higher PN orders, additional orbital elements appear in these equations that have no Newtonian analog and couple to higher harmonics of $u$ and $V$.

The mass and current-type multipole moments depend on the orbital trajectory $x_{j}$ and velocity $v_{j}$. Within the QK parametrization, these are simply
\begin{align}
\label{traj}
x_{j} &= \left[r \; {\rm cos}(\phi), r \; {\rm sin}(\phi), 0\right]\,,
\\
\label{veloc}
v_{j} &= \frac{d x_{j}}{dt}\,.
\end{align}
The trajectory $x_{j}$ can easily be expressed in terms of the eccentric anomaly through Eq.~\eqref{r-eqn} and the expressions for ${\rm cos}(\phi)$ and ${\rm sin}(\phi)$ in terms of $u$, which are given in Eq.~(3.30) in~\cite{PW}, with the additional replacement $e \rightarrow e_{\phi}$. The orbital velocity $v_{j}$ requires us to calculate $\dot{r}$ and $\dot{\phi}$, which we now detail.

The presence of the various eccentricities can often complicate the analysis of quantities in the QK parametrization, so we choose one to work with and express all of the results in terms of it. The one most commonly used in the literature is $e_{t}$ (see~\cite{Damour:2004bz, Arun:2007sg, Arun:2007rg, Arun:2009mc} for example), which is related to $e_{r}$ and $e_{\phi}$ via
\begin{align}
e_{r} &= e_{t} \left[1 + \frac{x}{2} (8 - 3 \eta)\right]\,,
\\
\label{ephi}
e_{\phi} &= e_{t} \left[1 + x (4 - \eta)\right]\,.
\end{align}
The expansion parameter in PN theory for eccentric binaries is defined via $x := (m \omega)^{2/3}$, with $\omega = K n$; 
at 1PN order, the PN expansion parameter becomes
\begin{equation}
\label{x-eqn}
x = \varepsilon - \frac{\varepsilon^{2}}{1 - e_{t}^{2}} \left[-\frac{3}{4} - \frac{\eta}{12} + e_{t}^{2} \left(-\frac{5}{4} + \frac{\eta}{12}\right)\right]\,,
\end{equation}
where $\varepsilon = -2 E/\mu > 0$ is the reduced orbital energy. We work in the regime $x \ll 1$ and expand all expressions accordingly, truncating at 1PN order. The reduced energy $\varepsilon$ is also related to the semi-major axis $a$ of the orbit through
\begin{align}
\label{a-eqn}
a = \frac{m}{\varepsilon} \left[1 + \frac{\varepsilon}{4} \left(-7 + \eta\right)\right]\,.
\end{align}
Similarly, the mean motion $n$ is
\begin{equation}
\label{n-eqn}
n = \frac{\varepsilon^{3/2}}{m} \left[1 + \frac{\varepsilon}{8} \left(-15 + \eta\right)\right]\,.
\end{equation}
By combining Eq.~\eqref{x-eqn} with Eqs.~\eqref{a-eqn} and~\eqref{n-eqn}, we can write $a$ and $n$ in terms of $x$, specifically
\begin{align}
a &= \frac{m}{x} \left[1 - x \frac{3 - \eta + e_{t}^{2} \left(-9 + \eta\right)}{1 - e_{t}^{2}}\right]\,,
\\
n &= \frac{x^{3/2}}{m} \left(1 - \frac{3 x}{1 - e_{t}^{2}}\right)\,.
\end{align}
The last variable we need to relate to $x$ is $k$, which is simply
\begin{equation}
\label{k-eqn}
k = \frac{3 x}{1 - e_{t}^{2}}\,.
\end{equation}

With all of the orbital variables expressed in terms of $e_{t}$ and $x$, we can write the QK equations in terms of those variables and $u$, specifically $r[e_{t}, x; u], \dot{r}[e_{t}, x; u]$, etc. Starting with the radial equation from Eq.~\eqref{r-eqn}, we obtain
\begin{align}
r &= \frac{m}{x} \left[1 - e_{t} \; {\rm cos}(u)\right]  - \frac{m}{6 (1 - e_{t}^{2})} \left\{6 - 2 \eta + e_{t}^{2} (-9 + \eta) 
\right.
\nn \\
&\left.
+ [18 - 7 \eta + e_{t}^{2} (-6 + 7 \eta)] e_{t} \; {\rm cos}(u)\right\}\,.
\end{align}
The temporal equation in Eq.~\eqref{t-eqn} is already expressed in terms of $e_{t}$ and $u$, so it does not require any re-writing. To obtain $\dot{r}$, we begin by taking a time derivative of Eq.~\eqref{t-eqn}, and solving for $\dot{u}$ to obtain
\begin{equation}
\dot{u} = \frac{n}{1 - e_{t} \; {\rm cos}(u)}\,.
\end{equation}
We then apply the chain rule to Eq.~\eqref{r-eqn} to obtain $\dot{r}$,
\begin{equation}
\dot{r} = \sqrt{x} \frac{e_{t} \; {\rm sin}(u)}{1 - e_{t} \; {\rm cos}(u)} \left[1 + \frac{x}{6} \frac{-7 \eta + e_{t}^{2} (-6 + 7 \eta)}{1 - e_{t}^{2}}\right]\,.
\end{equation}
The same procedure can be applied to Eq.~\eqref{phi-eqn} to obtain
\begin{equation}
\dot{\phi} = \frac{x^{3/2}}{m} \frac{\sqrt{1-e_{t}^{2}}}{[1 - e_{t} \; {\rm cos}(u)]^{2}} \left\{1 + x \frac{(-4 + \eta) e_{t} \left[e_{t} - {\rm cos}(u)\right]}{(1 - e_{t}^{2}) [1 - e_{t} \; {\rm cos}(u)]}\right\}
\end{equation}
With $\dot{r}$ and $\dot{\phi}$ specified, it is straight forward to express $v_{j}$ and $v^{2}$ as functions of $u$. This completes the calculation of all quantities necessary to evaluate the source multipole moments to the PN order required.

\subsection{Averaging and Enhancement Factors}

While we now have all of the expressions necessary in order to calculate the GW fluxes, there is still one bit of mathematical machinery that we still need to discuss. The effect of radiation reaction is to induce changes in the orbital elements of the binary. These changes can be broken down into \emph{secular} effects, which describe monotonic changes in the orbital elements, and \emph{oscillatory} effects. The oscillatory effects are typically small over one orbit, and thus one averages the fluxes over the orbital period in order to extract the more important secular changes of the orbital elements~\cite{Isaacson:1968ra, Isaacson:1968gw}:
\begin{align}
\label{P-avg}
\langle {\cal{P}}_{\infty} \rangle &= \frac{1}{P} \int_{0}^{P} {\cal{P}}_{\infty}(t) dt\,,
\\
\label{G-avg}
\langle {\cal{G}}_{\infty} \rangle &= \frac{1}{P} \int_{0}^{P} {\cal{G}}_{\infty}(t) dt\,.
\end{align}
Further, the orbit-averaged fluxes are invariant with respect to radiation reaction gauge transformations~\cite{PW}. These averaged fluxes will, in general, still depend on the coordinate system used to determine the conservative orbital dynamics. Of course, this can be averted by writing the fluxes in terms of gauge invariants of the conservative dynamics, such as the orbital energy and angular momentum, to make the averaged fluxes completely gauge invariant. As a result, it is very straightforward to obtain observables from the averaged fluxes, such as the amount of orbital energy and angular momentum that is lost per orbit.

Before we move on, however, their is one important contention that we should note about the above argument. As the eccentricity increases from the circular limit, most of the GW power gets emitted during pericenter passage. Once we move toward the parabolic limit, nearly all of the power is emitted in a short burst around pericenter and as a result, the timescale over which the orbit changes becomes much smaller than the orbital period. This is significantly different from the case described above, where the effect of radiation reaction occurs over a timescale much longer than the orbital period. As a result, orbit averaging is not really suitable when the eccentricity is sufficiently close to unity and the orbit averaged fluxes in Eqs.~\eqref{P-avg} and~\eqref{G-avg} break down~\cite{Loutrel-avg}. The eccentricity at which this occurs, however, is sufficiently close to the parabolic limit that the averaged fluxes are still a very accurate approximation of radiation reaction for a wide range of eccentricities. Indeed, we know this to be true up to at least $e_{t} = 0.7$, as can be seen from Figs.~(12.3) and (12.4) in~\cite{PW}.

The end result of the averaging procedure for an nPN order term in the fluxes generally has the form
\begin{align}
\label{eq:Efluxform}
\langle {\cal{P}}_{\infty}^{\rm (nPN)} \rangle &= \frac{32}{5} \eta^{2} x^{5+n} p_{n} f_{n}(e_{t}, \eta)\,,
\\
\label{eq:Lfluxform}
\langle {\cal{G}}^{{\rm (nPN)}}_{\infty} \rangle &= \frac{32}{5} m \eta^{2} x^{7/2 + n} l_{n} g_{n}(e_{t}, \eta)\,,
\end{align}
where $p_{n}$ and $l_{n}$ are rational numbers. When $n=0$, we obtain the Newtonian order fluxes, which were computed in~\cite{PetersMathews,Peters:1964zz}. The functions $f_{n}(e_{t}, \eta)$ and $g_{n}(e_{t}, \eta)$ are referred to as enhancement factors. In the circular limit $(e_{t} = 0)$, they are typically defined to be equal to one, with the exception of the 3PN order enhancement factors from the tail-squared and tail-of-tail terms, which are defined to be zero. As we go toward the parabolic limit $(e_{t} = 1)$, the enhancement factors diverge, thus ``enhancing" the fluxes.

The enhancement factors for the instantaneous fluxes can actually be evaluated within the time domain by performing a change of variables in the averaging integral from $t$ to $u$. The tail enhancement factors, on the other hand, cannot usually be evaluated for eccentric binaries by using the source multipole moments in the time domain due to the complicated structure of the hereditary integrals. The integrals can be done in the time domain for small eccentricities $(e_{t} \ll 1)$, but for arbitrary eccentricities, one typically has to evaluate the source multipoles in a Fourier series~\cite{Blanchet:1993ec, Arun:2007rg}. Finally, the memory enhancement factors, which only enter into the averaged angular momentum flux, involves an integral over the time varying orbital elements due to the presence of radiation reaction. In order to evaluate this, we would have to solve the equations for $\langle \dot{e}_{t} \rangle$ and $\langle \dot{x} \rangle$, which can be directly obtained from Eqs.~\eqref{P-avg} and~\eqref{G-avg}. However, due to the stated issue regarding the inapplicability of the orbit averaged procedure to sufficiently eccentric systems, such an approach would eventually breakdown. The remainder of this paper is dedicated to presenting new methods for evaluating the tail and memory enhancement factors completely analytically and in closed form to 3PN order.

\section{Fourier Decomposition of Multipole Moments}
\label{Fourier}

We here derive the Fourier decomposition of source multipole moments, which is necessary in order to evaluate the tail terms in the fluxes. We begin by reviewing how to perform the Fourier decomposition of the mass quadrupole to Newtonian order. We proceed to derive the Fourier decomposition of the mass quadrupole at 1PN order, which to our knowledge, has not been previously reported in the literature.

\subsection{Multipole Moments at Newtonian Order}

Before we derive the 1PN Fourier decomposition of the mass quadrupole, it is useful to review the Fourier decomposition at Newtonian order. At this order, the mass quadrupole is simply given by Eq.~\eqref{mass-quad}, where $x_{j}$ is the trajectory of $\mu$ for the system in Fig.~\ref{orbit}. The trajectory can easily be written in terms of the eccentric anomaly using the results from Section~\ref{Kep}.

The Fourier decomposition of the mass quadrupole at Newtonian order takes the form
\begin{equation}
I_{jk} = \sum_{p=-\infty}^{\infty} \underset{(p)}{{\cal{I}}_{jk}} e^{i p \ell}
\end{equation}
where we have truncated all expressions to leading order in $x$, used the fact that at Newtonian order $x=m/a$ and $e_{t} = e$, and where the Fourier coefficients are given by
\begin{equation}
\underset{(p)}{{\cal{I}}_{jk}} = \int_{0}^{2\pi} \frac{d\ell}{2\pi} I_{jk}(t) e^{-i p \ell}
\end{equation}
The integrals in the Fourier coefficients can be easily computed using Eq.~\eqref{t-eqn} to change the integration variable from $\ell$ to $u$, with $e_{t} = e$, and the integral representation of the Bessel function
\begin{equation}
\label{Bessel-def}
\int_{0}^{2\pi} \frac{d u}{2\pi} e^{i [p u - x\; {\rm sin}(u)]} = J_{p}(x)\,.
\end{equation}
The necessary integrals that need to be evaluated are of the form
\begin{align}
\label{cos-int}
\int_{0}^{2\pi} \frac{du}{2\pi} {\rm cos}(q u) e^{-i p [u - e \; {\rm sin}(u)]} &= \frac{1}{2} \left[J_{p-q}(p e) + J_{p+q}(p e)\right]\,,
\\
\label{sin-int}
\int_{0}^{2\pi} \frac{du}{2\pi} {\rm sin}(q u) e^{-i p [u - e \; {\rm sin}(u)]} &= \frac{1}{2 i} \left[J_{p-q}(p e) - J_{p+q}(p e)\right]\,,
\end{align}
where $q$ is a positive integer and $i$ is the imaginary number. 

There is one final step of simplification that will be crucial later on in our re-summation method: to order-reduce the Bessel functions $J_{p-q}(p e)$ and $J_{p+q}(p e)$ by using the recursion relation~\cite{NIST}
\begin{align}
\label{rec1}
J_{\nu - 1}(x) + J_{\nu +1}(x) = \frac{2 \nu}{x} J_{\nu}(x)
\end{align}
until we obtain the Fourier coefficients in terms of just $J_{p}(p e)$ and $J_{p \pm 1}(p e)$. We then replace the $J_{p \pm 1}(p e)$ terms with the derivative of $J_{p}(p e)$ with respect to its argument through~\cite{NIST}
\begin{align}
\label{rec2}
J'_{\nu}(x) &= J_{\nu - 1}(x) - \frac{\nu}{x} J_{\nu}(x)\,,
\\
\label{rec3}
J'_{\nu}(x) &= - J_{\nu + 1}(x) + \frac{\nu}{x} J_{\nu}(x)\,.
\end{align}
The end result for the Newtonian Fourier coefficients of the mass quadrupole are
\begin{align}
\label{I-Newt}
\underset{(p)}{\hat{\cal{I}}_{xx}} &= -\frac{2}{3} \frac{3 - e^{2}}{e^{2}} \frac{J_{p}(p e)}{p^{2}} + 2 \frac{1 - e^{2}}{e} \frac{J'_{p}(p e)}{p}\,,
\\
\underset{(p)}{\hat{\cal{I}}_{xy}} &= 2 i \sqrt{1 - e^{2}} \left[- \frac{1 - e^{2}}{e^{2}} \frac{J_{p}(p e)}{p} + \frac{1}{e} \frac{J'_{p}(p e)}{p^{2}}\right]\,,
\\
\underset{(p)}{\hat{\cal{I}}_{yy}} &= \frac{2}{3} \frac{3 - 2 e^{2}}{e^{2}} \frac{J_{p}(p e)}{p^{2}} - 2 \frac{1 - e^{2}}{e} \frac{J'_{p}(p e)}{p}\,,
\\
\underset{(p)}{\hat{\cal{I}}_{zz}} &= \frac{2}{3} \frac{J_{p}(p e)}{p^{2}}\,,
\end{align}
where we have introduced the reduced mass quadrupole 
\begin{equation}
\label{eq:reduced-mass-quad}
\hat{I}_{jk} \equiv \frac{x^{2}}{\mu m^{2}} I_{jk}\,.
\end{equation}
which reduces to $\hat{I}_{jk} = (\mu a^{2})^{-1} I_{jk}$ in the Newtonian limit.

When calculating the tail fluxes complete to 3PN order, one must also consider the 2.5PN order mass octopole and current quadrupole tails. The procedure for the Fourier decomposition of these multipoles, given by Eqs.~\eqref{mass-oct} and~\eqref{curr-quad}, is exactly the same at that presented above to Newtonian order. For completeness, we present the Fourier coefficients for these two multipoles in~\ref{Fourier-app}.

\subsection{Mass Quadruple at 1PN Order}

In order to complete the calculation of the fluxes to 3PN order, we need the Fourier decomposition of the mass quadrupole at 1PN order. The calculation at 1PN order is not as straightforward, due to the fact that relativistic precession modifies the structure of the Fourier decomposition. We begin by computing the mass quadrupole at 1PN order in terms of the eccentric anomaly $u$. To this order, the mass quadrupole is
\begin{equation}
\label{1PN-quad}
I_{jk} =  \mu A(v, r) x_{<j} x_{k>} + \mu r^{2} B v_{<j} v_{k>} + 2 \mu r \dot{r} C x_{<j} v_{k>}\,,
\end{equation}
where
\begin{align}
A(v, r) &= 1 + \left[v^{2} \left(\frac{29}{42} - \frac{29}{14} \eta\right) + \frac{m}{r} \left(-\frac{5}{7} + \frac{8}{7} \eta\right)\right]\,,
\\
B &= \frac{11}{21} - \frac{11}{7} \eta\,,
\qquad
C = -\frac{2}{7} + \frac{6}{7} \eta\,.
\end{align}

The first step is to write the mass quadrupole in terms of $r, v, \dot{r}, \dot{\phi},$ and $\phi$ which can be done through Eqs.~\eqref{traj} and~\eqref{veloc}. Although the variables $r, v, \dot{r},$ and $\dot{\phi}$ can all be directly related to $u$ using the results of Section~\ref{Kep}, we instead split the orbital phase $\phi$ into two terms: one with period $2\pi$ and one with period $2\pi K$, namely
\begin{equation}
\phi = K \ell + W(\ell)
\end{equation}
where at 1PN order $W = K (V - \ell)$. This allows us to separate the different harmonics of the ``magnetic type'' precession terms~\cite{Arun:2007rg}, i.e.
\begin{equation}
I_{jk} = \sum_{m=-2}^{2} \underset{(m)}{I_{jk}} \; e^{i m k \ell}
\end{equation}
where $k = K - 1$. The coefficients in this expansion, in turn, can be written in terms of STF tensors as
\begin{align}
\label{I-1PN-mag}
\underset{(m)}{I_{jk}} &= \underset{(m)}{I} \; \underset{(m)}{\textrm{M}_{jk}}
\end{align}
where 
\allowdisplaybreaks[4]
\begin{align}
\label{Ipm2}
\underset{(\pm2)}{I} &= \frac{1}{4} \mu r^{2} e^{\pm 2 i (\ell + W)} \left[1 - \frac{5 m}{7 r} + \frac{9 \dot{r}^{2}}{14} \pm \frac{10 i r \dot{r} \dot{\phi}}{21} + \frac{
 r^{2} \dot{\phi}^{2}}{6} + \eta \left(\frac{8 m}{7 r} - \frac{27 \dot{r}^{2}}{14} \mp \frac{10 i r \dot{r} \dot{\phi}}{7} - \frac{r^{2} \dot{\phi}^{2}}{2}\right)\right]
\\
\label{I0}
\underset{(0)}{I} &= \frac{1}{6} \mu r^{2} \left\{1 - \frac{5 m}{7 r} + \frac{1}{14} \left[9 \dot{r}^{2} + 17 r^{2} \dot{\phi}^{2}\right] + \eta \left[\frac{8 m}{7 r} - \frac{1}{14} \left(27 \dot{r}^{2} - 51 r^{2} \dot{\phi}^{2}\right)\right] \right\}\,,
\end{align}
and
\begin{align}
\label{eq:mSTF2}
\underset{(\pm 2)}{\textrm{M}_{jk}} &= \left[
\begin{matrix}
	1 & \mp i & 0 \\
	- & -1 & 0 \\
	- & - & 0 \\
\end{matrix}
\right]\,,
\qquad
\underset{(0)}{\textrm{M}_{jk}} = \left[
\begin{matrix}
	1 & 0 & 0 \\
	- & 1 & 0 \\
	- & - & -2 \\
\end{matrix}
\right]\,,
\end{align}
and all other coefficients vanish. We are now left with writing $\underset{(\pm2)}{I}$ and $\underset{(0)}{I}$ in terms of the eccentric anomaly $u$, which can be done straightforwardly with the results of Section~\ref{Kep}.

Before proceeding, let us discuss how to deal with the exponential terms in Eq.~\eqref{Ipm2}. The combination $\ell + W$ can be written as
\begin{align}
\ell + W &= \ell + (1 + k) (V - \ell)
\nn \\
&= V(e_{\phi}; u) + k [V(e_{\phi}; u) - u + e_{t} \; {\rm sin}(u)]\,,
\end{align}
where recall that $V$ is a function of $e_{\phi}$ and not $e_{t}$. In order to properly handle the exponentials at 1PN order, we PN expand about $k \ll 1$ (which we are free to do since $k$ is directly related to $x$) to obtain
\begin{align}
\label{lpW}
e^{\pm 2 i (\ell + W)} &= e^{\pm 2 i V(e_{\phi}; u)} \pm 2 i k e^{\pm 2 i V(e_{\phi}; u)} \left[V(e_{\phi}; u) - u + e_{t} \; {\rm sin}(u)\right]\,.
\end{align}
The quantity $k$ can be replaced with its expression in terms of $x$ without loss of accuracy. The exponential prefactor now only depends on $V$, and it can thus be immediately rewritten in terms of trigonometric functions. To 1PN order, these expressions are still given by their Newtonian counterpart, except with the replacement $e \rightarrow e_{\phi}$. After inserting these expressions into Eq.~\eqref{lpW}, we simply write $e_{\phi}$ in term of $e_{t}$ using Eq.~\eqref{ephi} and PN expand. For the 1PN terms in Eq.~\eqref{lpW} (i.e.~the terms proportional to $k$) that depend on $V$, we can immediately replace $V(e_{\phi}; u)$ with $V(e_{t}; u)$, since any correction to this generates 2PN terms.

With this at hand, we can now calculate the magnetic type quadrupole moments entirely in terms of $u$. Working with the reduced quadrupole moment of Eq.~\eqref{eq:reduced-mass-quad} to 1PN order, we have~\cite{Arun:2007rg}
\begin{align}
\label{PN-decomp}
\underset{(m)}{\hat{I}}(x, \eta; u) &= \underset{(m)}{\hat{I}^{00}}(u) + x \left[\;\underset{(m)}{\hat{I}^{01}}(u) + \eta \; \underset{(m)}{\hat{I}^{11}}(u)\right]\,.
\end{align}
By writing the moments in this way, we separate out the dependence on the system parameters $x$ and $\eta$, and are left with functions solely of $u$. We proceed by Fourier decomposing the magnetic coefficients via
\begin{align}
\label{eq:Fourier-decomp}
\underset{(m)}{\hat{I}^{ab}}(u) &= \sum_{p=-\infty}^{\infty} \;\; \underset{(p, m)}{\hat{\cal{I}}^{ab}} e^{i p \ell}\,,
\end{align}
where $(a,b) \in [0,1]$ and the Fourier coefficients are given by
\begin{equation}
\label{Fourier1PN}
\underset{(p, m)}{\hat{\cal{I}}^{ab}} = \int_{0}^{2\pi} \frac{d \ell}{2 \pi} \;\; \underset{(m)}{\hat{I}^{ab}}(u) \; e^{-i p \ell}\,.
\end{equation}
Most of these integrals can be easily computed using Eqs.~\eqref{cos-int} and~\eqref{sin-int}. The only terms that cannot be evaluated are those proportional to $V(e_{t}; u) - u$, where $V(e_{t}; u)$ is given by Eq.~\eqref{V-eqn} with the replacement $e_{\phi} \rightarrow e_{t}$. As far as we are aware, there is no closed-form expression for these integrals. We simply leave them as undetermined integrals for the time being, and detail how to evaluate them when they appear in the enhancement factors.

After evaluating the Fourier integrals, we can once again perform the reduction of order on the resulting Bessel functions using the recurrence relations in Eqs.~\eqref{rec1}-\eqref{rec3}. The end result gives the desired Fourier coefficients, specifically
\allowdisplaybreaks[4]
\begin{align}
\underset{(p, \pm2)}{\hat{\cal{I}}^{00}} &= \frac{J_{p}(p e_{t})}{2 e_{t}^{2} p^{2}} \left[-2 \pm 2 p \sqrt{1 - e_{t}^{2}} + e_{t}^{2} \left(1 \mp 2 p \sqrt{1 - e_{t}^{2}} \right)\right] \mp \frac{J'_{p}(p e_{t})}{e_{t} p^{2}} \left[\sqrt{1 - e_{t}^{2}} \mp p (1 - e_{t}^{2}) \right]
\nn \\
\label{Ipm201-Four}
\underset{(p, \pm2)}{\hat{\cal{I}}^{01}} &= - \frac{J_{p}(p e_{t})}{84 e_{t}^{2} (1 - e_{t}^{2}) p^{3}} \left[\mp 1512 - 222 p + 2268 p \sqrt{1 - e_{t}^{2}}  \mp 756 p^{2} \pm 226 p^{2} \sqrt{1 - e_{t}^{2}} - 4 p^{3} 
\right.
\nn \\
&\left.
+ 4 e_{t}^{6} p^{3} + e_{t}^{2} \left(\pm 756 + 333 p - 1260 p \sqrt{1 - e_{t}^{2}} \pm 1512 p^{2} \mp 270 p^{2} \sqrt{1 - e_{t}^{2}}  + 12 p^{3}\right) 
\right.
\nn \\
&\left.
+ e_{t}^{4} \left(57 p \mp 756 p^{2} \pm 44 p^{2} \sqrt{1 - e_{t}^{2}} - 12 p^{3}\right) \right] + \frac{J'_{p}(p e_{t})}{42 e_{t} (1 - e_{t}^{2})^{3/2} p^{3}} \left[756 \pm 111 p 
\right.
\nn \\
&\left.
\mp 1134 p \sqrt{1 - e_{t}^{2}} + 378 p^{2} - 113 p^{2} \sqrt{1 - e_{t}^{2}} \pm 2 p^{3} \mp 2 e_{t}^{6} p^{3} 
\right.
\nn \\
&\left.
+ e_{t}^{2} \left(-756 \mp 222 p \pm 819 p \sqrt{1 - e_{t}^{2}} - 756 p^{2} + 136 p^{2} \sqrt{1 - e_{t}^{2}} \mp 6 p^{3}\right) 
\right.
\nn \\
&\left. 
+ e_{t}^{4} \left(\pm 111 p + 378 p^{2} - 23 p^{2} \sqrt{1 - e_{t}^{2}} \pm 6 p^{3}\right) \right] + \frac{3}{4 (1 - e_{t}^{2})} \int_{0}^{2 \pi} \frac{d \ell}{2 \pi} e^{-i p \ell} \left[V(e_{t}; u) - u\right] 
\nn \\
&\times \left[ \pm 3 i e_{t}^{2} \mp 4 i e_{t} {\rm cos}(u) \mp i \left(-2 + e_{t}^{2}\right) {\rm cos}(2 u) + 4 e_{t} \sqrt{1 - e_{t}^{2}} \; {\rm sin}(u) - 2 \sqrt{1 - e_{t}^{2}} \; {\rm sin}(2 u)\right]\,,
\\
\underset{(p, \pm2)}{\hat{\cal{I}}^{11}} &= \frac{J_{p}(p e_{t})}{84 e_{t}^{2} p^{2}} \left[-134 \pm 146 p \sqrt{1 - e_{t}^{2}} - 12 p^{2} - 12 e_{t}^{4} p^{2} + e_{t}^{2} \left(-17 \pm 22 p \sqrt{1 - e_{t}^{2}} + 24 p^{2}\right)\right]
\nn \\
& + \frac{J'_{p}(p e_{t})}{42 e_{t} \sqrt{1 - e_{t}^{2}} p^{2}} \left[\mp 67 + 73 p \sqrt{1 - e_{t}^{2}} \mp 6 p^{2} \mp 6 e_{t}^{4} p^{2} + e_{t}^{2} \left(\pm 25 + 8 p \sqrt{1 - e_{t}^{2}} \pm 12 p^{2}\right)\right]\,,
\\
\label{Ipm011-Four}
\underset{(p, 0)}{\hat{\cal{I}}^{00}} &= - \frac{1}{3} \frac{J_{p}(p e_{t})}{p^{2}}\,,
\qquad
\underset{(p, 0)}{\hat{\cal{I}}^{01}} = - \frac{75 - 19 e_{t}^{2}}{42 (1 - e_{t}^{2}) p^{2}} J_{p}(p e_{t}) + \frac{26 e_{t}}{21 p} J'_{p}(p e_{t})\,,
\nn \\
\underset{(p, 0)}{\hat{\cal{I}}^{11}} &= \frac{17}{126 p^{2}} J_{p}(p e_{t}) - \frac{e_{t}}{21 p} J'_{p}(p e_{t})\,,
\end{align}
which completes the calculation of the 1PN mass quadrupole. How to obtain the Newtonian result in Eq.~\eqref{I-Newt} from the above expressions may not seem obvious. However, one simply has to go back to the separation of magnetic type precession terms in Eq.~\eqref{I-1PN-mag}, consider the (00) terms of the moments $\underset{(\pm2)}{I}$ and $\underset{(0)}{I}$, and take the limits $k \rightarrow 0$ and $e_{t} \rightarrow e$.

Let us conclude by summarizing the construction of the 1PN mass quadrupole moment. The latter is given by Eq.~\eqref{I-1PN-mag} in terms of STF tensors and the magnetic type quadrupole moment. The former are given in Eq.~\eqref{eq:mSTF2}, while the latter is PN decomposed in Eq.~\eqref{PN-decomp}. The coefficients of this expansion are Fourier decomposed in Eq.~\eqref{eq:Fourier-decomp}, with the Fourier coefficients given in Eqs.~\eqref{Ipm201-Four}-\eqref{Ipm011-Four}.  In the next section, we will use these 1PN mass quadrupole in the 3PN part of the fluxes. 

\section{Energy \& Angular Momentum Fluxes: Tail Effects}
\label{Tails}

We now move onto the tail energy and angular momentum fluxes. These quantities involve integrals over the entire past history of the source, i.e.~over the source's past light cone. This section presents the formal definition of these fluxes and the orbit-averaged expressions in terms of enhancement factors that depend on the Fourier decomposition of the multipole moments.   

\subsection{Integral Definitions and Orbit-Averages}

Let us begin by decomposing the tails terms in the energy and angular momentum fluxes [see Eq.~\eqref{eq:PNdecomp-E}-\eqref{eq:PNdecomp-L}] in terms of multipolar interactions:
\begin{align}
\label{eq:Ptails}
{\cal{P}}_{\infty}^{\rm tails} &= {\cal{P}}_{\infty}^{\rm MQ tail} + \left({\cal{P}}_{\infty}^{\rm MO tail} + {\cal{P}}_{\infty}^{\rm CQ tail} \right) + \left({\cal{P}}_{\infty}^{\rm MQ (tail)^{2}} + {\cal{P}}_{\infty}^{\rm MQ tail(tails)}\right)\,,
\\
\label{eq:Qtails}
{\cal{G}}_{\infty}^{\rm tails} & = {\cal{G}}_{\infty}^{{\rm MQ tail}} + \left( {\cal{G}}_{\infty}^{{\rm MO tail}} + {\cal{G}}_{\infty}^{{\rm CQ tail}}\right) + \left({\cal{G}}_{\infty}^{\rm (tail)^{2}} + {\cal{G}}_{\infty}^{\rm tail(tails)} \right).
\end{align} 
The first terms in both equations, the so-called mass quadrupole tails, first enters at 1.5PN order relative to the leading-order instantaneous fluxes. The second terms, the so-called mass octopole and current quadrupole tails, enter at 2.5PN order, while the last term, the so-called tails-squared and tails-of-tails terms, enter at 3PN order.  

The mass quadrupole tail, defined via\footnote{The quantity $r_{0}$ in the logarithm is a regularization factor that accounts for the zero point of retarded time being different between wave-zone and harmonic coordinates~\cite{Blanchet:1993ec, Arun:2007rg}; this regularization factor is unphysical and it drops out of any calculation that deals with observables.}
\begin{align}
\label{1.5-tail}
{\cal{P}}_{\infty}^{\rm MQ tail} &= \frac{4 M}{5} \dddot{I}^{jk}(t) \int_{0}^{\infty} d\tau \overset{(5)}{I}_{\!\!jk}(t - \tau) \left[{\rm ln}\left(\frac{\tau}{2 r_{0}}\right) + \frac{11}{12}\right]\,,
\\
{\cal{G}}_{\infty}^{\rm MQ tail} &= \frac{4 M}{5} \epsilon^{iab} \hat{L}_{i} \left\{\ddot{I}_{aj}(t) \int_{0}^{\infty}  d\tau \overset{(5)}{I^{j}}_{b}(t-\tau) \left[{\rm ln}\left(\frac{\tau}{2 r_{0}}\right) + \frac{11}{12}\right] 
\right.
\nn \\
&\left.
+ \dddot{I}_{bj}(t) \int_{0}^{\infty} d\tau \overset{(4)}{I^{j}}_{a}(t-\tau) \left[{\rm ln}\left(\frac{\tau}{2 r_{0}}\right) + \frac{11}{12}\right]\right\}\,,
\end{align}
is a quadratic non-linear interaction between the source quadrupole moment and the conserved mass monopole of the system, i.e.~the ADM mass
\begin{equation}
M = m \left[1 - \frac{1}{2} \eta x + {\cal{O}}({\rm 2PN})\right]\,.
\end{equation}
The mass octopole and current quadrupole tails are
\begin{align}
{\cal{P}}_{\infty}^{\rm MO tail} &= \frac{4M}{189} \overset{(4)}{I}_{\!\!jkl}(t) \int_{0}^{\infty} \!\!\!\!\!\! d\tau \overset{(6)}{I^{jkl}}(t - \tau) \left[{\rm ln}\left(\frac{\tau}{2 r_{0}}\right) + \frac{97}{60}\right]\!\!,
\\
{\cal{P}}_{\infty}^{\rm CQ tail} &= \frac{64 M}{45} \overset{(3)}{J}_{\!\!jk}(t) \int_{0}^{\infty} \!\!\!\! d\tau \overset{(5)}{J^{jk}}(t - \tau) \left[{\rm ln}\left(\frac{\tau}{2 r_{0}}\right) + \frac{7}{6}\right]\!\!,
\\
{\cal{G}}_{\infty}^{\rm MO tail} &=  \frac{2 M}{63} \epsilon^{iab} \hat{L}_{i} \left\{ \dddot{I}_{ajk} \int_{0}^{\infty} d\tau \overset{(6)}{I^{jk}}_{b}(t-\tau) \left[{\rm ln}\left(\frac{\tau}{2 r_{0}}\right) + \frac{97}{60}\right]
\right.
\nn \\
&\left.
+ \overset{(4)}{I}_{\!\!bjk}(t) \int_{0}^{\infty} d\tau \overset{(5)}{I^{jk}}_{a}(t-\tau) \left[{\rm ln}\left(\frac{\tau}{2 r_{0}}\right) + \frac{97}{60}\right]\right\}\,,
\\
{\cal{G}}_{\infty}^{\rm CQ tail} &= \frac{64 M}{45} \epsilon^{iab} \hat{L}_{i} \left\{ \ddot{J}_{aj}(t) \int_{0}^{\infty} d\tau \overset{(5)}{J^{j}}_{b}(t-\tau) \left[{\rm ln}\left(\frac{\tau}{2 r_{0}}\right) + \frac{7}{6}\right]
\right.
\nn \\
&\left.
+ \dddot{J}_{bj}(t) \int_{0}^{\infty} d\tau \overset{(4)}{J^{j}}_{a}(t-\tau) \left[{\rm ln}\left(\frac{\tau}{2 r_{0}}\right) + \frac{7}{6}\right]\right\}\,,
\end{align}
while the tail-squared and tail-of-tails terms are
\begin{align}
&{\cal{P}}_{\infty}^{\rm MQ (tail)^{2}}\!\! = \frac{4 M^{2}}{5} \left\{ \int_{0}^{\infty} d\tau \overset{(5)}{I}_{jk}(t - \tau) \left[{\rm ln}\left(\frac{\tau}{2 r_{0}}\right) + \frac{11}{12}\right]\right\}^{2},
\\
&{\cal{P}}_{\infty}^{\rm MQ tail(tails)} = \frac{4 M^{2}}{5} \overset{(3)}{I}_{jk}(t) \int_{0}^{\infty} d\tau \overset{(6)}{I^{jk}}(t-\tau)\left[{\rm ln}^{2}\left(\frac{\tau}{2 r_{0}}\right) + \frac{57}{50} {\rm ln}\left(\frac{\tau}{2 r_{0}}\right) + \frac{124627}{44100}\right],
\\
&{\cal{G}}_{\infty}^{\rm MQ (tail)^{2}} = \frac{8 M^{2}}{5} \epsilon^{iab} \hat{L}_{i} \left\{\int_{0}^{\infty} d\tau \overset{(4)}{I}_{aj}(t-\tau) \left[{\rm ln}\left(\frac{\tau}{2 r_{0}}\right) + \frac{11}{12}\right]\right\}
\nn \\
& \qquad \qquad \qquad \times \left\{\int_{0}^{\infty} d\tau \overset{(5)}{I^{j}}_{b}(t-\tau) \left[{\rm ln}\left(\frac{\tau}{2 r_{0}}\right) + \frac{11}{12}\right]\right\}\!\!,
\\
&{\cal{G}}_{\infty}^{\rm MQ tail(tail)} = \frac{4 M^{2}}{5} \epsilon^{iab} \hat{L}_{i} \left\{ \ddot{I}_{aj}(t) \int_{0}^{\infty} d\tau \overset{(6)}{I^{j}}_{b}(t-\tau) \left[ {\rm ln}^{2}\left(\frac{\tau}{2 r_{0}}\right) + \frac{57}{70} {\rm ln}\left(\frac{\tau}{2 r_{0}}\right) + \frac{124627}{44100}\right]
\right.
\nn \\
&\left.
\qquad \qquad \qquad + \dddot{I}_{bj}(t) \int_{0}^{\infty} d\tau \overset{(5)}{I^{j}}_{a}(t-\tau) \left[ {\rm ln}^{2}\left(\frac{\tau}{2 r_{0}}\right) + \frac{57}{70} {\rm ln}\left(\frac{\tau}{2 r_{0}}\right) + \frac{124627}{44100}\right]\right\}\!\!.
\end{align}

The calculation of the energy and angular momentum fluxes to 3PN order requires the mass quadrupole tail term to 1PN order, but all other tail terms to leading PN order. The former, in turn, requires the mass quadrupole moment to 1PN order, but all other multipole moments can be computed to leading PN order. In particular, we can replace any eccentricity parameter appearing in these multipoles by $e_{t}$, since PN corrections to these would enter at higher PN order. Similarly, 2PN order modifications to the orbital dynamics introduce corrections at 3.5 PN order and higher, so they can also be neglected. Radiation-reaction in the hereditary integrals can also be neglected because these effects enter at 2.5PN order in the orbital dynamics, and thus they introduce modifications in the tail fluxes at 4PN order~\cite{Blanchet:1993ec}. 

After orbit-averaging, the tail terms in the energy and angular momentum flux take the form of Eqs.~\eqref{eq:Efluxform}-\eqref{eq:Lfluxform} in terms of enhancement factors~\cite{Arun:2007rg}. In particular, the mass quadrupole tail terms are
\begin{align}
\label{P-tail}
\left<{\cal{P}}_{\infty}^{\rm MQ tail}\right> &= \frac{32}{5} \eta^{2} x^{5} \left\{4 \pi x^{3/2} \varphi(e_{t}) + \pi x^{5/2} \left[-\frac{428}{21} \alpha(e_{t}) + \frac{178}{21} \eta \theta(e_{t})\right]\right\}\,,
\\
\langle {\cal{G}}_{\infty}^{{\rm MQ tail}} \rangle &= \frac{32}{5} m \eta^{2} x^{7/2} \left\{4 \pi x^{3/2} \tilde{\varphi}(e_{t}) + \pi x^{5/2} \left[-\frac{428}{21} \tilde{\alpha}(e_{t}) + \frac{178}{21} \eta \tilde{\theta}(e_{t})\right]\right\}\,,
\end{align}
which depend on six enhancement factors: the first two $[\varphi(e_{t}),\tilde{\varphi}(e_{t})]$ come from the Newtonian contribution to the mass quadrupole and enter at 1.5PN order; the other four $[\alpha(e_{t}),\tilde{\alpha}(e_{t})]$ and $[\theta(e_{t}),\tilde{\theta}(e_{t})]$ come from the 1PN contribution to the mass quadrupole and enter at 2.5PN order. The mass octopole and current quadrupole tail terms are
\begin{align} 
\langle {\cal{P}}_{\infty}^{\rm MO tail} \rangle &= \frac{32}{5} \eta^{2} x^{5} \left[\frac{16403}{2016} \pi (1 - 4 \eta) x^{5/2} \beta(e_{t})\right]\,,
\\
\langle {\cal{P}}_{\infty}^{\rm CQ tail} \rangle &= \frac{32}{5} \eta^{2} x^{5} \left[\frac{\pi}{18} (1 - 4 \eta) x^{5/2} \gamma(e_{t})\right]\,,
\\
\langle {\cal{G}}_{\infty}^{{\rm MO tail}} \rangle &= \frac{32}{5} m \eta^{2} x^{7/2} \left[\frac{16403}{2016} \pi (1 - 4 \eta) x^{5/2} \tilde{\beta}(e_{t})\right]\,,
\\
\langle {\cal{G}}_{\infty}^{{\rm CQ tail}} \rangle &= \frac{32}{5} m \eta^{2} x^{7/2} \left[\frac{\pi}{18} (1 - 4 \eta) x^{5/2} \tilde{\gamma}(e_{t})\right]\,, 
\end{align}
which enter at 2.5PN order and depend on four enhancement factors: the first two $[\beta(e_{t}),\tilde{\beta}(e_{t})]$ come from the mass octopole and the second two $[\gamma(e_{t}),\tilde{\gamma}(e_{t})]$ come from the current quadrupole. Finally, the tail-squared and tail-of-tails terms enter at 3PN order and are given by
\begin{align} 
\langle {\cal{P}}_{\infty}^{\rm (tail)^{2}+ tail(tails)} \rangle &= \frac{32}{5} \eta^{2} x^{8} \left\{- \frac{1712}{105} \chi(e_{t}) 
\right. 
\nn \\
& \left.
+ \left[-\frac{116761}{3675} + \frac{16}{3} \pi^{2}- \frac{1712}{105} \left({\rm ln}(4 \omega r_{0}) + \gamma_{E}\right)\right] F(e_{t}) \right\}
\\
\label{G-tail}
\langle {\cal{G}}_{\infty}^{\rm (tail)^{2}+ tail(tails)} \rangle &= \frac{32}{5} m \eta^{2} x^{13/2} \left\{- \frac{1712}{105} \tilde{\chi}(e_{t}) 
\right. 
\nn \\
& \left.
+ \left[-\frac{116761}{3675} + \frac{16}{3} \pi^{2}- \frac{1712}{105} \left({\rm ln}(4 \omega r_{0}) + \gamma_{E}\right)\right] \tilde{F}(e_{t}) \right\}\,,
\end{align}
which depends on two enhancement factors $[\chi(e_{t}),\tilde{\chi}(e_{t})]$.

\subsection{Tail Enhancement Factors}
\label{enhance}

The tail terms of the orbit-averaged energy and angular momentum fluxes depend on enhancement factors that must be evaluated to calculate how compact binaries inspiral. These enhancement factors can be expressed as sums over the Fourier components of the multipole moments. The enhancement factors that enter the mass quadrupole tail terms are\linebreak
\begin{align}
\label{varphi}
\varphi(e_{t}) &= \frac{1}{32} \sum_{p=1}^{\infty} p^{7} | \underset{\!\!\!\!\!\!(p)}{{\cal{I}}_{jk}^{00}} |^{2}\,,
\qquad
\tilde{\varphi}(e_{t}) = - \frac{i}{16} \epsilon^{ijk} \hat{L}_{i} \sum_{p=1}^{\infty} p^{5} \; \underset{\!\!\!\!\!\!(p)}{{\cal{I}}_{ja}^{00}} \;\; \underset{\!\!\!\!\!\!(p)}{{\cal{I}}_{ka}^{00}}^{*}\,,
\\
\label{alpha}
\alpha(e_{t}) &= \frac{441}{3424 (1 - e_{t}^{2})} \sum_{p=1}^{\infty} p^{7} | \underset{\!\!\!\!\!\!(p)}{{\cal{I}}_{jk}^{00}} |^{2} 
- \frac{63}{3424 (1 - e_{t}^{2})} \sum_{m,s=-2}^{2} \sum_{p=1}^{\infty} p^{6} (4 s + 3 m) \; \underset{\!\!\!\!\!\!(p,m)}{{\cal{I}}_{jk}^{00}}^{*} \; \underset{\!\!\!\!\!\!(p,s)}{{\cal{I}}_{00}^{jk}}
\nn \\
&- \frac{21}{3424} \sum_{p=1}^{\infty} p^{7} \left[\; \underset{\!\!\!\!\!\!(p)}{{\cal{I}}_{jk}^{01}}^{*} \underset{\!\!\!\!\!\!(p)}{{\cal{I}}_{00}^{jk}} + \underset{\!\!\!\!\!\!(p)}{{\cal{I}}_{jk}^{00}}^{*} \underset{\!\!\!\!\!\!(p)}{{\cal{I}}_{01}^{jk}}\right]\,,
\\
\tilde{\alpha}(e_{t}) &= - \frac{189 i}{856 (1 - e_{t}^{2})} \epsilon^{ijk} \hat{L}_{i} \sum_{p=1}^{\infty} p^{6} \; \underset{\!\!\!\!\!\!(p)}{{\cal{I}}_{ja}^{00}} \; \underset{\!\!\!\!\!\!(p)}{{{\cal{I}}_{k}^{a}}^{00}}^{*} + \frac{189 i}{1712 (1 - e_{t}^{2})} \epsilon^{ijk} \hat{L}_{i} \sum_{m,s = -2}^{2} \sum_{p=1}^{\infty} p^{5} (m+s) \; \underset{\!\!\!\!\!\!(p,s)}{{\cal{I}}_{ja}^{00}} \; \underset{\!\!\!\!\!\!(p,m)}{{{\cal{I}}_{k}^{a}}^{00}}^{*}
\nn \\
&+ \frac{21 i}{1712} \epsilon^{ijk} \hat{L}_{i} \sum_{p=1}^{\infty} p^{6} \left[ \; \underset{\!\!\!\!\!\!(p)}{{\cal{I}}_{ja}^{00}} \; \underset{\!\!\!\!\!\!(p)}{{{\cal{I}}_{k}^{a}}^{01}}^{*} + \underset{\!\!\!\!\!\!(p)}{{\cal{I}}_{ja}^{01}} \; \underset{\!\!\!\!\!\!(p)}{{{\cal{I}}_{k}^{a}}^{00}}^{*}\right]
\\
\label{theta}
\theta(e_{t}) &= - \frac{21}{2848} \sum_{p=1}^{\infty} p^{7} | \underset{\!\!\!\!\!\!(p)}{{\cal{I}}_{jk}^{00}} |^{2}
+ \frac{21}{1424} \sum_{p=1}^{\infty} p^{7} \left[\; \underset{\!\!\!\!\!\!(p)}{{\cal{I}}_{jk}^{11}}^{*} \underset{\!\!\!\!\!\!(p)}{{\cal{I}}_{00}^{jk}} + \underset{\!\!\!\!\!\!(p)}{{\cal{I}}_{jk}^{00}}^{*} \underset{\!\!\!\!\!\!(p)}{{\cal{I}}_{11}^{jk}}\right]\,.
\\
\tilde{\theta}(e_{t}) &= \frac{21 i}{1424} \epsilon^{ijk} \hat{L}_{i} \sum_{p=1}^{\infty} p^{6} \underset{\!\!\!\!\!\!(p)}{{\cal{I}}_{ja}^{00}} \; \underset{\!\!\!\!\!\!(p)}{{{\cal{I}}_{k}^{a}}^{00}}^{*} - \frac{21 i}{712} \epsilon^{ijk} \hat{L}_{i} \sum_{p=1}^{\infty} p^{6} \left[\; \underset{\!\!\!\!\!\!(p)}{{\cal{I}}_{ja}^{00}} \; \underset{\!\!\!\!\!\!(p)}{{{\cal{I}}_{k}^{a}}^{11}}^{*} + \underset{\!\!\!\!\!\!(p)}{{\cal{I}}_{ja}^{11}} \; \underset{\!\!\!\!\!\!(p)}{{{\cal{I}}_{k}^{a}}^{00}}^{*}\right]
\end{align}
Those that enter the mass octopole and current quadrupole terms are
\begin{align}
\beta(e_{t}) &= \frac{20}{49209} \sum_{p=1}^{\infty} p^{9} | \underset{\!\!\!\!\!\!\!\!(p)}{\hat{\cal{I}}_{jkl}} |^{2}\,,
\\
\tilde{\beta}(e_{t}) &= - \frac{20 i}{16403} \epsilon^{ijk} \hat{L}_{i} \sum_{p=1}^{\infty} p^{8} \; \underset{\!\!\!\!\!\!\!\!(p)}{{\cal{I}}_{jab}} \; \underset{\!\!\!\!\!\!\!(p)}{{{\cal{I}}}_{k}^{ab}}^{*}\,,
\\
\gamma(e_{t}) &= 4 \sum_{p=1}^{\infty} p^{7} | \underset{\!\!\!\!\!\!(p)}{\hat{\cal{J}}_{jk}} |^{2}\,,
\\
\tilde{\gamma}(e_{t}) &= - 8 i \epsilon^{ijk} \hat{L}_{i} \sum_{p=1}^{\infty} p^{6} \underset{\!\!\!\!\!\!\!\!(p)}{{\cal{J}}_{ja}} \underset{\!\!\!\!\!\!\!\!(p)}{{\cal{J}}_{ka}}^{*}\,,
\end{align}
and those that enter the tail-squared and tail-of-tails terms are
\begin{align}
\chi(e_{t}) &= \frac{1}{64} \sum_{p=1}^{\infty} p^{8} {\rm ln}\left(\frac{p}{2}\right) | \underset{\!\!\!\!\!\!(p)}{\hat{\cal{I}}_{jk}} |^{2}\,,
\\
\tilde{\chi}(e_{t}) &= - \frac{i}{32} \epsilon^{ijk} \hat{L}_{i} \sum_{p=1}^{\infty} p^{7} {\rm ln}\left(\frac{p}{2}\right) \underset{\!\!\!\!\!\!\!(p)}{{\cal{I}}_{ja}} \underset{\!\!\!\!\!\!\!(p)}{{\cal{I}}_{ka}}^{*}\,.
\end{align}
In these expressions, a superscript asterisk stands for complex conjugation, $i$ is the imaginary number, $\hat{L}^{i}$ is the unit orbital angular momentum, and $\epsilon^{ijk}$ is the three-dimensional Levi-Civita symbol. 
 
Let us now explain how these expressions are derived. Much of this was initially done in~\cite{Arun:2007rg}, so we will not repeat the analysis here in detail, but rather sketch how the calculation is done for the 1PN mass quadrupole tail term in the energy flux (the mass quadrupole tail term in the angular momentum flux follows the exact same procedure.) Following the notation of~\cite{Arun:2007rg}, the Fourier decomposition of the 1PN mass quadrupole is
\begin{equation}
I_{jk}(t) = \sum_{m=-2}^{2} \sum_{p=-\infty}^{\infty} \;\; \underset{(p,m)}{{\cal{I}}^{jk}} e^{i (p + m k) \ell}\,,
\end{equation}
where the Fourier coefficients are
\begin{equation}
\underset{(p,m)}{{\cal{I}}^{jk}} = \underset{(p,m)}{{\cal{I}}} \; \underset{(m)}{\textrm{M}^{jk}}\,.
\end{equation}
After taking time derivatives on $I_{jk}(t)$ and inserting these into Eq.~\eqref{1.5-tail}, one finds
\begin{align}
\langle {\cal{P}}_{\infty}^{\rm MQ tail} \rangle &= \frac{4 M n^{8}}{5} \sum_{p,q=-\infty}^{\infty} \sum_{m,s=-2}^{2} (p + m k)^{3} (q + s k)^{5} \underset{(p,m)}{{\cal{I}}^{jk}} \;\; \underset{(q,s)}{{\cal{I}}_{jk}} \langle e^{i [p + q + (m + s) k] \ell} \rangle 
\nn \\
&\times \int_{0}^{\infty} d\tau e^{-i (q + s k) n \tau} \left[{\rm ln}\left(\frac{\tau}{2r_{0}}\right) + \frac{11}{12} \right]
\end{align}

We have two integrals that we have to evaluate in this expression, the orbital average and the hereditary integral. The orbital average is defined via
\begin{equation}
\langle e^{i [p + q + (m + s) k] \ell} \rangle = \int_{0}^{2\pi} \frac{d \ell}{2\pi} e^{i [p + q + (m + s) k] \ell}
\end{equation}
where $(p,q,m,s)$ are integers. Normally, this would evaluate to the discrete Kronecker delta, but the presence of $k$ in the exponential complicates things. In general, $k$ is not an integer and the end result will not take the simple form of a the Kronecker delta. To evaluate the integral, we simply expand about $k \ll 1$, to obtain
\begin{align}
\label{exp-avg}
\langle e^{i [p + q + (m + s) k] \ell} \rangle &= 
\begin{cases}
	\frac{m + s}{p + q} k & p \neq -q \\
	1 + i \pi (m + s) k & p = -q \\
\end{cases}\,,
\end{align}

The hereditary integral on the other hand is a little more involved. To evaluate this integral, we rotate the mean motion into the complex plane using $n = -i \nu$. By doing this, we replace the complex exponential in the hereditary integral with a decaying real exponential, which regularizes the behavior of the integrand when $\tau \rightarrow \infty$. The integral can then be performed using an integral table or \texttt{Mathematica}. To obtain the final answer, we rotate back using $\nu = i n$ and PN expand about $k$ to obtain
\begin{align}
\int_{0}^{\infty} d\tau &e^{-i (q + s k) n \tau} {\rm ln}\left(\frac{\tau}{2 r_{0}}\right) = \frac{i s k}{q^{2} n} + \left(1 - \frac{s k}{q}\right) \left\{-\frac{1}{qn} \left[\frac{\pi}{2} {\rm sign}(q) - i \left( {\rm ln}\left(2 n |q| r_{0}\right) + \gamma_{E}\right)\right]\right\}\,,
\end{align}
where $\gamma_{E} = 0.5772...$ is the Euler constant.

The structure of the average of the exponential in Eq.~\eqref{exp-avg} indicates that we need to be cautious when evaluating the summation over $p$. For convenience, we write the summation as
\begin{align}
\sum_{p=-\infty}^{\infty} \langle e^{i [p + q + (m + s) k] \ell} \rangle &= \sum_{p=-\infty}^{-q-1} \frac{m+s}{p+q} k + \underset{p \rightarrow -q}{\rm lim}\left[1 + i \pi (m + s) k\right] + \sum_{p=-q+1}^{\infty} \frac{m+s}{p+q} k\,.
\end{align}
The terms that depend on summations over $p$ lead to terms in $\langle {\cal{P}}_{\infty}^{\rm MQ tail} \rangle$ of the form
\begin{align}
S_{1} &= - \frac{k}{n} \sum_{m,s = -2}^{2} \sum_{q=-\infty}^{\infty} \sum_{p=-\infty}^{-q-1} \frac{m+s}{p+q} \; p^{3} q^{4} \underset{(p,m)}{{\cal{I}}^{jk}_{00}} \underset{(q,s)}{{\cal{I}}^{00}_{jk}} \left\{\frac{\pi}{2} {\rm sign}(q) - i \left[{\rm ln}(2 n |q| r_{0})+\gamma_{E}\right]\right\}\,,
\\
S_{2} &= - \frac{k}{n} \sum_{m,s = -2}^{2} \sum_{q=-\infty}^{\infty} \sum_{p=-q+1}^{\infty} \frac{m+s}{p+q} \; p^{3} q^{5} \underset{(p,m)}{{\cal{I}}^{jk}_{00}} \underset{(q,s)}{{\cal{I}}^{00}_{jk}} \left\{\frac{\pi}{2} {\rm sign}(q) - i \left[{\rm ln}(2 n |q| r_{0})+\gamma_{E}\right]\right\}\,,
\end{align}
but these vanish by noting that
\begin{equation}
\sum_{m,s = -2}^{2} (m + s) \underset{(p,m)}{{\cal{I}}^{jk}_{00}} \underset{(q,s)}{{\cal{I}}^{00}_{jk}} = 0\,.
\end{equation}
The terms that depend on the limit lead to six different terms in $\langle {\cal{P}}_{\infty}^{\rm MQ tail} \rangle$ after PN expanding in $x$. The first of these terms is
\begin{align}
L_{1} &= \frac{1}{n} \sum_{m,s=-2}^{2} \sum_{q=-\infty}^{\infty} q^{7} \underset{(-q,m)}{{\cal{I}}_{jk}^{00}} \underset{(q,s)}{{\cal{I}}_{00}^{jk}} \left\{\frac{\pi}{2} {\rm sign}(q) - i \left[{\rm ln}(2 n |q| r_{0})+\gamma_{E}\right]\right\}\,.
\end{align}
Splitting the summation over $q$ into separate sums (over positive and negative values) allows us to evaluate the sign function of $q$ individually in each sum. Next, we make the transformation $q \rightarrow -q$ in the sum over negative values, $m \rightarrow -m$ in the sum over positive values, and $s \rightarrow -s$ in the sum over negative values. After regrouping terms, we have
\begin{align}
L_{1} &= \frac{\pi}{2 n} \sum_{m,s=-2}^{2} \sum_{q=1}^{\infty} q^{7} \left[\underset{(-q,-m)}{{\cal{I}}_{jk}^{00}} \underset{(q,s)}{{\cal{I}}_{00}^{jk}} + \underset{(q,m)}{{\cal{I}}_{jk}^{00}} \underset{(-q,-s)}{{\cal{I}}_{00}^{jk}}\right] 
\nn \\
&+ i \sum_{m,s=-2}^{2} \sum_{q=1}^{\infty} q^{7} \left[\underset{(-q,-m)}{{\cal{I}}_{jk}^{00}} \underset{(q,s)}{{\cal{I}}_{00}^{jk}} - \underset{(q,m)}{{\cal{I}}_{jk}^{00}} \underset{(-q,-s)}{{\cal{I}}_{00}^{jk}}\right] \left[{\rm ln}(2 n q r_{0})+\gamma_{E}\right]\,.
\end{align}
Taking $m \leftrightarrow s$ in the second term in both of the square brackets, which we are free to do since $m$ and $s$ run over the same values, makes the term proportional to ${\rm ln}(2 n q r_{0})$ vanish, as it must since it depends on the unphysical regularization scale $r_{0}$. Reconstructing the Fourier components via 
\begin{equation}
\underset{(q)}{{\cal{I}}_{jk}^{00}} = \sum_{m=-2}^{2} \underset{(q,m)}{{\cal{I}}_{jk}^{00}}\,.
\end{equation}
one finds
\begin{equation}
L_{1} = \frac{\pi}{n} \sum_{p=1}^{\infty} p^{7} | \underset{(p)}{{\cal{I}}_{jk}^{00}} |^{2}\,,
\end{equation}
where we have replaced $q$ with $p$ and used the fact that the mass quadrupole is a real valued function, so the Fourier coefficients satisfy
\begin{equation}
\underset{(p,m)}{{\cal{I}}_{jk}^{00}}^{*} =  \underset{(-p,-m)}{{\cal{I}}_{jk}^{00}}\,.
\end{equation}

The analysis of the remaining terms follows the exact same procedure. The non-vanishing terms are
\begin{align}
L_{2} &= \frac{3 \pi x}{(1 - e_{t}^{2}) n} \sum_{m,s=-2}^{2} \sum_{p=1}^{\infty} p^{6} (4s + 3m) \underset{(p,m)}{{\cal{I}}_{jk}^{00}}^{*} \underset{(p,s)}{{\cal{I}}_{00}^{jk}}\,,
\\
L_{3} &= \frac{\pi x}{n} \sum_{p=1}^{\infty} p^{7} \left[\underset{(p)}{{\cal{I}}_{jk}^{01}}^{*} \underset{(p)}{{\cal{I}}_{00}^{jk}} + \underset{(p)}{{\cal{I}}_{jk}^{00}}^{*} \underset{(p)}{{\cal{I}}_{01}^{jk}}\right]\,,
\\
L_{4} &= \frac{\pi x \eta}{n} \sum_{p=1}^{\infty} p^{7} \left[\underset{(p)}{{\cal{I}}_{jk}^{11}}^{*} \underset{(p)}{{\cal{I}}_{00}^{jk}} + \underset{(p)}{{\cal{I}}_{jk}^{00}}^{*} \underset{(p)}{{\cal{I}}_{11}^{jk}}\right]\,.
\end{align}
The remaining two terms vanish either directly from the simplification procedure or due to the summations over $m$ and $s$, just like $S_{1}$ and $S_{2}$ did. To obtain the result in Eq.~\eqref{P-tail}, one simply has to regroup terms to obtain the desired enhancement factors.

\section{Tail Fluxes: Resummation of Asymptotic Enhancement Factors}
\label{Bessel}

The enhancement factors are given in terms of infinite sums over the Fourier components of the multipole moments, which is is not a practical representation for evaluation of a compact binary inspiral. In particular, if the binary is highly elliptical, a very large number of terms would have to be kept to obtain an accurate representation of the fluxes.  This section details how to resum these infinite sums through uniform asymptotic expansions. The truncation of these will lead to superasymptotic series (i.e.~an optically truncated asymptotic expansion), and by correcting the behavior of these series at small eccentricity, we will arrive at hyperasymptotic series. We conclude this section by comparing our resummed results to numerically-evaluated tail terms in the fluxes.

\subsection{Asymptotic Resummation Method for the Enhancement factors}

The structure of the Fourier coefficients in Section~\ref{Fourier} shows that they depend on the Bessel functions $J_{p}(p e_{t})$ and $J'_{p}(p e_{t})$. Sums that involve these particular Bessel functions are referred to as Kapteyn series~\cite{Watson}. There are a host of techniques, both exact and approximate, for re-summing Kapteyn series, many of which are detailed in~\cite{Nikishov:2013}. We focus on one particular method which relies on the asymptotic properties of $J_{p}(p e_{t})$. The re-summation procedure is the following:
 \begin{enumerate}
 	\item Replace the Bessel functions $J_{p}(p e_{t})$ and $J'_{p}(p e_{t})$ in the tail enhancement factors with their uniform asymptotic expansions.
	\item Replace the summation over $p$ with an integral.
	\item Series expand the result of the integrals about $\epsilon = 1 - e_{t}^{2} \ll 1$.
\end{enumerate}

The first step requires the uniform asymptotic expansion of the Bessel functions as $p \to \infty$~\cite{NIST}, namely
\begin{align}
\label{J-eqn}
J_{p}(p e_{t}) &= \left(\frac{4 \zeta}{1 - e_{t}^{2}}\right)^{1/4} \left[\frac{{\rm Ai}\left(p^{2/3} \zeta\right)}{p^{1/3}} \sum_{k=0}^{\infty} \frac{a_{k}\left(\zeta\right)}{p^{2k}} + \frac{{\rm Ai}'\left(p^{2/3} \zeta\right)}{p^{5/3}} \sum_{k=0}^{\infty} \frac{b_{k}(\zeta)}{p^{2k}}\right]\,,
\\
\label{JP-eqn}
J'_{p}(p e_{t}) &= - \frac{2}{e_{t}} \left(\frac{1 - e_{t}^{2}}{4 \zeta}\right)^{1/4} \left[\frac{{\rm Ai}\left(p^{2/3} \zeta\right)}{p^{4/3}} \sum_{k=0}^{\infty} \frac{c_{k}\left(\zeta\right)}{p^{2k}} + \frac{{\rm Ai}'\left(p^{2/3} \zeta\right)}{p^{2/3}} \sum_{k=0}^{\infty} \frac{d_{k}(\zeta)}{p^{2k}}\right]\,,
\end{align}
which is valid uniformly for $e_{t} \in (0,1)$, where ${\rm Ai}$ and ${\rm Ai'}$ are the Airy function and its derivative, and 
\begin{equation}
\zeta = \left[\frac{3}{2} {\rm ln}\left(\frac{1 + \sqrt{1 - e_{t}^{2}}}{e_{t}}\right) - \frac{3}{2}  \sqrt{1 - e_{t}^{2}}\right]^{2/3}\,.
\end{equation}
We can replace the Airy functions with their representations as modified Bessel functions of the second kind, specifically~\cite{NIST}
\begin{align}
{\rm Ai}(x) &= \frac{1}{\pi} \sqrt{\frac{x}{3}} K_{1/3}\left(\frac{2}{3} x^{3/2}\right)\,,
\\
{\rm Ai}'(x) &= - \frac{1}{\pi} \frac{x}{\sqrt{3}} K_{2/3}\left(\frac{2}{3} x^{3/2}\right)\,.
\end{align}
The functions $(a_{k}, b_{k}, c_{k}, d_{k})$ are given by
\begin{align}
a_{k} &= \sum_{s=0}^{2k} \mu_{s} \zeta^{-3s/2} u_{2 k - s}\left[(1 - e_{t}^{2})^{-1/2}\right]
\\
b_{k} &= -\zeta^{-1/2} \sum_{s=0}^{2k+1} \lambda_{s} \zeta^{-3s/2} u_{2k-s+1}\left[(1 - e_{t}^{2})^{-1/2}\right]
\\
c_{k} &= -\zeta^{1/2} \sum_{s=0}^{2k+1} \mu_{s} \zeta^{-3s/2} v_{2k-s+1}\left[(1 - e_{t}^{2})^{-1/2}\right]
\\
d_{k} &= \sum_{s=0}^{2k} \lambda_{s} \zeta^{-3s/2} v_{2k-s}\left[(1 - e_{t}^{2})^{-1/2}\right]
\end{align}
with the coefficients
\begin{align}
\mu_{s} &= - \frac{6s+1}{6s-1} \lambda_{s}\,,
\\
\lambda_{s} &= \frac{1}{(144)^{s} s!} \prod_{j=0}^{2s-1} (2s + 2j +1)\,,
\end{align}
and the functions $u_{k}$ and $v_{k}$ are defined as
\begin{align}
u_{k+1}(t) &= \frac{1}{2} t^{2} \left(1 - t^{2}\right) \frac{d u_{k}(t)}{d t} + \frac{1}{8} \int_{0}^{t} dz \left(1 - 5 z^{2}\right) u_{k}(z)\,,
\\
v_{k+1}(t) &= u_{k+1}(t) + t \left(t^{2} - 1\right)\left[\frac{1}{2} u_{k}(t) + t \frac{d u_{k}(t)}{d t}\right]\,.
\end{align}

Are we justified in replacing Bessel functions by an expansion about $p = \infty$? The accuracy of typical Taylor series is usually very poor when evaluating the series far from its expansion point. However, the series in question is asymptotic, and thus, its properties are very different from standard convergent power series. Asymptotic series generally approximate a function faster than a convergent power series provided one keeps the right number of terms. This is sometimes summarized by Carrier's rule: ``divergent series converge faster than convergent series because they do not have to converge''~\cite{Boyd}. What is meant by this is that we need only keep a few terms in the asymptotic expansion to obtain a good approximation to the function in question (but if we keep more terms than optimal, then the series will typically diverge). 

\begin{figure}[ht]
\centering
\includegraphics[clip=true,scale=0.34]{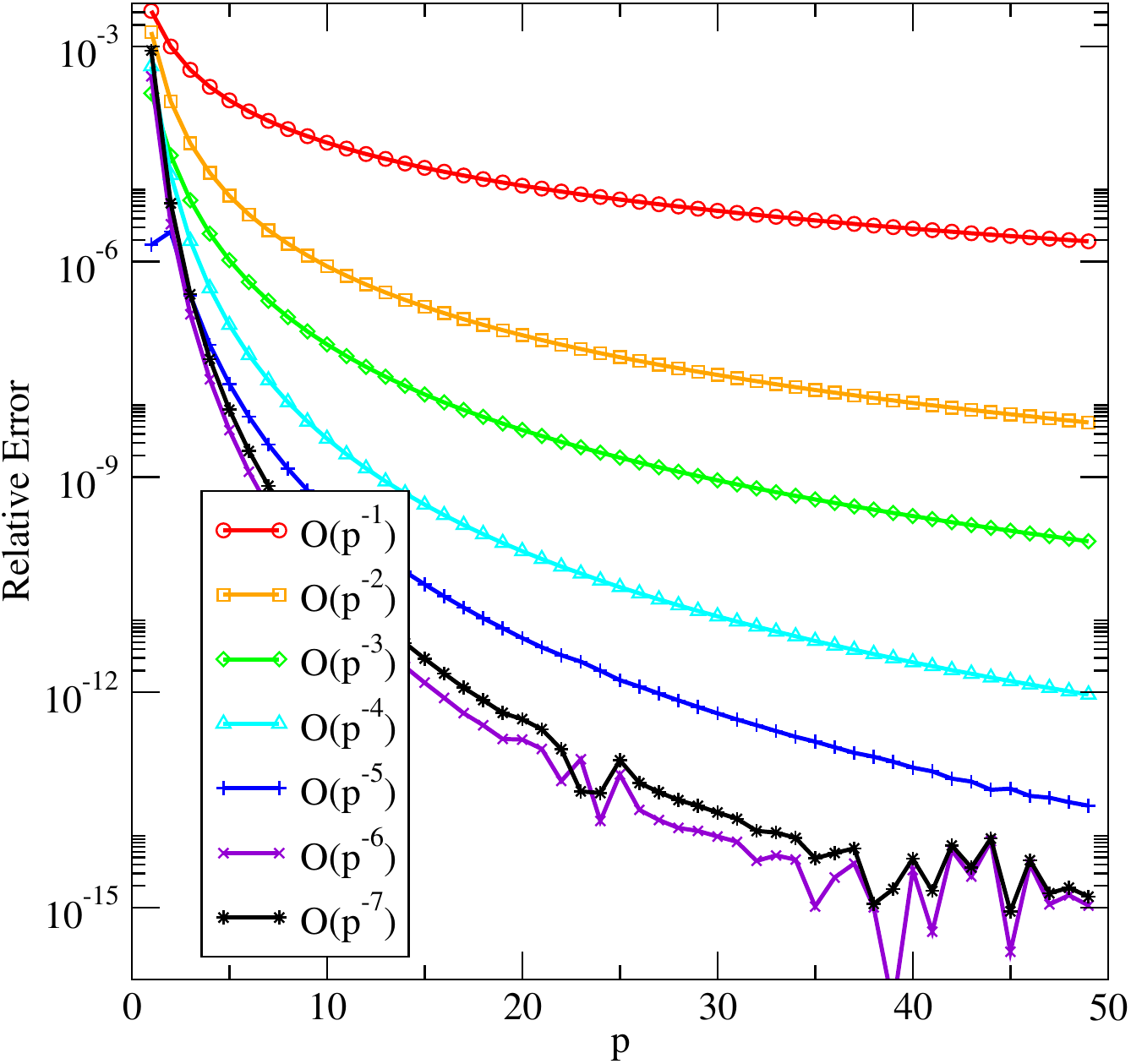}
\caption{\label{converge} Accuracy of the uniform asymptotic expansion of the Bessel function $J_{p}(p e_{t})$ as a function of $p$ and for eccentricity $e_{t} = 0.9$. Going to higher order in the asymptotic expansion causes the series to become more accurate compared to the Bessel function, but at sixth order in $1/p$, the relative error achieve a minimum.}
\end{figure}

For our purposes, this can be shown by comparing $J_{p}(p e_{t})$ with its asymptotic expansion at different orders. Figure~\ref{converge} shows the relative error between the Bessel function and its asymptotic expansion, up to seventh order, using an eccentricity of $e_{t} = 0.9$. Observe that at sixth and seventh order, the accuracy of the asymptotic series rapidly approaches machine precision for $p>30$, while still being highly accurate for values $p \le 30$. Observe also that the series achieves a minimum error at sixth order and that going to higher order only results in worse accuracy. This is a typical feature of divergent asymptotic series, which signals the order of an optimal asymptotic expansion or a \emph{superasymptotic expansion} for short. It is worth noting that these properties change depending on the value of $e_{t}$. The smaller $e_{t}$ is, the higher the order in the expansion that is needed to obtain the superasymptotic expansion. Further the value of $p$ where the error becomes comparable to machine precision is also dependent on $e_{t}$. For the case considered in Fig.~\ref{converge}, the properties do not change provided $e_{t} \in (0.85,0.95)$, but one would need more (less) terms if one wished to consider smaller (larger) values of the eccentricity. As we will show below, the order to which we keep the asymptotic expansion of the Bessel function is not crucial, provided it is high enough to allow for a robust calculation of the superasymptotic expansions for the enhancement factors. 

One may wonder whether one can replace the summations with integrals of the form  
\begin{equation}
\sum_{p=1}^{\infty} \rightarrow \int_{1}^{\infty} dp\,,
\end{equation}
when the asymptotic expansion is formally only valid as $p \to \infty$.  The Fourier decomposition can be thought of as a way to modify circular orbits through the introduction of epicycles of frequency $\omega = n p$ in such a way to generate an elliptical orbit. If the eccentricity is small, then only a finite number of terms contribute to the overall Fourier series. However, in the limit as $e_{t} \rightarrow 1$, an infinite number of epicycles contributes. The spacing between consecutive values of the Fourier index $\delta p =1$ becomes infinitesimal in this limit and the summation collapses to an integral without loss of accuracy. Hence, we are justified in taking the summation to an integral provided we work in the high-eccentricity limit. This is the reason for the third step in the re-summation procedure, i.e.~expanding the result of the integral about small eccentricity $\epsilon = 1 - e_{t}^{2} \ll 1$.

A final issue of practice, not principle, is whether the integrals that are obtained after replacing the sums can actually be evaluated in closed form. As it turns out, the integrals take the form
\begin{align}
\label{integral}
\int_{1}^{\infty} dp \; p^{n} K_{a}\left(\frac{2}{3} \zeta^{3/2} p\right) K_{b}\left(\frac{2}{3} \zeta^{3/2} p\right)\,,
\end{align}
where $(a,b)$ are either $1/3$ or $2/3$. This structure occurs in all of the enhancement factors except for those at 3PN order, which also contain a logarithmic term in the integrands. In all cases, these integrals can be evaluated in closed form in terms of hypergeometric functions with integral tables~\cite{Gradshteyn} or \texttt{Mathematica}. 

The high-eccentricity expansion of the closed-form expression of these integrals depends on the sign of $n$. At low order in the asymptotic expansion of the Bessel functions [Eqs.~\eqref{J-eqn} and~\eqref{JP-eqn}] one typically generates terms with $n > 0$, but at sufficiently high order, terms with $n<0$ are also generated. When $n$ is positive, the integrand resembles a Gaussian that peaks at a value $n > 1$, and the larger the eccentricity, the higher in $p$ the integrand peaks at, as expected from the epicycle nature of the Fourier decomposition. In such a case, the lower $n$ is, the higher order in $\epsilon$ the integrated result becomes, as shown in Table~\ref{lo}; this table also shows that the leading-order in $\epsilon$ behavior of the integrated result is independent of $a$ and $b$. 

When $n$ is negative, the integrand peaks at $p=1$ for all values of eccentricity, and to leading order in $1/p$, the integrand becomes
\begin{align}
p^{-|n|} K_{a}\left(\frac{2}{3} \zeta^{3/2} p\right) K_{b}\left(\frac{2}{3} \zeta^{3/2} p\right) &\sim \frac{3^{a+b}\Gamma(a) \Gamma(b)}{4 p^{a+b+|n|} \zeta^{(3/2)(a + b)}}\,,
\end{align}
where observe that the power of $\zeta$ does not depend on $n$. Since $\zeta$ is the only quantity that contains any eccentricity dependence in the above expression, all $n<0$ terms contribute at the same eccentricity order. In fact, when $e_{t} \sim 1$, $\zeta \sim \epsilon$, and the leading order expansion in $\epsilon$ of the integral is
\begin{equation}
\int_{1}^{\infty} dp \; p^{-|n|} K_{a}\left(\frac{2}{3} \zeta^{3/2} p\right) K_{b}\left(\frac{2}{3} \zeta^{3/2} p\right) \sim \epsilon^{-(3/2)(a+b)}\,.
\end{equation}
Thus, when $n<0$, all terms with the same values of $(a,b)$ in the asymptotic expansion enter at the same order in $\epsilon$, effectively generating an infinite number of terms starting at orders $\epsilon^{-2}, \epsilon^{-3/2}$ and $\epsilon^{-1}$, depending on the term considered in the asymptotic expansion. We refer to the lowest order in $\epsilon$ at which this happens for a given term in the asymptotic expansion as the \emph{breakdown order}. 

{\renewcommand{\arraystretch}{1.2}
\begin{table}
\centering
\begin{centering}
\begin{tabular}{ccccc}
\hline
\hline
\noalign{\smallskip}
	 $ n=8 $  & $ n=7 $ & $ n=6 $ & $ n=5 $ & $ n=4 $ \\ 
\hline
\noalign{\smallskip}
	$\epsilon^{-27/2}$ & $\epsilon^{-12}$ & $\epsilon^{-21/2}$  & $\epsilon^{-9}$ & $\epsilon^{-15/2}$  \\ 
\noalign{\smallskip}	
\hline
\hline
\end{tabular}
\end{centering}
\caption{\label{lo} Leading order dependence on $\epsilon = 1 - e_{t}^{2}$ of the integrals in Eq.~\eqref{integral} for various positive powers of $n$.}
\end{table}}

This might seem like an unsurmountable problem for the re-summation procedure. The previous discussion, however, has left out a few important considerations that ameliorate the problem. The first is that the above analysis neglected the pre-factors of the integral in Eq.~\eqref{integral}, which also depend on eccentricity. For the enhancement factors being considered, these pre-factors take the breakdown order to higher order in $\epsilon$ [usually ${\cal{O}}(\epsilon^{0})$, but sometimes higher order in $\epsilon$ for example in the 3PN enhancement factor $\tilde{\chi}(e_{t})$.] Thus, if we can obtain accurate superasymptotic series expressions for the enhancement factors that terminate at an order lower than the breakdown order, then we will not need to concern ourselves with this problem. As we show in the next section, this is in fact the case for the enhancement factors considered in this work. 

A second argument for why this apparent problem is not present in practice is the following. The uniform asymptotic expansion of the Bessel functions is actually divergent, so at the breakdown order, the more terms one keeps, the worse the approximation becomes. One could thus determine the superasymptotic expansion by comparing the asymptotic approximation to a numerical result, thus finding the optimal number of terms at the breakdown order that would need to be kept. As said in the previous paragraph, however, one does not need to worry about this to the order in $\epsilon$ we work. For the analysis of the enhancement factors to 3PN order, it is sufficient to cut the expansion of the Bessel functions in Eqs.~\eqref{J-eqn} and~\eqref{JP-eqn} at seventh order in $1/p$ ($k = 3$).

\subsection{Superasymptotic Enhancement Factors}
\label{super}

The resummation method detailed above generates asymptotic series in $\epsilon$ for the enhancement factors that must be optimally truncated to obtain the best approximation to the exact enhancement factors. The truncation of the asymptotic series at its optimal order generates a superasymptotic series for the enhancement factors. Let us then construct such superasymptotic series by first carrying out a uniform asymptotic expansion of the Bessel functions to seventh order, then performing the integrals over the Fourier index and series expanding the resulting expressions about $\epsilon \ll 1$, and finally truncating the resulting asymptotic series in $\epsilon$ (at a given order in $\epsilon$) to compare them to the numerical enhancement factors and determine the optimal truncation order.

The numerical enhancement factors that we compare the asymptotic expansions to are computed by directly evaluating the Bessel sums numerically, thus breaking from previous methods used in~\cite{Arun:2007rg}. To do this, we define a numerical tolerance $\delta$ and mandate that for any enhancement factor $E(e_{t})$,
\begin{equation}
\left| \frac{E_{Q+1}(e_{t})}{\sum_{q=1}^{Q} E_{q}(e_{t})} \right| < \delta\,,
\end{equation}
where $E_{q}$ is the summand in the enhancement factors of Sec.~\ref{Fourier} (see e.g.~Eq.~\eqref{varphi}). For example, for $\varphi(e_{t})$,
\begin{align}
\varphi_{q}(e_{t}) &= \frac{q^{3}}{12 e_{t}^{4}} J^{2}_{q}(q e_{t}) \left[-3 e_{t}^{6} q^{2} + 3 (1 + q^{2}) - 3 e_{t}^{2} (1 + 3 q^{2}) + e_{t}^{4} (1 + 9 q^{2})\right] 
\nn \\
&+ \frac{(1 - e_{t}^{2}) q^{3}}{4 e_{t}^{2}} J'^{2}_{q}(q e_{t}) \left[1 + (1 - e_{t}^{2}) q^{2}\right] - \frac{q^{4}}{4 e_{t}^{3}} J_{q}(q e_{t}) J'_{q}(q e_{t}) \left[4 - 7 e_{t}^{2} + 3 e_{t}^{4}\right]\,.
\end{align}
We adopt $\delta = 10^{-15}$ for our computation. We have verified that these expressions are consistent with the data of~\cite{Arun:2009mc}. 

\begin{figure}[ht]
\centering
\includegraphics[clip=true,scale=0.34]{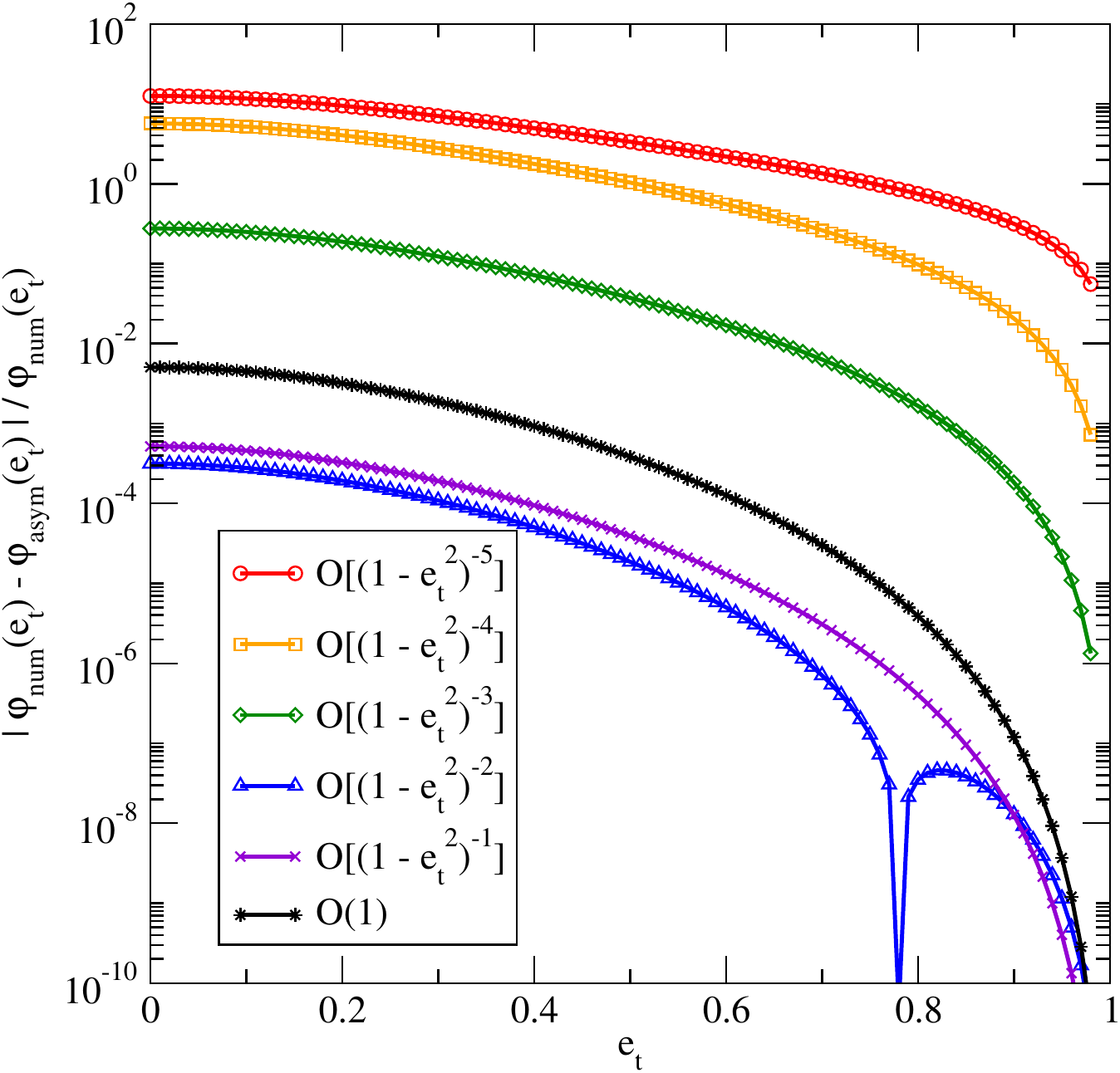}
\caption{\label{phi-err} Accuracy of the asymptotic series for the enhancement factor $\varphi(e_{t})$. The relative error to numerical results increases as we go to higher order in the series, but reaches a minimum at ${\cal{O}}\left[(1-e_{t}^{2})^{-2}\right]$. The series can thus be optimally truncated at this order, thus generating the superasymptotic series for $\varphi(e_{t})$.}
\end{figure}

Let us begin by comparing the numerical and the asymptotic series for the $\varphi(e_{t})$ enhancement factor in Figure~\ref{phi-err}. Observe that the minimum error occurs when truncating the series at ${\cal{O}}(\epsilon^{-2})$ for $e_{t} \lesssim 0.9$ and ${\cal{O}}(\epsilon^{-1})$ for $e_{t} \gtrsim 0.9$. This change in the order of the optimally truncated series can sometimes happen with asymptotic series; in our case, however, the turnover occurs at a relatively small overall error $< 10^{-8}$, and thus the difference generated by switching orders is negligible. As a result, we choose the $\varphi(e_{t})$ superasymptotic series to include terms up to ${\cal{O}}(\epsilon^{-2})$ for all values of $e_{t} < 1$: 
\begin{align}
\label{varphi-super}
\varphi_{\rm super}(e_{t}) &= \frac{1}{\Gamma \left(\frac{1}{3}\right) \Gamma \left(\frac{2}{3}\right)} \left[\frac{1328}{27 (1 - e_{t}^{2})^{5}} - \frac{992}{15 (1 - e_{t}^{2})^{4}} + \frac{33982}{1575 (1 - e_{t}^{2})^{3}} - \frac{1577}{1575 (1 - e_{t}^{2})^{2}}\right]\,.
\end{align}
Observe that although the superasymptotic series is most accurate when $e_{t} \sim 1$ (with an accuracy of ${\cal{O}}(10^{-10})$), it can be as accurate as $10^{-3} - 10^{-4}$ in the limit $e_{t} \rightarrow 0$, a key property of asymptotic series. 

This analysis can be repeated for the remaining enhancement factors to obtain superasymptotic series. For the 1.5PN order enhancement factor for the mass quadrupole tail in the angular momentum flux, we find
\begin{align}
\label{varphit-super}
\tilde{\varphi}_{\rm super}(e_{t}) &= \frac{1}{\Gamma \left(\frac{1}{3}\right) \Gamma \left(\frac{2}{3}\right)} \left[\frac{16}{(1 - e_{t}^{2})^{7/2}} - \frac{206}{15 (1 - e_{t}^{2})^{5/2}} + \frac{47}{35 (1 - e_{t}^{2})^{3/2}} + \frac{1}{50 (1 - e_{t}^{2})^{1/2}}\right]\,.
\end{align}
Notice that the powers of $1 - e_{t}^{2}$ are half-integer powers, while the powers in $\varphi(e_{t})$ are integer powers. This property occurs for all enhancement factors that enter the angular momentum flux at half integer PN orders. Similarly, for the 2.5PN enhancement factors coming from the mass octopole and current quadrupole tails, we find
\begin{align}
\beta_{\rm super}(e_{t}) &= \frac{1}{\Gamma \left(\frac{1}{3}\right) \Gamma \left(\frac{2}{3}\right)} \left[ \frac{7244800}{49209 (1 - e_{t}^{2})^{6}} - \frac{39162880}{147627 (1 - e_{t}^{2})^{5}} + \frac{16731136}{114821 (1 - e_{t}^{2})^{4}} 
\right.
\nn \\
&\left.
- \frac{14146304}{574105 (1 - e_{t}^{2})^{3}} + \frac{1052528}{1722315 (1 - e_{t}^{2})^{2}} - \frac{260144}{2052425375 (1 - e_{t}^{2})}\right]\,,
\\
\gamma_{\rm super}(e_{t}) &= \frac{1}{\Gamma \left(\frac{1}{3}\right) \Gamma \left(\frac{2}{3}\right)} \left[\frac{1280}{3 (1 - e_{t}^{2})^{6}} - \frac{7808}{9 (1 - e_{t}^{2})^{5}} + \frac{97472}{175 (1 - e_{t}^{2})^{4}} - \frac{20368}{175 (1 - e_{t}^{2})^{3}} 
\right.
\nn \\
&\left.
+ \frac{113228}{28875 (1 - e_{t}^{2})^{2}} - \frac{122}{625625 (1 - e_{t}^{2})}\right]
\end{align}
for the energy flux and
\begin{align} 
\tilde{\beta}_{\rm super}(e_{t}) &= \frac{1}{\Gamma \left(\frac{1}{3}\right) \Gamma \left(\frac{2}{3}\right)} \left[ \frac{6732800}{147627 (1 - e_{t}^{2})^{9/2}} - \frac{988160}{16403 (1 - e_{t}^{2})^{7/2}} + \frac{2192128}{114821 (1 - e_{t}^{2})^{5/2}} 
\right. 
\nn \\
&\left.
- \frac{40640}{49209 (1 - e_{t}^{2})^{3/2}} - \frac{93424}{31575775 (1 - e_{t}^{2})^{1/2}}\right]
\\
\label{vargamma-super}
\tilde{\gamma}_{\rm super}(e_{t}) &= \frac{1}{\Gamma \left(\frac{1}{3}\right) \Gamma \left(\frac{2}{3}\right)} \left[\frac{640}{9 (1 - e_{t}^{2})^{9/2}} - \frac{512}{5 (1 - e_{t}^{2})^{7/2}} + \frac{6448}{175 (1 - e_{t}^{2})^{5/2}} - \frac{1024}{525 (1 - e_{t}^{2})^{3/2}} 
\right.
\nn \\
&\left.
+ \frac{14}{1375 (1 - e_{t}^{2})^{1/2}} + \frac{568}{56875} (1 - e_{t}^{2})^{1/2}\right]\,,
\end{align}
for the angular momentum flux.  Notice that the optimal truncation order can change when considering different enhancement factors, and recall that this optimal order is here determined by comparing each of the asymptotic enhancement factors to numerical factors, as explained in the $\varphi(e_{t})$ case. 

For the 2.5PN order enhancement factors that come from the mass quadrupole tail, specifically $[\alpha(e_{t}),\tilde\alpha(e_{t})]$ and $[\theta(e_{t}),\tilde\theta(e_{t})]$, it is easiest to break each of them into the distinct sums shown in Eq.~\eqref{alpha} and~\eqref{theta}, and consider the re-summation of each summation individually. Let us consider $[\theta(e_{t}),\tilde\theta(e_{t})]$ first, which can be split as follows
\begin{align}
\theta(e_{t}) &= \theta_{1}(e_{t}) + \theta_{2}(e_{t})\,,
\\
\tilde{\theta}(e_{t}) &= \tilde{\theta}_{1}(e_{t}) + \tilde{\theta}_{2}(e_{t})\,,
\\
\theta_{1}(e_{t}) &= - \frac{21}{2848} \sum_{p=1}^{\infty} p^{7} | \underset{(p)}{{\cal{I}}_{jk}^{00}} |^{2}
\\
\theta_{2}(e_{t}) &= \frac{21}{1424} \sum_{p=1}^{\infty} p^{7} \left[ \underset{(p)}{{\cal{I}}_{jk}^{11}}^{*} \underset{(p)}{{\cal{I}}_{00}^{jk}} + \underset{(p)}{{\cal{I}}_{jk}^{00}}^{*} \underset{(p)}{{\cal{I}}_{11}^{jk}}\right]\,,
\\
\tilde{\theta}_{1}(e_{t}) &= \frac{21 i}{1424} \epsilon^{ijk} \hat{L}_{i} \sum_{p=1}^{\infty} p^{6} \underset{(p)}{{\cal{I}}_{ja}^{00}} \; \underset{(p)}{{{\cal{I}}_{k}^{a}}^{00}}^{*}\,,
\\
\tilde{\theta}_{2}(e_{t}) &= - \frac{21 i}{712} \epsilon^{ijk} \hat{L}_{i} \sum_{p=1}^{\infty} p^{6} \left[ \underset{(p)}{{\cal{I}}_{ja}^{00}} \; \underset{(p)}{{{\cal{I}}_{k}^{a}}^{11}}^{*} + \underset{(p)}{{\cal{I}}_{ja}^{11}} \; \underset{(p)}{{{\cal{I}}_{k}^{a}}^{00}}^{*}\right]\,.
\end{align}
The terms $[\theta_{1}(e_{t}),\tilde\theta_{1}(e_{t})]$ are simply re-scalings of $[\varphi(e_{t}),\tilde\varphi(e_{t})]$ respectively, so the superasymptotic expansions of those sums are simply re-scalings of Eq.~\eqref{varphi-super} and~\eqref{varphit-super}. We apply the re-summation procedure on $[\theta_{2}(e_{t}),\tilde\theta_{2}(e_{t})]$ to obtain superasymptotic expressions for them. To obtain the superasymptotic series for $[\theta(e_{t}),\tilde\theta(e_{t})]$, we simply combine the superasymptotics of the individual terms to obtain
\begin{align}
\label{theta-super}
\theta_{\rm super}(e_{t}) &= \frac{1}{\Gamma\left(\frac{1}{3}\right) \Gamma\left(\frac{2}{3}\right)} \left[\frac{34240}{801 (1 - e_{t}^{2})^{6}}
 - \frac{132560}{2403 (1 - e_{t}^{2})^{5}} + \frac{794344}{46725 (1 - e_{t}^{2})^{4}} 
\right.
\nn \\
&\left.
- \frac{141994}{140175 (1 - e_{t}^{2})^{3}} + \frac{465188}{7709625 (1 - e_{t}^{2})^{2}} - \frac{500627}{334083750 (1 - e_{t}^{2})}\right]
\\
\label{tildetheta-super}
\tilde{\theta}_{\rm super}(e_{t}) &= \frac{1}{\Gamma\left(\frac{1}{3}\right) \Gamma\left(\frac{2}{3}\right)} \left[\frac{4672}{267 (1 - e_{t}^{2})^{9/2}} - \frac{1452}{89 (1 - e_{t}^{2})^{7/2}} + \frac{119776}{46725 (1 - e_{t}^{2})^{5/2}} 
\right.
\nn \\
&\left.
- \frac{266}{2225 (1 - e_{t}^{2})^{3/2}} - \frac{21}{4450 (1 - e_{t}^{2})^{1/2}}\right]
\end{align}

Let us now consider $[\alpha(e_{t}),\tilde\alpha(e_{t})]$, which we can break down in a similar manner:
\begin{align}
\alpha(e_{t}) &= \alpha_{1}(e_{t}) + \alpha_{2}(e_{t}) + \alpha_{3}(e_{t})\,,
\\
\tilde{\alpha}(e_{t}) &= \tilde{\alpha}_{1}(e_{t}) + \tilde{\alpha}_{2}(e_{t}) + \tilde{\alpha}_{3}(e_{t})\,,
\\
\alpha_{1}(e_{t}) &= \frac{441}{3424 (1 - e_{t}^{2})} \sum_{p=1}^{\infty} p^{7} | \underset{(p)}{{\cal{I}}_{jk}^{00}} |^{2}\,,
\\
\alpha_{2}(e_{t}) &= \frac{63}{3424 (1 - e_{t}^{2})} \sum_{m,s=-2}^{2} \sum_{p=1}^{\infty} (4 s + 3 m) p^{6} \underset{(p,m)}{{\cal{I}}_{jk}^{00}}^{*}\underset{(p,s)}{{\cal{I}}_{00}^{jk}}\,,
\\
\alpha_{3}(e_{t}) &= - \frac{21}{3424} \sum_{p=1}^{\infty} p^{7} \left[ \underset{(p)}{{\cal{I}}_{jk}^{01}}^{*} \underset{(p)}{{\cal{I}}_{00}^{jk}} + \underset{(p)}{{\cal{I}}_{jk}^{00}}^{*} \underset{(p)}{{\cal{I}}_{01}^{jk}}\right]\,,
\\
\tilde{\alpha}_{1}(e_{t}) &= - \frac{189 i}{856 (1 - e_{t}^{2})} \epsilon^{ijk} \hat{L}_{i} \sum_{p=1}^{\infty} p^{6} \underset{(p)}{{\cal{I}}_{ja}^{00}} \; \underset{(p)}{{{\cal{I}}_{k}^{a}}^{00}}^{*}\,,
\\
\tilde{\alpha}_{2}(e_{t}) &= \frac{189 i}{1712 (1 - e_{t}^{2})} \epsilon^{ijk} \hat{L}_{i} \!\!\!\!\! \sum_{m,s = -2}^{2} \sum_{p=1}^{\infty} p^{5} (m+s) \underset{(p,s)}{{\cal{I}}_{ja}^{00}} \; \underset{(p,m)}{{{\cal{I}}_{k}^{a}}^{00}}^{*}\,,
\\
\tilde{\alpha}_{3}(e_{t}) &= \frac{21 i}{1712} \epsilon^{ijk} \hat{L}_{i} \sum_{p=1}^{\infty} p^{6} \left[ \underset{(p)}{{\cal{I}}_{ja}^{00}} \; \underset{(p)}{{{\cal{I}}_{k}^{a}}^{01}}^{*} + \underset{(p)}{{\cal{I}}_{ja}^{01}} \; \underset{(p)}{{{\cal{I}}_{k}^{a}}^{00}}^{*}\right]\,.
\end{align}
Like with $[\theta(e_{t}),\tilde\theta(e_{t})]$, the first terms $[\alpha_{1}(e_{t}),\tilde\alpha_{1}(e_{t})]$ are once again re-scalings of $[\varphi(e_{t}),\tilde\varphi(e_{t})]$ and we thus do not need to apply the re-summation procedure again on them. The second terms $[\alpha_{2}(e_{t}),\tilde\alpha_{2}(e_{t})]$, however, do need to be resummed, and we thus apply the same techniques discussed above.

The last two terms $[\alpha_{3}(e_{t}),\tilde\alpha_{3}(e_{t})]$ require some additional handling, since they contain undetermined integrals involving $V(e_{t}; u) - u$ [see e.g.~Eq.~\eqref{Ipm201-Four}]. To evaluate $[\alpha_{3}(e_{t}),\tilde\alpha_{3}(e_{t})]$, we separate out the terms that do not depend on these integrals, i.e.~those with summands that are proportional to $[J_{p}(p e_{t})^{2}, J_{p}'(p e_{t})^{2}, J_{p}(p e_{t}) J_{p}'(p e_{t})]$, which we call $[\alpha_{3}^{(1)}(e_{t}), \tilde{\alpha}_{3}^{(1)}(e_{t})]$, specifically 
\begin{align}
\alpha_{3}^{(1)}(e_{t}) &= \frac{8 J_{p}(p e_{t})^{2} p^{3}} {21 e_{t}^{4} (1 - e_{t}^{2})^{3/2}} \left[1890 - 4221 e_{t}^{2} + 3024 e_{t}^{4} - 693 e_{t}^{6} 
\right.
\nn \\
&\left.
+ \sqrt{1 - e_{t}^{2}} \left(-111 + 222 e_{t}^{2} - 36 e_{t}^{4} - 19 e_{t}^{6}\right)\right] + \frac{8 J_{p}(p e_{t})^{2} p^{5}} {21 e_{t}^{4}} (1 - e_{t}^{2})^{3/2} \left[378 - 378 e_{t}^{2} 
\right.
\nn \\
&\left.
 + \sqrt{1 - e_{t}^{2}} \left(-115 + 23 e_{t}^{2}\right)\right] + \frac{8 J'_{p}(p e_{t})^{2}p^{3}}{7 e_{t}^{2} \sqrt{1 - e_{t}^{2}}} \left[630 - 525 e_{t}^{2} + \sqrt{1 - e_{t}^{2}} \left(-37 + 37 e_{t}^{2}\right)\right] 
\nn \\
&+ \frac{8 J'_{p}(p e_{t})^{2}p^{5}}{21 e_{t}^{2}} (1 - e_{t}^{2}) \left[-115 + 25 e_{t}^{2} + 378 \sqrt{1 - e_{t}^{2}} \right] + \frac{288 J_{p}(p e_{t}) J'_{p}(p e_{t}) p^{2}}{e_{t}^{3} \sqrt{1 - e_{t}^{2}}} \left(-2 + e_{t}^{2}\right) 
\nn \\
&+ \frac{32 J_{p}(p e_{t}) J'_{p}(p e_{t}) p^{6}}{21 e_{t}^{3}} \left(1 - e_{t}^{2}\right)^{3} + \frac{8 J_{p}(p e_{t}) J'_{p}(p e_{t}) p^{4}}{21 e_{t}^{3} \sqrt{1 - e_{t}^{2}}} \left[-3024 + 5040 e_{t}^{2} - 2016 e_{t}^{4} 
\right.
\nn \\
&\left.
+ \sqrt{1 - e_{t}^{2}} \left(448 - 603 e_{t}^{2} + 103 e_{t}^{4}\right)\right]\,,
\\
\tilde{\alpha}_{3}^{(1)}(e_{t}) &= - \frac{144 J_{p}(p e_{t})^{2} p}{e_{t}^{4} (1 - e_{t}^{2})} \left(-2 + e_{t}^{2}\right)^{2} - \frac{8 J_{p}(p e_{t})^{2} p^{3}}{3 e_{t}^{4}} \sqrt{1 - e_{t}^{2}} \left[-64 + 70 e_{t}^{2} + 5 e_{t}^{4} 
\right.
\nn \\
&\left.
+ \sqrt{1 - e_{t}^{2}} \left(432 - 234 e_{t}^{2}\right)\right] + \frac{32 J_{p}(p e_{t})^{2} p^{5}}{21 e_{t}^{4}} \left(1 - e_{t}^{2}\right)^{7/2} - \frac{576 J'_{p}(p e_{t})^{2}p}{e_{t}^{2}} 
\nn \\
&+ \frac{32 J_{p}(p e_{t})^{2}p^{5}}{21 e_{t}^{4}} (1 - e_{t}^{2})^{5/2} + \frac{16 J'_{p}(p e_{t})^{2} p^{3}}{21 e_{t}^{2} \sqrt{1 - e_{t}^{2}}} \left[224 - 358 e_{t}^{2} + 134 e_{t}^{4} 
\right.
\nn \\
&\left.
- \sqrt{1 - e_{t}^{2}} \left(-1512 + 1197 e_{t}^{2}\right)\right] + \frac{8 J_{p}(p e_{t}) J'_{p}(p e_{t}) p^{2}}{7 e_{t}^{3} (1 - e_{t}^{2})^{3/2}} \left[-148 + 370 e_{t}^{2} - 240 e_{t}^{4} + 18 e_{t}^{6} 
\right.
\nn \\
&\left.
+ \sqrt{1 - e_{t}^{2}} \left(2520 - 3864 e_{t}^{2} + 1449 e_{t}^{4}\right)\right] + \frac{32 J_{p}(p e_{t}) J'_{p}(p e_{t}) p^{4}}{21 e_{t}^{3}} \left(-115 + 24 
\right.
\nn \\
&\left.
+ 378 \sqrt{1 - e_{t}^{2}}\right)\,.
\end{align}
Since these expressions are of the form of the other tail enhancement factors, we simply apply our re-summation procedure to them and find their superasymptotic expressions. The remaining terms, which we denote $[\alpha_{3}^{(2)}(e_{t}), \tilde{\alpha}_{3}^{(2)}(e_{t})]$, do contain the undetermined integrals. Although these terms are complex, the enhancement factors $[\alpha_{3}^{(2)}(e_{t}), \tilde{\alpha}_{3}^{(2)}(e_{t})]$ are real valued. Hence the imaginary part vanishes upon integration, and we are left with solely the real part, specifically
\begin{align}
\alpha_{3}^{(2)}(e_{t}) &= \frac{12 p^{5} J_{p}(p e_{t})}{\pi e_{t}^{2} \sqrt{1 - e_{t}^{2}}} \int_{0}^{2 \pi} du \; [1 - e_{t} {\rm cos}(u)] \;  {\rm arctan}\left[\frac{\beta_{t} \; {\rm sin}(u)}{1 - \beta_{t} \; {\rm cos}(u)}\right] \left\{2 (-2 + e_{t}^{2}) [e_{t} - {\rm cos}(u)] 
\right.
\nn \\
&\left.
\times {\rm cos}\left\{p [u - e_{t} {\rm sin}(u)]\right\} {\rm sin}(u) + (-1 + e_{t}^{2}) p [-3 e_{t}^{2} + 4 e_{t} {\rm cos}(u) + (-2 + e_{t}^{2}) {\rm cos}(2 u)] 
\right.
\nn \\
&\left.
\times {\rm sin}\left\{p [u - e_{t} {\rm sin}(u)]\right\} \right\} - \frac{12 p^{5} J'_{p}(p e_{t})}{\pi e_{t} \sqrt{1 - e_{t}^{2}}} \int_{0}^{2 \pi} du \; [1 - e_{t} {\rm cos}(u)] {\rm arctan}\left[\frac{\beta_{t} \; {\rm sin}(u)}{1 - \beta_{t} \; {\rm cos}(u)}\right] 
\nn \\
&\times
\left\{4 (-1 + e_{t}^{2}) p [e_{t} - {\rm cos}(u)] {\rm cos}\left\{p [u - e_{t} {\rm sin}(u)]\right\} {\rm sin}(u) 
\right.
\nn \\
&\left.
+ [3 e_{t}^{2} - 4 e_{t} {\rm cos}(u) - (-2 + e_{t}^{2}) {\rm cos}(2 u)]  {\rm sin}\left\{p [u - e_{t} {\rm sin}(u)]\right\}\right\}\,,
\\
\tilde{\alpha}_{3}^{(2)}(e_{t}) &= - \frac{12 p^{4} J_{p}(p e_{t})}{\pi e_{t}^{2} (1 - e_{t}^{2})} \int_{0}^{2 \pi} du \; [1 - e_{t} {\rm cos}(u)] \; {\rm arctan}\left[\frac{\beta_{t} \; {\rm sin}(u)}{1 - \beta_{t} \; {\rm cos}(u)}\right] \left\{ 8 (-1 + e_{t}^{2})^{2} p [-e_{t} + {\rm cos}(u)] 
\right.
\nn \\
&\left.
\times {\rm cos}\left\{p [u - e_{t} {\rm sin}(u)]\right\} {\rm sin}(u) + (-2 + e_{t}^{2}) [-3 e_{t}^{2} + 4 e_{t} {\rm cos}(u) + (-2 + e_{t}^{2}) {\rm cos}(2 u)] 
\right.
\nn \\
&\left.
\times
{\rm sin}\left\{p (u - e_{t} {\rm sin}(u)]\right\}\right\} + \frac{24 p^{4} J'_{p}(p e_{t})}{\pi e_{t}} \int_{0}^{2 \pi} du \; [1 - e_{t} {\rm cos}(u)] {\rm arctan}\left[\frac{\beta_{t} \; {\rm sin}(u)}{1 - \beta_{t} \; {\rm cos}(u)}\right]
\nn \\
&\times
 \left\{4 [-e_{t} + {\rm cos}(u)] {\rm cos}\left\{n [u - e_{t} {\rm sin}(u)]\right\} {\rm sin}(u) 
\right.
\nn \\
&\left.
+ p [3 e_{t}^{2} - 4 e_{t} {\rm cos}(u) - (-2 + e_{t}^{2}) {\rm cos}(2 u)] {\rm sin}\left\{p [u - e_{t} {\rm sin}(u)]\right\}\right\}\,.
\end{align}
To obtain analytic expressions for these terms, we follow the procedure of~\cite{Forseth:2015oua} and begin by factoring out the high eccentricity dependence, which can be determined by comparison to other 2.5PN order enhancement factors. For the energy flux, the controlling factor is $(1-e_{t}^{2})^{-6}$, while for the angular momentum flux it is $(1-e_{t}^{2})^{-9/2}$. After factoring out this dependence, we perform an expansion about $e_{t} \ll 1$ to ${\cal{O}}(e_{t}^{30})$, which allows us to evaluate the necessary integrals, and obtain approximants of the form
\begin{align}
\alpha_{3}^{(2)}(e_{t}) &= \frac{A(e_{t})}{(1 - e_{t}^{2})^{6}}\,,
\\
\tilde{\alpha}_{3}^{(2)}(e_{t}) &= \frac{\tilde{A}(e_{t})}{(1 - e_{t}^{2})^{9/2}}\,,
\end{align}
where $[A(e_{t}), \tilde{A}(e_{t})]$ are polynomials to ${\cal{O}}(e_{t}^{30})$. Finally, we perform a Pad\'{e} re-summation of the resulting polynomial in $e_{t}$ by writing
\begin{align}
\label{A-PD}
A^{\rm PD}(e_{t}) = \frac{\sum_{n=0}^{N} A^{(n)} e_{t}^{n}}{1 + \sum_{m=1}^{M} A_{(m)} e_{t}^{m}}
\end{align}
and likewise for $\tilde{A}(e_{t})$. We obtain Pad\'{e} approximants of order $(M,N) = (14,16)$ for each of these enhancement factors. We chose the order of the Pad\'{e} approximants such that they are the most accurate compared to numerical results given the order we work to in the small eccentricity expansion. We list the coefficients of these Pad\'{e} approximants in~\ref{Pade}. To obtain the final superasymptotic expressions for $[\alpha(e_{t}),\tilde\alpha(e_{t})]$, we simply combine all of the terms back together to find
\begin{align}
\label{alphasuper}
\alpha_{\rm super}(e_{t}) &= \frac{1}{\Gamma\left(\frac{1}{3}\right) \Gamma\left(\frac{2}{3}\right)} \left[\frac{77776}{321 (1 - e_{t}^{2})^{6}} - \frac{15904}{107 (1 - e_{t}^{2})^{11/2}} - \frac{300512}{963 (1 - e_{t}^{2})^{5}} + \frac{19530}{107 (1 - e_{t}^{2})^{9/2}} 
\right.
\nn \\
&\left.
+ \frac{4871974}{56175 (1 - e_{t}^{2})^{4}} - \frac{26952}{535 (1 - e_{t}^{2})^{7/2}}  + \frac{111533}{56175 (1 - e_{t}^{2})^{3}}  + \frac{8313}{5350 (1 - e_{t}^{2})^{5/2}} 
\right.
\nn \\
&\left.
- \frac{2280749}{6179250 (1 - e_{t}^{2})^{2}} - \frac{4293}{294250 (1 - e_{t}^{2})^{3/2}} \right] - \frac{21}{3424} \frac{A^{\rm PD}(e_{t})}{(1 - e_{t}^{2})^{6}}
\\
\tilde{\alpha}_{\rm super}(e_{t}) &= \frac{1}{\Gamma\left(\frac{1}{3}\right) \Gamma\left(\frac{2}{3}\right)} \left[\frac{67688}{963 (1 - e_{t}^{2})^{9/2}} - \frac{4893}{107 (1 - e_{t}^{2})^{4}} - \frac{5996}{107 (1 - e_{t}^{2})^{7/2}} + \frac{38157}{1070 (1 - e_{t}^{2})^{3}} 
\right.
\nn \\
&\left.
+ \frac{89699}{56175 (1 - e_{t}^{2})^{5/2}} - \frac{5877}{2140 (1 - e_{t}^{2})^{2}} + \frac{944}{1605 (1 - e_{t}^{2})^{3/2}}\right] - \frac{21}{3424} \frac{\tilde{A}^{\rm PD}(e_{t})}{(1 - e_{t}^{2})^{9/2}}
\end{align}

Finally, we apply the re-summation procedure to the 3PN tail-of-tails and tail-squared enhancement factors $[\chi(e_{t}), \tilde\chi(e_{t})]$ to obtain the following superasymptotic expressions,
\begin{align}
\chi_{\rm super}(e_{t}) &= \frac{1}{(1 - e_{t}^{2})^{13/2}} \left[\frac{421543}{1536} - \frac{52745 \gamma_{E}}{1024} - \frac{52745 {\rm ln}(2)}{256} - \frac{52745 {\rm ln}(3)}{2048} - \frac{158235 {\rm ln}(1 - e_{t}^{2})}{2048}\right]
\nn \\
& + \frac{1}{(1 - e_{t}^{2})^{11/2}} \left[-\frac{2777339}{5120} + \frac{24717 \gamma_{E}}{256} + \frac{24717 {\rm ln}(2)}{64} + \frac{24717 {\rm ln}(3)}{512} + \frac{74151 {\rm ln}(1 - e_{t}^{2})}{512} \right]
\nn \\
& + \frac{1}{(1 - e_{t}^{2})^{9/2}} \left[\frac{10449133}{30720} - \frac{86065 \gamma_{E}}{1536} - \frac{86065 {\rm ln}(2)}{384} - \frac{86065 {\rm ln}(3)}{3072} - \frac{86065 {\rm ln}(1 - e_{t}^{2})}{1024}\right]
\nn \\
& + \frac{1}{(1 - e_{t}^{2})^{7/2}} \left[-\frac{1090519}{15360} + \frac{7895 \gamma_{E}}{768} + \frac{7895 {\rm ln}(2)}{192} + \frac{355271 {\rm ln}(3)}{69120} + \frac{{\rm ln}(243)}{86400} 
\right.
\nn \\
&\left.
+ \frac{7895 {\rm ln}(1 - e_{t}^{2})}{512}\right] + \frac{1}{(1 - e_{t}^{2})^{5/2}} \left[\frac{760247221}{275968000} - \frac{297 \gamma_{E}}{1024} - \frac{297 {\rm ln}(2)}{256} - \frac{1024063 {\rm ln}(3)}{7096320}
\right.
\nn \\
&\left.
 - \frac{2521 {\rm ln}(243)}{17740800} - \frac{891 {\rm ln}(1 - e_{t}^{2})}{2048}\right] + \frac{1}{(1 - e_{t}^{2})^{3/2}} \left[\frac{568287127}{67267200000} + \frac{71 {\rm ln}(2)}{682500} + \frac{1327283 {\rm ln}(3)}{2882880000} 
\right.
\nn \\
&\left.
- \frac{19843 {\rm ln}(243)}{221760000} - \frac{71 {\rm ln}(768)}{5460000}\right] + \frac{1}{(1 - e_{t}^{2})^{1/2}} \left[-\frac{4896210901}{4708704000000} + \frac{12270499 {\rm ln}(2)}{24324300000}
\right.
\nn \\
&\left.
 + \frac{423525727 {\rm ln}(3)}{5448643200000} - \frac{410009 {\rm ln}(243)}{139708800000} - \frac{12270499 {\rm ln}(768)}{194594400000}\right]\,,
\\
\label{tildechisuper}
\tilde{\chi}_{\rm super}(e_{t}) &= \frac{1}{(1 - e_{t}^{2})^{5}} \left[\frac{35583}{512} - \frac{3465 \gamma_{E}}{256} - \frac{3465 {\rm ln}(2)}{64} - \frac{3465 {\rm ln}(3)}{512} - \frac{10395 {\rm ln}(1 - e_{t}^{2})}{512} \right] 
\nn \\
&+ \frac{1}{(1 - e_{t}^{2})^{4}} \left[-\frac{51359}{512} + \frac{4655 \gamma_{E}}{256} + \frac{4655 {\rm ln}(2)}{64} + \frac{4655 {\rm ln}(3)}{512} + \frac{13965 {\rm ln}(1 - e_{t}^{2})}{512}\right]
\nn \\
& + \frac{1}{(1 - e_{t}^{2})^{3}} \left[\frac{47481}{1280} - \frac{1515 \gamma_{E}}{256} - \frac{1515 {\rm ln}(2)}{64} - \frac{1519 {\rm ln}(3)}{512} + \frac{{\rm ln}(243)}{640} - \frac{4545 {\rm ln}(1 - e_{t}^{2})}{512}\right]
\nn \\
& + \frac{1}{(1 - e_{t}^{2})^{2}} \left[-\frac{53091}{22400} + \frac{69 \gamma_{E}}{256} + \frac{69 {\rm ln}(2)}{64} + \frac{3041 {\rm ln}(3)}{23040} + \frac{{\rm ln}(243)}{1800} + \frac{207 {\rm ln}(1 - e_{t}^{2})}{512}\right]
\nn \\
& + \frac{1}{(1 - e_{t}^{2})} \left[-\frac{956569}{68992000} + \frac{{\rm ln}(2)}{700} - \frac{6269 {\rm ln}(3)}{4435200} + \frac{7061 {\rm ln}(243)}{22176000} - \frac{{\rm ln}(768)}{5600}\right]
\nn \\
& + \frac{15822507}{22422400000} - \frac{553 {\rm ln}(2)}{6435000} - \frac{553 {\rm ln}(3)}{51480000} + \frac{553 {\rm ln}(768)}{51480000}
\end{align}
This completes the calculation of the superasymptotic expansion of the enhancement factors.
 
\subsection{Hyperasymptotic Enhancement Factors}
\label{super}

The main concern with the superasymptotic series for the enhancement factors is their relatively poor accuracy when the eccentricity is small, especially compared to other methods for evaluating the enhancement factors. Post-circular like methods, like those in~\cite{Tanay:2016zog, Forseth:2015oua}, give better accuracy when $e_{t} \ll 1$, whereas the superasymptotic series are better for $\epsilon = 1 - e_{t}^{2} \ll 1$. We can resolve this concern by producing a \emph{hyperasymptotic series} through a combination of the superasymptotic expansions we already derived with the post-circular expansion of the enhancement factors in~\cite{Tanay:2016zog, Forseth:2015oua} 

Consider an enhancement factor $E(e_{t})$ with a superasymptotic expression $E_{\rm super}(e_{t})$. The simplest way of improve the accuracy of the superasymptotic series in the small eccentricity limit is to consider the remainder functional
\begin{equation}
E_{\rm remainder}(e_{t}) = E(e_{t}) - E_{\rm super}(e_{t})\,.
\end{equation}
To study the behavior of the remainder when the eccentricity is small, we create a Taylor series of order $N$ for the remainder by
\begin{align}
\delta E^{N}(e_{t}) = \hat{{\cal{T}}}^{N}_{e_{t}} E_{\rm remainder}(e_{t})\,,
\end{align}
where $\hat{{\cal{T}}}^{N}_{e_{t}}$ is a differential operator that generates the Taylor series. We could have chosen a different representation for the remainder, such as a Pad\'e resummation of the resulting Taylor expansion, but as we will see below, a simple Taylor expansion is sufficiently accurate and the most appropriate in the $e_{t} \ll 1$ limit. With an accurate representation of the small eccentricity behavior, we can then generate a hyperasymptotic series for the enhancement factor $E(e_{t})$ via
\begin{equation}
E_{\rm hyper}(e_{t}) = E_{\rm super}(e_{t}) + \delta E^{N}(e_{t})\,.
\end{equation}
These hyperasymptotic expressions have two important properties: (1) they have the right limiting behavior at small eccentricity and (2) we have analytic control over the remainder.

\begin{figure}[ht]
\centering
\includegraphics[clip=true,scale=0.34]{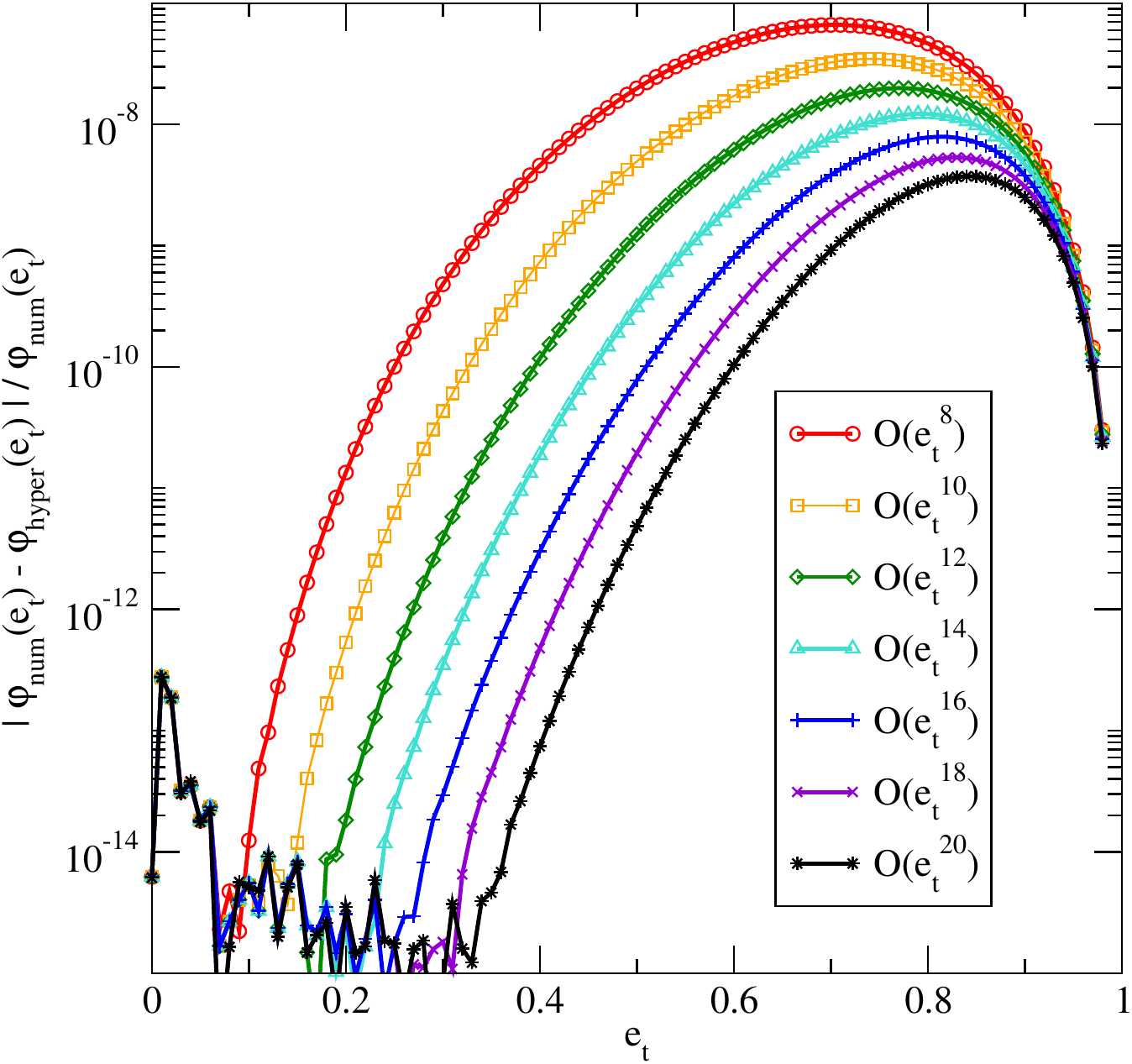}
\caption{\label{M0} Comparison of the numerical results for $\varphi(e_{t})$ with its hyperasymptotic series, $\varphi_{\rm hyper}(e_{t})$, at different orders in $e_{t}$. The addition of the remainder improves the accuracy at small eccentricity by about nine orders of magnitude.}
\end{figure}

Figure~\ref{M0} compares the numerical $\varphi(e_{t})$ and the hyperasymptotic $\varphi_{\rm hyper}(e_{t})$. Observe that the hyperasymptotic series is well-behaved in the small eccentricity limit as expected, with a relative error that collapses to the level of the tolerance $\delta$ in the circular limit. Further, the error can be increasingly improved by going to higher order in $e_{t}$ in the remainder. The remainder functionals at twentieth order for $[\varphi(e_{t}),\tilde{\varphi}(e_{t})]$ are 
\begin{align}
\label{phi-hyper}
\delta \varphi^{20}(e_{t}) &= 1 - \frac{3427}{945 \Gamma(\frac{1}{3}) \Gamma(\frac{2}{3})} + e_{t}^{2} \left[\frac{2335}{192} - \frac{208456}{4725 \Gamma(\frac{1}{3}) \Gamma(\frac{2}{3})}\right] + e_{t}^{4} \left[\frac{42955}{768} - \frac{319561}{1575 \Gamma(\frac{1}{3}) \Gamma(\frac{2}{3})}\right] 
\nn \\
&+ e_{t}^{6} \left[\frac{6204647}{36864} - \frac{2884936}{4725 \Gamma(\frac{1}{3}) \Gamma(\frac{2}{3})}\right] + e_{t}^{8} \left[\frac{352891481}{884736} - \frac{1367347}{945 \Gamma(\frac{1}{3}) \Gamma(\frac{2}{3})}\right] 
\nn \\
&+ e_{t}^{10} \left[\frac{286907786543}{353894400} - \frac{61760}{21 \Gamma(\frac{1}{3}) \Gamma(\frac{2}{3})}\right] + e_{t}^{12} \left[\frac{6287456255443}{4246732800} - \frac{1208431}{225 \Gamma(\frac{1}{3}) \Gamma(\frac{2}{3})}\right] 
\nn \\
&
+ e_{t}^{14} \left[\frac{5545903772613817}{2219625676800} - \frac{4758512}{525 \Gamma(\frac{1}{3}) \Gamma(\frac{2}{3})}\right] + e_{t}^{16} \left[\frac{422825073954708079}{106542032486400} 
\right.
\nn \\
&\left.
- \frac{7558199}{525 \Gamma(\frac{1}{3}) \Gamma(\frac{2}{3})}\right] + e_{t}^{18} \left[\frac{1659160118498286776339}{276156948204748800} - \frac{20596024}{945 \Gamma(\frac{1}{3}) \Gamma(\frac{2}{3})}\right]
\nn \\
&+ e_{t}^{20} \left[\frac{724723372042305454448081}{82847084461424640000} - \frac{29987903}{945 \Gamma(\frac{1}{3}) \Gamma(\frac{2}{3})}\right]
\\
\label{tildephi-hyper}
\delta \tilde{\varphi}^{20}(e_{t}) &= 1 - \frac{3811}{1050 \Gamma(\frac{1}{3}) \Gamma(\frac{2}{3})} + e_{t}^{2} \left[\frac{209}{32} - \frac{49751}{2100 \Gamma(\frac{1}{3}) \Gamma(\frac{2}{3})}\right] + e_{t}^{4} \left[\frac{2415}{128} - \frac{574913}{8400 \Gamma(\frac{1}{3}) \Gamma(\frac{2}{3})}\right]
\nn \\
&+ e_{t}^{6} \left[\frac{730751}{18432} - \frac{23011}{160 \Gamma(\frac{1}{3}) \Gamma(\frac{2}{3})}\right] + e_{t}^{8} \left[\frac{10355719}{147456} - \frac{326097}{1280 \Gamma(\frac{1}{3}) \Gamma(\frac{2}{3})}\right] 
\nn \\
&+ e_{t}^{10} \left[\frac{6594861233}{58982400} - \frac{5191733}{12800 \Gamma(\frac{1}{3}) \Gamma(\frac{2}{3})}\right] + e_{t}^{12} \left[\frac{23422887967}{141557760} - \frac{30732361}{51200 \Gamma(\frac{1}{3}) \Gamma(\frac{2}{3})}\right] 
\nn \\
&+ e_{t}^{14} \left[\frac{51535146547541}{221962567680} - \frac{603727553}{716800 \Gamma(\frac{1}{3}) \Gamma(\frac{2}{3})}\right] + e_{t}^{16} \left[\frac{16666910315347223}{53271016243200} 
\right.
\nn \\
&\left.
- \frac{2603342599}{2293760 \Gamma(\frac{1}{3}) \Gamma(\frac{2}{3})}\right] + e_{t}^{18} \left[\frac{8055842533080274417}{19725496300339200} - \frac{20389261321}{13762560 \Gamma(\frac{1}{3}) \Gamma(\frac{2}{3})}\right]
\nn \\
&+ e_{t}^{20} \left[\frac{1024885995293794354963}{1972549630033920000} - \frac{74113622297}{39321600 \Gamma(\frac{1}{3}) \Gamma(\frac{2}{3})}\right]
\end{align}
Obviously, the more terms one keeps in the expansion of the remainder, the more accurate the approximation will be at small eccentricity; in practice, how many terms to keep will depend on the accuracy desired in this limit.  

This procedure can be applied for the remaining enhancement factors through 3PN order. For the 2.5PN order enhancement factors, it is simplest to consider the total 2.5PN order fluxes
\begin{align}
\langle {\cal{P}}_{\infty}^{\rm 2.5PN tails} \rangle &= \frac{32}{5} \pi \eta^{2} x^{15/2} \left[-\frac{8191}{672} \psi(e_{t}) - \frac{583}{24} \eta \zeta(e_{t})\right]\,,
\\
\langle {\cal{G}}_{\infty}^{\rm 2.5PN tails} \rangle &= \frac{32}{5} \pi \eta^{2} x^{15/2} \left[-\frac{8191}{672} \tilde{\psi}(e_{t}) - \frac{583}{24} \eta \tilde{\zeta}(e_{t})\right]\,,
\end{align}
where the new enhancement factors $\psi(e_{t})$ and $\zeta(e_{t})$ are
\begin{align}
\psi(e_{t}) &= \frac{13696}{8191}\alpha(e_{t}) - \frac{16403}{24573} \beta(e_{t}) - \frac{112}{24573} \gamma(e_{t})\,,
\\
\zeta(e_{t}) &= -\frac{1424}{4081} \theta(e_{t}) + \frac{16403}{12243} \beta(e_{t}) + \frac{16}{1749} \gamma(e_{t})\,,
\end{align}
The analogs $\tilde\psi(e_{t})$ and $\tilde\zeta(e_{t})$ have the same definitions as those above except with $[\tilde\alpha(e_{t}), \tilde\theta(e_{t}), \tilde\beta(e_{t}), \tilde\gamma(e_{t})]$. The superasymptotic expressions for these new enhancement factors are simply the superasymptotics of the individual enhancement factors from which they are constructed. Applying the scheme to generate hyperasymptotic expression for these new enhancement factors, we find the remainders
\allowdisplaybreaks[4]
\begin{align}
\label{psi-hyper}
\delta \psi^{20}(e_{t}) &= 1 - \frac{241580064}{66551875 \Gamma(\frac{1}{3}) \Gamma(\frac{2}{3})} + e_{t}^{2} \left[\frac{188440}{8191} - \frac{1283135619824}{15373483125 \Gamma(\frac{1}{3}) \Gamma(\frac{2}{3})}\right] 
\nn \\
&+ e_{t}^{4} \left[\frac{78746077}{524224} - \frac{8377507600624}{15373483125 \Gamma(\frac{1}{3}) \Gamma(\frac{2}{3})}\right] + e_{t}^{6} \left[\frac{2769593143}{4718016} - \frac{10912663062368}{5124494375 \Gamma(\frac{1}{3}) \Gamma(\frac{2}{3})}\right] 
\nn \\
&+ e_{t}^{8} \left[\frac{1038414910159}{603906048} - \frac{8717702789819}{1397589375 \Gamma(\frac{1}{3}) \Gamma(\frac{2}{3})}\right] + e_{t}^{10} \left[\frac{10517947248419}{2516275200} 
\right.
\nn \\
&\left.
- \frac{233112310241024}{15373483125 \Gamma(\frac{1}{3}) \Gamma(\frac{2}{3})}\right] + e_{t}^{12} \left[\frac{51677468559131363}{5797498060800} - \frac{1325626967291149}{40995955000 \Gamma(\frac{1}{3}) \Gamma(\frac{2}{3})}\right]
\nn \\
&+ e_{t}^{14} \left[\frac{14698793962256164697}{852232214937600} - \frac{366425963194427}{5856565000 \Gamma(\frac{1}{3}) \Gamma(\frac{2}{3})}\right] 
\nn \\
&
+ e_{t}^{16} \left[\frac{13508357018274827128409}{436342894048051200} - \frac{147327907838689583}{1311870560000 \Gamma(\frac{1}{3}) \Gamma(\frac{2}{3})}\right] 
\nn \\
&
+ e_{t}^{18} \left[\frac{462509893308626646120797}{8835943604473036800} - \frac{124551349162839959}{655935280000 \Gamma(\frac{1}{3}) \Gamma(\frac{2}{3})}\right]
\nn \\
&+ e_{t}^{20} \left[\frac{4766936073001835060207935793}{56550039068627435520000} - \frac{9627820057257367657}{31484893440000 \Gamma(\frac{1}{3}) \Gamma(\frac{2}{3})}\right]\,,
\\
\delta \tilde\psi^{20}(e_{t}) &= 1 - \frac{55624031984}{15373483125 \Gamma(\frac{1}{3}) \Gamma(\frac{2}{3})} + e_{t}^{2} \left[\frac{102536}{8191} - \frac{698208327368}{15373483125 \Gamma(\frac{1}{3}) \Gamma(\frac{2}{3})}\right] 
\nn \\
&+ e_{t}^{4} \left[\frac{27975523}{524224} - \frac{2976133354982}{15373483125 \Gamma(\frac{1}{3}) \Gamma(\frac{2}{3})}\right] + e_{t}^{6} \left[\frac{709642057}{4718016} - \frac{8388221641661}{15373483125 \Gamma(\frac{1}{3}) \Gamma(\frac{2}{3})}\right] 
\nn \\
&+ e_{t}^{8} \left[\frac{203853989947}{603906048} - \frac{4302911627633}{3513939000 \Gamma(\frac{1}{3}) \Gamma(\frac{2}{3})}\right] + e_{t}^{10} \left[\frac{4944184758677}{7548825600} 
\right.
\nn \\
&\left.
- \frac{83488815643601}{35139390000 \Gamma(\frac{1}{3}) \Gamma(\frac{2}{3})}\right] + e_{t}^{12} \left[\frac{6658083547039409}{5797498060800} - \frac{17744668624161}{4259320000 \Gamma(\frac{1}{3}) \Gamma(\frac{2}{3})}\right] 
\nn \\
&+ e_{t}^{14} \left[\frac{35397103550602159}{18938493665280} - \frac{404305552234341}{59630480000 \Gamma(\frac{1}{3}) \Gamma(\frac{2}{3})}\right] + e_{t}^{16} \left[\frac{179071486944184743991}{62334699149721600} 
\right.
\nn \\
&\left.
- \frac{764822533448511}{73391360000 \Gamma(\frac{1}{3}) \Gamma(\frac{2}{3})}\right] + e_{t}^{18} \left[\frac{7457576214411997508197}{1767188720894607360} - \frac{1348218095132281}{88069632000 \Gamma(\frac{1}{3}) \Gamma(\frac{2}{3})}\right] 
\nn \\
&+ e_{t}^{20} \left[\frac{337929898617545561543145703}{56550039068627435520000} - \frac{38182917211753667}{1761392640000 \Gamma(\frac{1}{3}) \Gamma(\frac{2}{3})}\right]\,,
\\
\delta \zeta^{20}(e_{t}) &= 1 - \frac{250066100248}{68935741875 \Gamma(\frac{1}{3}) \Gamma(\frac{2}{3})} + e_{t}^{2} \left[\frac{1011565}{48972} - \frac{5165477150408}{68935741875 \Gamma(\frac{1}{3}) \Gamma(\frac{2}{3})}\right] 
\nn \\
&+ e_{t}^{4} \left[\frac{106573021}{783552} - \frac{1619660334008}{3282654375 \Gamma(\frac{1}{3}) \Gamma(\frac{2}{3})}\right] + e_{t}^{6} \left[\frac{456977827}{854784} - \frac{133691089979528}{68935741875 \Gamma(\frac{1}{3}) \Gamma(\frac{2}{3})}\right] 
\nn \\
&+ e_{t}^{8} \left[\frac{128491074157}{82059264} - \frac{55938524367784}{9847963125 \Gamma(\frac{1}{3}) \Gamma(\frac{2}{3})}\right] + e_{t}^{10} \left[\frac{342306246988373}{90265190400} 
\right.
\nn \\
&\left.
- \frac{105369692129672}{7659526875 \Gamma(\frac{1}{3}) \Gamma(\frac{2}{3})}\right] + e_{t}^{12} \left[\frac{69677817044303231}{8665458278400} - \frac{8704718214568}{298423125 \Gamma(\frac{1}{3}) \Gamma(\frac{2}{3})}\right] 
\nn \\
&+ e_{t}^{14} \left[\frac{2386244038997979551}{154402711142400} - \frac{429418866068552}{7659526875 \Gamma(\frac{1}{3}) \Gamma(\frac{2}{3})}\right] + e_{t}^{16} \left[\frac{5987988065386963552943}{217399017288499200} 
\right.
\nn \\
&\left.
- \frac{765322594645592}{7659526875 \Gamma(\frac{1}{3}) \Gamma(\frac{2}{3})}\right] + e_{t}^{18} \left[\frac{61448938675545297383797}{1329005313235353600} - \frac{1651784262696184}{9847963125 \Gamma(\frac{1}{3}) \Gamma(\frac{2}{3})}\right] 
\nn \\
&+ e_{t}^{20} \left[\frac{6249104916857243979130332809}{84524737921768488960000} - \frac{18488329739373848}{68935741875 \Gamma(\frac{1}{3}) \Gamma(\frac{2}{3})}\right]\,,
\\
\delta \tilde\zeta^{20}(e_{t}) &= 1 - \frac{83380049048}{22978580625 \Gamma(\frac{1}{3}) \Gamma(\frac{2}{3})} + e_{t}^{2} \left[\frac{102371}{8162} - \frac{348496717732}{7659526875 \Gamma(\frac{1}{3}) \Gamma(\frac{2}{3})}\right] 
\nn \\
&+ e_{t}^{4} \left[\frac{14250725}{261184} - \frac{1516042558243}{7659526875 \Gamma(\frac{1}{3}) \Gamma(\frac{2}{3})}\right] + e_{t}^{6} \left[\frac{722230667}{4701312} - \frac{8537054301053}{15319053750 \Gamma(\frac{1}{3}) \Gamma(\frac{2}{3})}\right] 
\nn \\
&+ e_{t}^{8} \left[\frac{102744533069}{300883968} - \frac{4337433374609}{3501498000 \Gamma(\frac{1}{3}) \Gamma(\frac{2}{3})}\right] + e_{t}^{10} \left[\frac{9843430194463}{15044198400} 
\right.
\nn \\
&\left.
- \frac{83109477430673}{35014980000 \Gamma(\frac{1}{3}) \Gamma(\frac{2}{3})}\right] + e_{t}^{12} \left[\frac{43605309737981}{38513147904} - \frac{52296280129859}{12732720000 \Gamma(\frac{1}{3}) \Gamma(\frac{2}{3})}\right]
\nn \\
&+ e_{t}^{14} \left[\frac{19069924628449467}{10484134707200} - \frac{392069990496293}{59419360000 \Gamma(\frac{1}{3}) \Gamma(\frac{2}{3})}\right] + e_{t}^{16} \left[\frac{600336343160521814159}{217399017288499200} 
\right.
\nn \\
&\left.
- \frac{732589863363103}{73131520000 \Gamma(\frac{1}{3}) \Gamma(\frac{2}{3})}\right] + e_{t}^{18} \left[\frac{28244435149543337941721}{7043728160147374080} - \frac{3829628554688939}{263273472000 \Gamma(\frac{1}{3}) \Gamma(\frac{2}{3})}\right] 
\nn \\
&+ e_{t}^{20} \left[\frac{158264167343831506620212273}{28174912640589496320000} - \frac{35764742675504291}{1755156480000 \Gamma(\frac{1}{3}) \Gamma(\frac{2}{3})}\right]\,.
\end{align}

Finally, for the 3PN order enhancement factors from the tail-of-tails and tail-squared terms, we have to tenth order in $e_{t}$ 
\begin{align}
\delta \chi^{10}(e_{t}) &= -\frac{36718454998853}{9417408000000} + \gamma_{E} + \frac{{\rm ln}(3)}{2} + {\rm ln}(16) + e_{t}^{2} \left[-\frac{131766997689301}{1448832000000} + \frac{62 \gamma_{E}}{3} + 57 {\rm ln}(2) 
\right.
\nn \\
&\left.
+ \frac{27619 {\rm ln}(3)}{768}\right]  + e_{t}^{4} \left[-\frac{17257920310633973}{25113088000000} + \frac{9177 \gamma_{E}}{64} + \frac{182657 {\rm ln}(2)}{192} - \frac{51243 {\rm ln}(3)}{1024}\right] 
\nn \\
&+ e_{t}^{6} \left[-\frac{92129506724738033}{30135705600000} + \frac{76615 \gamma_{E}}{128} - \frac{296449 {\rm ln}(2)}{384} + \frac{8680309 {\rm ln}(3)}{16384} + \frac{244140625 {\rm ln}(5)}{147456}\right] 
\nn \\
&+ e_{t}^{8} \left[-\frac{343678592520953093}{34440806400000} + \frac{1903055 \gamma_{E}}{1024} + \frac{59103559 {\rm ln}(2)}{3072} + \frac{1180327577 {\rm ln}(3)}{131072}
\right.
\nn \\
&\left.
 - \frac{10498046875 {\rm ln}(5)}{1179648}\right] + e_{t}^{10} \left[-\frac{3051437850147557459}{114802688000000} + \frac{9732723 \gamma_{E}}{2048} - \frac{55480099157 {\rm ln}(2)}{1382400} 
\right.
\nn \\
&\left.
- \frac{4787048773551 {\rm ln}(3)}{104857600} + \frac{2342041015625 {\rm ln}(5)}{113246208} + \frac{33232930569601 {\rm ln}(7)}{943718400}\right]\,,
\\
\label{chi-hyper}
\delta \tilde\chi^{10}(e_{t}) &= -\frac{6244218863}{1601600000} + \gamma_{E} + \frac{{\rm ln}(3)}{2} + {\rm ln}(16) + e_{t}^{2} \left[-\frac{3744713821}{68992000} + \frac{389 \gamma_{E}}{32} + \frac{1007 {\rm ln}(2)}{32} 
\right.
\nn \\
&\left.
+ \frac{2965 {\rm ln}(3)}{128}\right] + e_{t}^{4} \left[-\frac{18889494241}{68992000} + \frac{3577 \gamma_{E}}{64} + \frac{27699 {\rm ln}(2)}{64} - \frac{27245 {\rm ln}(3)}{512}\right] 
\nn \\
&+ e_{t}^{6} \left[-\frac{60790558061}{68992000} + \frac{43049 \gamma_{E}}{256} - \frac{1804397 {\rm ln}(2)}{2304} + \frac{1946309 {\rm ln}(3)}{8192} + \frac{48828125 {\rm ln}(5)}{73728}\right] 
\nn \\
&+ e_{t}^{8} \left[-\frac{151724371031}{68992000} + \frac{102005 \gamma_{E}}{256} + \frac{15307525 {\rm ln}(2)}{2304} + \frac{184515253 {\rm ln}(3)}{65536} - \frac{2099609375 {\rm ln}(5)}{589824}\right] 
\nn \\
&+ e_{t}^{10} \left[-\frac{160776312313}{34496000} + \frac{207311 \gamma_{E}}{256} - \frac{107117837 {\rm ln}(2)}{6400} - \frac{815824256293 {\rm ln}(3)}{52428800} 
\right.
\nn \\
&\left.
+ \frac{156103515625 {\rm ln}(5)}{18874368} + \frac{4747561509943 {\rm ln}(7)}{471859200}\right]\,.
\end{align}
One can easily extend these expressions to higher order in eccentricity, but then they become quite lengthy.

\begin{figure*}[ht]
\centering
\includegraphics[clip=true,scale=0.34]{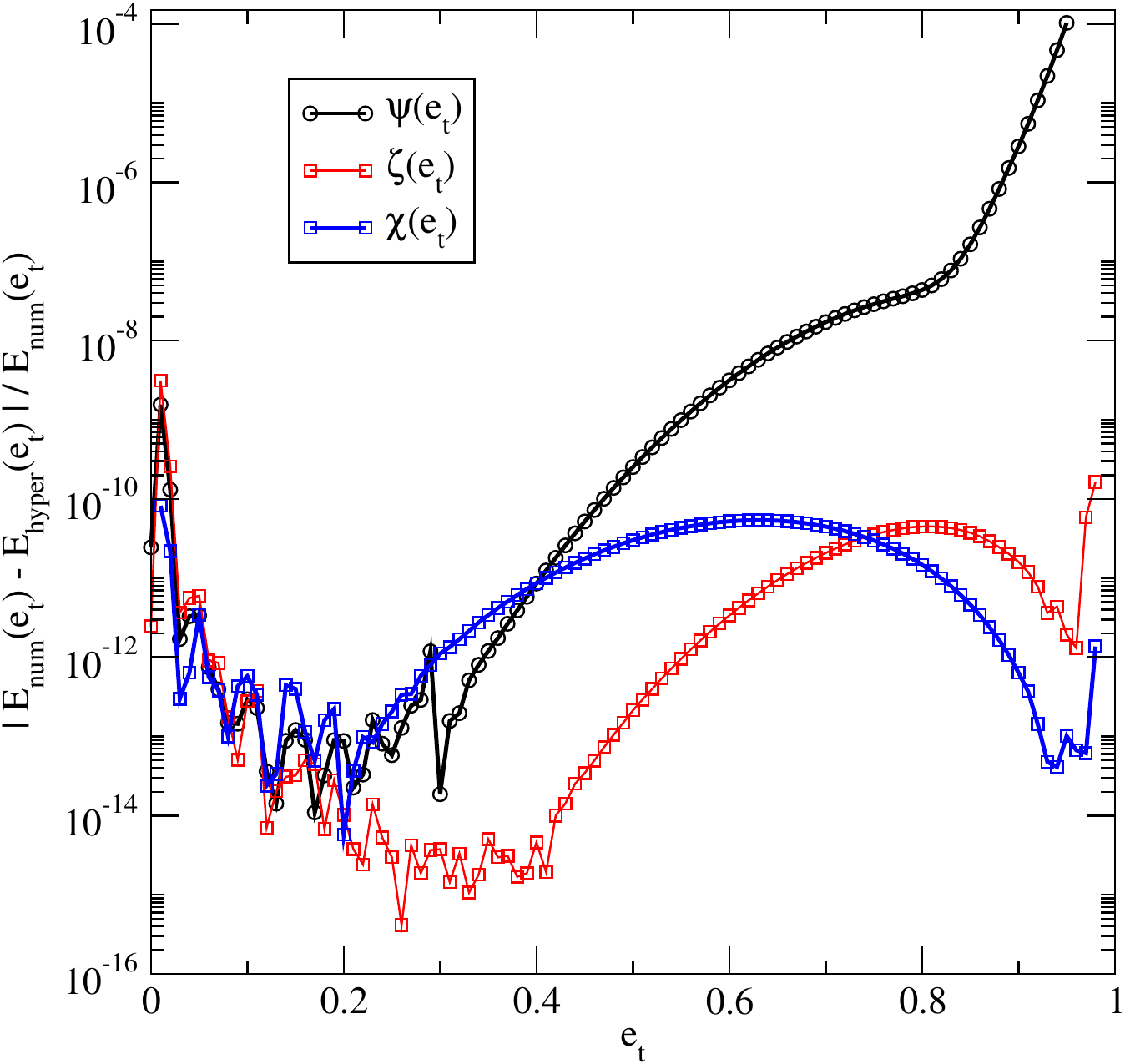}
\includegraphics[clip=true,scale=0.34]{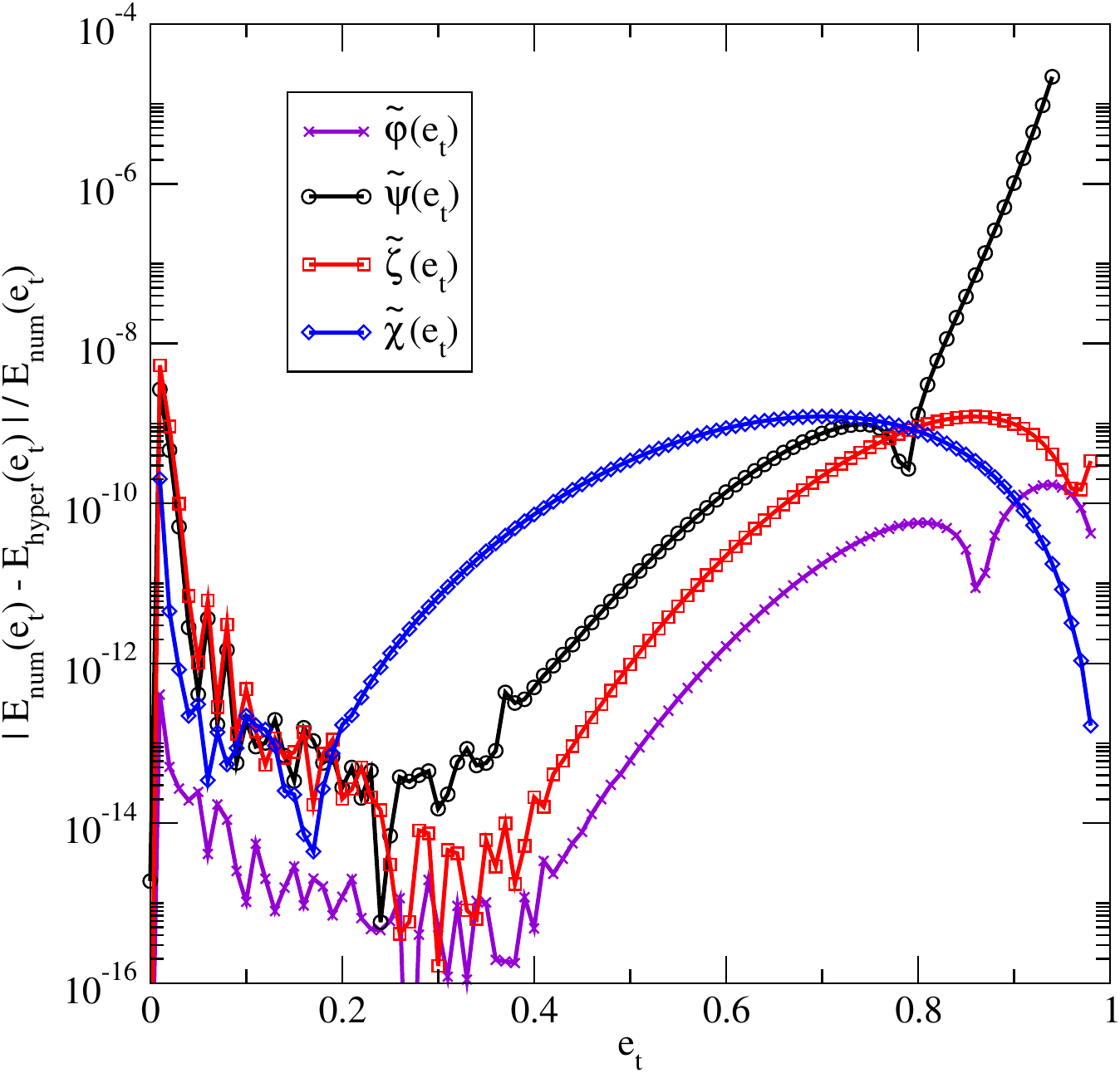}
\caption{\label{err-hyper} Comparison of numerical results for tail enhancement factors with their hyperasymptotic expressions for the energy flux (left) and angular momentum flux (right). With the exception of the 2.5PN order enhancement factors $[\psi(e_{t}), \tilde{\psi}(e_{t})]$, the hyperasymptotic expressions are accurate to relative errors better than $10^{-8}$ for all eccentricities. For $[\psi(e_{t}), \tilde{\psi}(e_{t})]$, the relative errors at the last data point, which is $e_{t} = 0.95$ for these enhancement factors, are approximately $[1\times10^{-4}, 5\times10^{-5}]$, respectively.}
\end{figure*}

For comparison, we display the relative error in all hyperasymptotic series relative to the numerical enhancement factors in the two panels of Fig.~\ref{err-hyper}. As can be seen from Fig.~\ref{M0} and Fig.~\ref{err-hyper}, the error for the enhancement factors that we were able to complete the superasymptotic analysis on are typically less that $10^{-8}$ for arbitrary eccentricity. The exceptions to this are the 2.5PN order enhancement factors $[\psi(e_{t}), \tilde{\psi}(e_{t})]$, which have relative errors of $\sim [1\times10^{-4}, 5\times10^{-5}]$ at $e_{t} = 0.95$. The reason for the poor accuracy in these enhancement factors is that the high eccentricity dependence is very sensitive to $[\alpha_{3}^{(2)}(e_{t}), \tilde{\alpha}_{3}^{(2)}(e_{t})]$, whose re-summation does not follow the usual procedure of obtaining superasymptotic expressions (and rather we had to Pad\'e a small eccentricity expansion for these). 

%

\section{Energy \& Angular Momentum Fluxes: Memory Terms}
\label{memory}

The memory introduces a DC offset in the waveform that depends on how the orbit evolves under radiation reaction. This section will present the formal structure of the memory terms in the fluxes and the averaging of the DC and oscillatory component.

\subsection{Integral Definitions and Harmonic Decomposition}
\label{harmonic}
The memory contributions enter into the radiative moments as time anti-derivatives of the source multipoles at 2.5PN order. For the radiative mass quadrupole, the memory contribution is given in Eq.~\eqref{U-mem}. To see how this enters the fluxes, it is convenient to perform a change of variable using $T=t_{R} - \tau$, which gives
\begin{equation}
U^{\rm mem}_{jk}(t_{R}) = \int_{-\infty}^{t_{R}} dT \overset{(3)}{I}_{a<j}(T) \; \overset{(3)}{I^{a}}_{k>}(T)\,.
\end{equation}
To obtain the memory contributions to the fluxes, we simply need to combine this with Eqs.~\eqref{sw-P} and~\eqref{sw-G} and separate out the pieces that depend on the above integrals. For the energy flux, Eq.~\eqref{sw-P} only contains derivatives with respect to $t_{R}$, which when acting on the above memory contribution cancel out the time antiderivative, converting the memory contribution to the radiative quadrupole moment into an instantaneous term. As a result, there is no explicit memory term in the energy flux to 3PN order~\cite{Arun:2009mc}.

The angular momentum flux given in Eq.~\eqref{sw-G}, on the other hand, has terms that do not depend on derivatives with respect to $t_{R}$. As a result, the anti-derivative in Eq.~\eqref{U-mem} will not cancel and the memory integrals will enter explicitly into the angular momentum flux. The contribution to the memory from the mass quadrupole enters the angular momentum flux at 2.5PN order and is given by~\cite{Arun:2009mc}
\begin{equation}
\label{G-mem}
{\cal{G}}_{\infty}^{\rm memory} = \frac{4}{35} \hat{L}^{i} \epsilon_{iab} \overset{(3)}{I^{a}}_{j}(t_{R}) \int_{-\infty}^{t_{R}} dT \; \overset{(3)}{I^{<b}}_{c}(T) \; \overset{(3)}{I^{j>c}}(T)\,.
\end{equation}
Notice that the hereditary integral formally extends infinitely far into the past. Physically, however, the binary will have formed at some instant in the past, which we call $t_{0}$, and will evolve up to the current retarded time $t_{R}$. As a result, the lower limit of the integral can be cut at $t_{0}$ without loss of accuracy. In fact, if one wanted to compute the memory numerically, the integral would have to be cut at the instant at which the numerical computation is initiated.

Since the memory enters into the fluxes at 2.5PN order, and we are working consistently to 3PN order in the fluxes, we simply have to evaluate Eq.~\eqref{G-mem} on Newtonian orbits. Following~\cite{Arun:2009mc}, we will express Eq.~\eqref{G-mem} in terms of the relative coordinates of the orbital dynamics. The Newtonian mass quadrupole is given in Eq.~\eqref{mass-quad}, and taking the necessary number of time derivatives, we obtain
\begin{equation}
\overset{(3)}{I}_{jk} = \frac{6 \mu m}{r^{4}} \left(\dot{r} \; x_{<j} x_{k>} - \frac{4}{3} r \; x_{<j} v_{k>}\right)\,.
\end{equation}
Inserting this into Eq.~\eqref{G-mem}, we obtain
\begin{align}
\label{G-mem-rel}
{\cal{G}}_{\infty}^{\rm memory} &= \frac{64}{105} \frac{\eta^{3} m^{6}}{r_{R}} \int_{t_{0}}^{t_{R}} dT \left[\left(\dot{\phi}_{R} \frac{\dot{r}^{2}}{r^{4}} - \frac{\dot{r}_{R}}{r_{R}} \frac{\dot{r}\dot{\phi}}{r^{3}}\right) {\rm cos}\left(2 \phi - 2 \phi_{R}\right) 
\right.
\nn \\
&\left.
+ \left(-4 \dot{\phi}_{R} \frac{\dot{r} \dot{\phi}}{r^{3}} - \frac{1}{4} \frac{\dot{r}_{R}}{r_{R}} \frac{\dot{r}^{2}}{r^{4}}\right) {\rm sin}\left(2 \phi - 2 \phi_{R}\right)\right]
\end{align}
where we have used the fact that $\hat{L}_{i}$ is aligned with the z-axis, and the subscript $R$ indicates that the relative coordinates are evaluated at $t_{R}$, i.e. the relative coordinates of the orbit that the binary is currently on. Relative coordinates without the subscript are functions of the integration variable $T$, and depend on the past lifetime of the source. This expression is equivalent to Eq.~(A2) in~\cite{Arun:2009mc}, with the exception that we have not set $\phi_{R}$ (which is equivalent to $\phi_{0}$ in their notation) to zero. We are concerned with the average over the current orbit the binary is on, while~\cite{Arun:2009mc} was concerned with the instant at which $t_{R}$ corresponds to pericenter passage.

Let us now express the memory of Eq.~\eqref{G-mem-rel} in terms of harmonics of $e^{i \phi}$. To do this, we break with the notation of Sec.~\ref{Kep}, and express the relative coordinates in terms of the orbital phase using
\begin{align}
\label{r-of-phi}
r &= \frac{p}{1 + e \; {\rm cos}(\phi)}\,,
\\
\label{phidot-of-phi}
\dot{\phi} &= \sqrt{\frac{m}{p^{3}}} \left[1 + e \; {\rm cos}(\phi)\right]^{2}\,,
\\
\label{rdot-of-phi}
\dot{r} &= \sqrt{\frac{m}{p}} \; e \; {\rm sin}(\phi)\,,
\end{align}
where $p = a (1 - e^{2})$ is the semi-latus rectum of the orbit and where we have dropped the subscript on $e$, since we are on Newtonian orbits. Inserting these into Eq.~\eqref{G-mem-rel}, we obtain
\begin{align}
\label{eq:memTot}
{\cal{G}}_{\infty}^{\rm memory} = \sum_{q = -8}^{8} \int_{t_{0}}^{t_{R}} dT \; F_{q}\left[p(T), e(T), p_{R}, e_{R}, \phi_{R}\right] e^{i q \phi(T)}\,,
\end{align}
where $F_{q}$ are functions of the orbital elements $p$ and $e$, which we have written as explicit functions of $T$ to indicate that they are changing under radiation reaction. This expression is equivalent to Eq.~(A4) in~\cite{Arun:2009mc}, except that now the functions $F_{q}$ also depend on $\phi_{R}$. 

Let us first consider the DC term, i.e. $q = 0$, which takes the form
\begin{align}
\label{eq:memDC}
{\cal{G}}_{\infty}^{\rm DC\;mem} &= -\frac{16}{5} \frac{m^{7} \eta^{3}}{r_{R}^{2}} \left[r_{R} \dot{\phi}_{R} {\rm cos}(2 \phi_{R}) + \frac{1}{4} \dot{r}_{R} {\rm sin}(2 \phi_{R})\right] {\cal{M}}_{0}(t_{R})\,,
\\
\label{eq:memDC-int}
{\cal{M}}_{0}(t_{R}) &= \int_{t_{0}}^{t_{R}} dT \frac{e^{2}}{p^{5}} \left(1 + \frac{20}{21} e^{2} + \frac{19}{336} e^{4}\right)\,.
\end{align}
where we refer to ${\cal{M}}_{0}$ as the DC enhancement integral. To evaluate this further, we require an approximation for the evolution of the orbital elements under radiation reaction. However, notice that to leading order in small eccentricity the integrand scales as $e^{2}$. This implies that for quasi-circular binaries, the enhancement integral will necessarily be small. Thus, the memory contribution to the angular momentum flux will be most important for binaries with non-negligible eccentricity. The remainder of this paper will be dedicated to approximating these integrals for binaries with high eccentricity.

\subsection{Osculating Approximation and Orbit Averaging}
\label{osculate}
How do we model such an evolution for highly eccentric systems? We begin by reviewing how this is done in the burst approximation of~\cite{Loutrel:2014vja}. At Newtonian order, the conservative orbital dynamics of the binary are governed by two conserved quantities, the orbital energy and angular momentum, specifically
\begin{align}
E &= - \frac{\eta m^{2} (1 - e^{2})}{2 p}\,,
\\
L &= \eta \sqrt{m^{3} p}\,.
\end{align}
These expressions can be inverted to obtain the functions $p(E,L)$ and $e(E,L)$. The binary is, however, emitting gravitational radiation, which carries energy and angular momentum away from the system, creating a radiation reaction force that causes the binary to inspiral. To all orders in the PN formalism, this radiation reaction is governed by the balance laws in Eq.~\eqref{balance}. To obtain the evolution of $p$ and $e$, specifically $\dot{p}$ and $\dot{e}$, we simply combine $p(E,L)$ and $e(E,L)$ with the balance laws. Working to leading order, radiation reaction forces the orbital elements to evolve as
\begin{align}
\label{pdot}
\dot{p}(p,e,\phi) &= - 14 \left(\frac{m}{p}\right)^{3} \Psi(\phi)^{3} \bigg\{ \frac{18}{35} \left(1 - e^{2}\right) - \frac{8}{7} \Psi(\phi) + \frac{6}{7} \Psi(\phi)^{2} + \left(\frac{a}{3} - \frac{b}{7}\right)
\nn \\
&
\times \left[2 (1 - e^{2}) - 5 \Psi(\phi) + 3 \Psi(\phi)^{2}\right] \bigg\}
\\
\label{edot}
\dot{e}(p,e,\phi) &= \frac{2}{5} \frac{m^{3}}{p^{4}} \frac{\Psi(\phi)^{3}}{e} \left[1 - e^{2} - \Psi(\phi)^{2}\right] \bigg\{ 9 (1 - e^{2}) - 20 \Psi(\phi) + 15 \Psi(\phi)^{2} + \left(\frac{35}{6} a - \frac{5}{2} b\right)
\nn \\
&\times \left[2 (1 - e^{2}) - 5 \Psi(\phi) + 3 \Psi(\phi)^{2}\right] \bigg\} - \frac{4}{15} \frac{m^{3}}{p^{4}} \frac{\Psi(\phi)^{3}}{e} \left[\frac{d\Psi(\phi)}{d\phi}\right]^{2} \bigg\{ 3 (1 - e^{2}) - 10 \Psi(\phi) 
\nn \\
&- 15 \Psi(\phi)^{2} + \left(\frac{5}{4} a + \frac{15}{8} b\right) \left[2 (1 - e^{2}) -7 \Psi(\phi) + 9 \Psi(\phi)^{2}\right]\bigg\}
\end{align}
where $\Psi(\phi) = 1 + e \; {\rm cos}(\phi)$, and $(a,b)$ are the parameters that specify the radiation reaction gauge~\cite{PW}. Note that these expressions are not orbit averaged, as they depend on the orbital phase $\phi$. Since we are working with binaries that have eccentricity $e\sim1$, it is not suitable to use the orbit averaged radiation reaction equations $\langle \dot{p} \rangle$ and $\langle \dot{e} \rangle$ for the orbit evolution~\cite{Loutrel-avg}.

The above radiation reaction equations have no closed-form general solution, yet we require the solutions $p(t)$ and $e(t)$. How does one solve such equations in the limit $e\rightarrow1$? The answer is to exploit the nature of GW emission for such binaries. Highly eccentric binaries have strikingly different GW emission from quasi-circular binaries. Rather than being a continuous signal, the GW emission resembles a set of discrete bursts emitted during pericenter passage~\cite{Turner:1977}. This implies that the orbital energy and angular momentum (or $p$ and $e$) are effectively constant throughout the orbit, except at pericenter where they change rapidly. As a result, the orbital evolution resembles a set of Keplerian ellipses that \emph{osculate} onto one another~\cite{Loutrel:2014vja}.

To describe such an evolution analytically, consider an eccentric binary that starts on an orbit with semi-latus rectum $p_{i-1}$ and eccentricity $\delta e_{i-1} = 1 - e_{i-1}$. Note that we have introduced the parameter $\delta e$, which we will consider to be small and will use as an expansion parameter. The reason for this is that the osculating nature of the orbits is really only present when the eccentricity is sufficiently close to unity. We will further take the changes to $p$ and $e$ to be effectively instantaneous at pericenter. The binary will thus move on a Keplerian ellipse with $p_{i-1}$ and $e_{i-1}$ from $\phi_{i-1} = 0$ up until it reaches $\phi_{i} = 2 \pi$. At this point, the binary changes due to radiation reaction to a new orbit with
\begin{align}
\label{p-next}
p_{i} &= p_{i-1} + \int_{0}^{2\pi} d\phi \; \frac{\dot{p}(p_{i-1},\delta e_{i-1}, \phi)}{\dot{\phi}(p_{i-1},\delta e_{i-1},\phi)}\,,
\\
\label{de-next}
\delta e_{i} &= \delta e_{i-1} - \int_{0}^{2\pi} d\phi \; \frac{\dot{e}(p_{i-1},\delta e_{i-1}, \phi)}{\dot{\phi}(p_{i-1},\delta e_{i-1},\phi)}\,.
\end{align}
Since the orbital elements $p_{i-1}$ and $\delta e_{i-1}$ are constant between the bounds of integration, the integral in the above expressions can be evaluated analytically, at which point we obtain
\begin{align}
p_{i} &= p_{i-1} - 48 \pi \eta \frac{m^{5/2}}{p_{i-1}^{3/2}} \left[1 - \frac{14}{15} \delta e_{i-1} + {\cal{O}}\left(\delta e_{i-1}^{2}\right)\right]\,,
\\
\delta e_{i} &= \delta e_{i-1} + \frac{170}{3} \pi \eta \left(\frac{m}{p}\right)^{5/2} \left[1 - \frac{667}{425} \delta e_{i-1} + {\cal{O}}\left(\delta e_{i-1}^{2}\right)\right]\,.
\end{align}
where we have expanded in $\delta e_{i-1}$.

The above equations provide a recursive algorithm that describes the orbital evolution as a set of discrete steps, by which the orbit osculates from one Keplerian ellipse to another due to the emission of GW bursts at pericenter. Note that we have worked solely to leading PN order in the orbital dynamics and in radiation reaction. The expressions in Eq.~\eqref{p-next} and~\eqref{de-next} are, however, purely general and apply to all orders in the PN formalism. Further, we have expanded all expressions in $\delta e \ll 1$, keeping only the first order corrections. One could easily keep higher order corrections, however, since we are interested in binaries formed with $e\sim1$, we will neglect them here.

We now apply this to the DC enhancement integral in Eq.~\eqref{eq:memDC-int}. Consider a binary that is initially on an unbound orbit with orbital elements $p_{0}$ and $e_{0} \ge 1$. As the binary moves through pericenter at time $t_{0}$, the binary will become bound due to the emission of GWs or through dynamical friction\footnote{The exact mechanism of formation does not matter at this level, only that it causes a change in the orbital energy that forces the binary to become bound.}. As a result, the binary's orbit becomes a Keplerian ellipse with $p_{1}$ and $\delta e_{1} \ll 1$. The binary will then evolve due to the emission of GW bursts at the instant of every pericenter passage at times $\{t_{i}\}$, from the time of formation $t_{0}$ up until the current time $t_{R}$, such that $\{t_{0} < t_{i} < t_{R}; \forall i\}$. From time $t_{0}$ to time $t_{N-1}$, we will say that the binary has gone through $N-1$ orbits and is currently on the $N$-th orbit. The current time will then be $t_{N-1} < t_{R} < t_{N}$.

The above description indicates that we can split the DC enhancement integral into discrete parts that correspond to the $N-1$ orbits that the binary has evolved through and the current orbit of the binary. More specifically, Eq.~\eqref{eq:memDC-int} becomes
\begin{equation}
\label{M0-tot}
{\cal{M}}_{0}(t_{R}) = \int_{t_{N-1}}^{t_{R}} dT \frac{e^{2}}{p^{5}} \left(1 + \frac{20}{21} e^{2} + \frac{19}{336} e^{4}\right) + \sum_{i=1}^{N-1} \int_{t_{i-1}}^{t_{i}} dT \frac{e^{2}}{p^{5}} \left(1 + \frac{20}{21} e^{2} + \frac{19}{336} e^{4}\right)\,.
\end{equation}
Notice that the past contribution involves the summations of $N-1$ orbits. In general, the memory will be larger for binaries that have evolved longer, or alternatively, have formed farther in the past. Thus, we focus on the past contribution and take the limit $N>>1$. Recall that in our osculating approximation, the orbital elements $p$ and $e$ are effectively constant between consecutive pericenter passages. Thus, within our approximation, we are free to take $p \rightarrow p_{i}[p_{i-1}, \delta e_{i-1}]$ and $e \rightarrow e_{i}[p_{i-1}, \delta e_{i-1}]$, and pull the dependence on these variables outside of the integral, by which we obtain
\begin{equation}
\label{M0-past}
{\cal{M}}_{0}(t_{R}) = \sum_{i=1}^{N-1} \frac{e_{i}^{2}}{p_{i}^{5}} \left(1 + \frac{20}{21} e_{i}^{2} + \frac{19}{336} e_{i}^{4} \right) \left(t_{i} - t_{i-1}\right)\,.
\end{equation}
We perform one final step of simplification, which is to recall that the time between consecutive pericenter passages, specifically $(t_{i} - t_{i-1})$, is simply the orbital period. Applying this, we finally obtain
\begin{equation}
\label{M0-past-final}
{\cal{M}}_{0}(t_{R}) = \frac{2 \pi}{m^{1/2}} \sum_{i=1}^{N-1} \frac{e_{i}^{2}}{p_{i}^{7/2}} \frac{\left(1 + \frac{20}{21} e_{i}^{2} + \frac{19}{336} e_{i}^{4}\right)}{(1 - e_{i}^{2})^{3/2}}\,.
\end{equation}

The last step is to perform the orbit average of Eq.~\eqref{eq:memDC} to find the averaged angular momentum flux. However, it is easily verified that this average vanishes upon using the result in Eq.~\eqref{M0-past-final}. In fact, this analysis can be repeated for the oscillatory terms in the memory and it does not change. Within this lowest order osculating approximation, the memory terms in the angular momentum flux all vanish. The reason for this is explained in~\cite{Arun:2009mc}: if the orbital elements of the binary are held constant, then the orbit average of the memory terms simply involves $\langle \overset{(3)}{I_{jk}} \rangle$, which vanishes since the quadrupole moment is periodic. 

There are two differences in our analysis relative to that of~\cite{Arun:2009mc}. The first is that the DC term also vanishes in our case, whereas previously it was non-vanishing. Here, we have not assumed $\phi_{R} = 0$ as was done previously, which causes the DC component to become periodic over the current orbit. This is more general than previous results since averaging after setting $\phi_{R} = 0$ is not really appropriate. The second difference is that here we have applied a model for the evolution of the binary in order to evaluate the hereditary integral, whereas the discussion in Sec.~5.C of~\cite{Arun:2009mc} concerns the idealized situation where the orbital elements are constant over the entire lifetime of the binary. Regardless, the end result is the same: if the orbital elements are held constant, whether over single orbits or over the entire history, the averaged memory terms will vanish.

Does this mean that the memory vanishes within the averaged angular momentum flux? The answer is no, specifically because we have assumed that the evolution of the binary is effectively instantaneous at pericenter. Thus, the result obtained here is actually a symptom of the inaccuracy of the osculating approximation applied here. The approximation is of course only a lowest order approximation; a higher order computation would incorporate the fact that the changes are not instantaneous, but are instead rapid and continuous around pericenter passage. This is actively being worked on~\cite{Loutrel-avg}, and the memory calculated using this higher order approximation will be included in a follow up study.

Finally, we comment on the contribution from the current orbit. If we apply the same osculating approximation here to this contribution, we obtain a term that scales linearly with $t_{R}$. It is not difficult then to see that averaging this contribution will not vanish, as it is not just simply a periodic function in $\phi_{R}$, but also involves the secularly growing $t_{R}$. However, as mentioned previously, we expect this term to be suppressed compared to the past contribution by $N^{-1}$, where $N$ is the number of orbits. In lieu of the difficulty of calculating the past contribution, we leave the details of the current contribution to future work. 

\section{Discussion}

We have here constructed analytic approximations to all non-linear hereditary effects in compact binary inspirals to 3PN order. For the tail terms, we constructed superasymptotic and hyperasymptotic series that are typically accurate to better than $10^{-8}$ relative to numerical PN results, except for two 2.5PN factors that are accurate to better than $10^{-4}$. 
We have checked that the numerical calculation of the orbital phase with the hereditary fluxes prescribed through our analytic, closed-form expressions is within double precision of the numerical phase obtained with the fluxes prescribed through infinite Bessel sums. The advantage of these analytic approximations is not only that they are accurate, but that they are controlled (meaning one can in principle go to higher order in perturbation theory to obtain more accurate expressions) and they are fast (meaning they can be directly evaluated without need for additional numerical simulations).   

With these analytic results at hand, we can now begin to consider the creation of closed-form, analytic waveform models and burst models for GW data analysis of binaries with generic eccentricity. For waveform-based searches, one can construct analytic time and Fourier domain waveforms through the stationary phase approximation to the Fourier integrals, allowing for the development of eccentric TaylorF2 waveforms. At high eccentricities, the stationary phase approximation may not be valid, and one may need to resort to the shifted uniform asymptotic method of~\cite{Klein:2014bua} to compute Fourier integrals. For burst models, one can construct 3PN accurate time-frequency tracks, following for example the method of~\cite{Loutrel:2014vja}. 

These calculations would allow us to determine more precisely whether the accuracy of the fluxes calculated here is enough for GW data analysis or whether one needs to go to higher order in perturbation theory. To do so, one would carry out a Bayesian parameter estimation study (whether using waveforms and matched filtering or a burst model with an informative prior) to estimate whether models constructed with analytic fluxes are efficient and faithful at recovering numerical signals. Such an analysis would reveal, for example, whether the posterior probability distribution obtained with models that use analytic fluxes biases the recovery of parameters.  One thing is clear, however, the analytic calculations presented here have moved us one step close to the creation of generic models for the GWs emitted in the inspiral of binary systems.

\section*{Acknowledgements}

We would like to thank K. G. Arun for several useful discussions. N.Y. acknowledges support from the NSF CAREER Grant PHY-1250636. N. L. acknowledges support from the NSF EAPSI Fellowship Award No. 1614203.

\appendix
\section{Fourier Coefficients of the Mass Octopole and Current Quadrupole}
\label{Fourier-app}

We here present the Fourier coefficients of the source mass octopole and current quadrupole necessary to complete the re-summation of the 2.5PN enhancement factors $\beta(e_{t})$ and $\gamma(e_{t})$. For the mass octopole, we have
\begin{align}
\underset{(p)}{\hat{\cal{I}}_{xxx}} &= -\frac{3}{5} \left(-5 e^{5} p^{2} + 10 e^{3} p^{2} + 6 e^{3} - 5 e p^{2} - 10 e\right) \frac{J'_{p}(p e)}{p^{3} e^{3}}  
\nn \\
&- \frac{3}{5} \left(5 e^{4} - 20 e^{2} + 15\right) \frac{J_{p}(p e)}{p^{2} e^{3}}\,,
\\
\underset{(p)}{\hat{\cal{I}}_{xxy}} &= -\frac{i}{5} \sqrt{1 - e^{2}} (25 e^{3} - 45 e ) \frac{J'_{p}(p e)}{p^{2} e^{3}} 
\nn \\
&- \frac{i}{5} \sqrt{1 - e^{2}} (15 e^{4} p^{2} - 30 e^{2} p^{2} - 6 e^{2} + 15 p^{2} + 30) \frac{J_{p}(p e)}{p^{3} e^{3}}\,,
\\
\underset{(p)}{\hat{\cal{I}}_{xyy}} &= \frac{1}{5} (-15 e^{5} p^{2} + 30 e^{3} p^{2} + 24 e^{3} - 15 e p^{2} - 30 e) \frac{J'_{p}(p e)}{p^{3} e^{3}} 
\nn \\
&+ \frac{1}{5} (20 e^{4} - 65 e^{2} + 45) \frac{J_{p}(p e)}{p^{2} e^{3}}\,,
\\
\underset{(p)}{\hat{\cal{I}}_{yyy}} &= \frac{3 i}{5} \sqrt{1 - e^{2}} (10 e^{3} - 15 e) \frac{J'_{p}(p e)}{p^{2} e^{3}} 
\nn \\
&+ \frac{3 i}{5} \sqrt{1 - e^{2}} (5 e^{4} p^{2} - 10 e^{2} p^{2} - 4 e^{2} + 5 p^{2} + 10) \frac{J_{p}(p e)}{p^{3} e^{3}}\,,
\\
\underset{(p)}{\hat{\cal{I}}_{zzx}} &= -\frac{1}{5} (5 e^{2}  - 5) \frac{J_{p}(p e)}{e p^{2}} - \frac{6}{5} \frac{J'_{p}(p e)}{p^{3}}\,,
\\
\underset{(p)}{\hat{\cal{I}}_{zzy}} &= \frac{6 i}{5} \sqrt{1 - e^{2}} \frac{J_{p}(p e)}{e p^{3}} - i \sqrt{1 - e^{2}} \frac{J'_{p}(p e)}{p^{2}}\,,
\end{align}
and for the current quadrupole
\begin{align}
\underset{(p)}{\hat{\cal{J}}_{zx}} &= \frac{1}{2} \sqrt{1 - e^{2}} \frac{J'_{p}(p e)}{p}\,,
\\
\underset{(p)}{\hat{\cal{J}}_{zy}} &= \frac{i}{2} (1 - e^{2}) \frac{J_{p}(p e)}{p e}\,.
\end{align}
%

\section{Pad\'{e} Approximant Coefficients}
\label{Pade}

The analytic expressions for the 2.5PN enhancement factors $[\alpha_{3}^{(2)}(e_{t}), \tilde{\alpha}_{3}^{(2)}(e_{t})]$ that we obtain take the form 
\begin{align}
\alpha_{3}^{(2)}(e_{t}) &= \frac{A^{\rm PD}(e_{t})}{(1 - e_{t}^{2})^{6}}\,,
\\
\tilde{\alpha}_{3}^{(2)}(e_{t}) &= \frac{\tilde{A}^{\rm PD}(e_{t})}{(1 - e_{t}^{2})^{9/2}}\,,
\end{align}
where $[A^{\rm PD}(e_{t}), \tilde{A}^{\rm PD}(e_{t})]$ are Pad\'{e} approximants of the form in Eq.~\eqref{A-PD}. We here display the non-zero coefficients of these Pad\'{e} approximants. The exact fractional form of the coefficients are too lengthy to show here, so we will display their numerical value as given by \texttt{Mathematica}. For $A^{\rm PD}(e_{t})$, we have
\begin{align}
A^{(2)} &= 2517\,, 
\qquad
A^{(4)} = -542.748497072893\,,
\nn \\
A^{(6)} &= -8106.76855568073\,,
\qquad
A^{(8)} = 8413.80043346759\,,
\nn \\
A^{(10)} &= -1797.84373379634\,,
\qquad
A^{(12)} = -645.518003878721\,,
\nn \\
A^{(14)} &= 196.536287243258\,,
\\
A_{(2)} &= -2.73321354671152\,,
\qquad
A_{(4)} = 2.72445722875519\,,
\nn \\
A_{(6)} &= -1.18148275600057\,,
\qquad
A_{(8)} = 0.198189499682584\,,
\nn \\
A_{(10)} &= -0.0052713663843431\,,
\qquad
A_{(12)} = -0.000104564833689961\,,
\nn \\
A_{(14)} &= -0.00001735320614776194\,,
\qquad
A_{(16)} = -2.68866627647753\times10^{-6}\,.
\end{align}
Similarly, for $\tilde{A}^{\rm PD}(e_{t})$ we have
\begin{align}
\tilde{A}^{(2)} &= 1428\,,
\qquad
\tilde{A}^{(4)} = -3082.0825943902\,,
\nn \\
\tilde{A}^{(6)} &= 1251.7504530861\,,
\qquad
\tilde{A}^{(8)} = 1534.29667420639\,,
\nn \\
\tilde{A}^{(10)} &= -1557.48894312597\,,
\qquad
\tilde{A}^{(12)} = 474.00656591547\,,
\nn \\
\tilde{A}^{(14)} &= -44.7331690120333\,,
\\
\tilde{A}_{(2)} &= -3.15201862352255\,,
\qquad
\tilde{A}_{(4)} = 3.82292854565682\,,
\nn \\
\tilde{A}_{(6)} &= -2.22448315038514\,,
\qquad
\tilde{A}_{(8)} = 0.6262354256498\,,
\nn \\
\tilde{A}_{(10)} &= -0.0738364336330394\,,
\qquad
\tilde{A}_{(12)} = 0.0021699571134637\,,
\nn \\
\tilde{A}_{(14)} &= 0.0000142157331173306\,,
\qquad
\tilde{A}_{(16)} = -3.8287501968869\times10^{-6}\,.
\end{align}
\bibliography{master}

\providecommand{\newblock}{}
\begin{thebibliography}{10}
\expandafter\ifx\csname url\endcsname\relax
  \def\url#1{{\tt #1}}\fi
\expandafter\ifx\csname urlprefix\endcsname\relax\def\urlprefix{URL }\fi
\providecommand{\eprint}[2][]{\url{#2}}

\bibitem{Harry:2010zz}
Harry G~M (LIGO Scientific Collaboration) 2010 {\em Class.Quant.Grav.\/} {\bf
  27} 084006

\bibitem{GW150914}
{Abbott, B P et al} (LIGO Scientific Collaboration and Virgo Collaboration)
  2016 {\em Phys. Rev. Lett.\/} {\bf 116}(6) 061102
  \urlprefix\url{http://link.aps.org/doi/10.1103/PhysRevLett.116.061102}

\bibitem{TheVirgo:2014hva}
Acernese F {\em et~al.\/} (VIRGO) 2015 {\em Class.Quant.Grav.\/} {\bf 32}
  024001 (\textit{Preprint} \eprint{1408.3978})

\bibitem{Uchiyama:2004vr}
Uchiyama T, Kuroda K, Ohashi M, Miyoki S, Ishitsuka H {\em et~al.\/} 2004 {\em
  Class.Quant.Grav.\/} {\bf 21} S1161--S1172

\bibitem{Unnikrishnan:2013qwa}
Unnikrishnan C 2013 {\em Int.J.Mod.Phys.\/} {\bf D22} 1341010

\bibitem{Yunes:2016jcc}
Yunes N, Yagi K and Pretorius F 2016 {\em Phys. Rev.\/} {\bf D94} 084002
  (\textit{Preprint} \eprint{1603.08955})

\bibitem{Hulse:1974eb}
Hulse R~A and Taylor J~H 1975 {\em Astrophys. J.\/} {\bf 195} L51--L53

\bibitem{2009MNRAS.395.2127O}
{O'Leary} R~M, {Kocsis} B and {Loeb} A 2009 {\em Mon. Not. R. Astron. Soc.\/}
  {\bf 395} 2127--2146 (\textit{Preprint} \eprint{0807.2638})

\bibitem{lee2010}
{Lee} W~H, {Ramirez-Ruiz} E and {van de Ven} G 2010 {\em {The Astrophysical
  Journal}\/} {\bf 720} 953--975 (\textit{Preprint} \eprint{0909.2884})

\bibitem{2003ApJ...598..419W}
{Wen} L 2003 {\em {The Astrophysical Journal}\/} {\bf 598} 419--430
  (\textit{Preprint} \eprint{astro-ph/0211492})

\bibitem{Kushnir:2013hpa}
Kushnir D, Katz B, Dong S, Livne E and Fern\'{a}ndez R 2013 {\em
  Astrophys.J.\/} {\bf 778} L37 (\textit{Preprint} \eprint{1303.1180})

\bibitem{2013PhRvL.111f1106S}
{Seto} N 2013 {\em Physical Review Letters\/} {\bf 111} 061106
  (\textit{Preprint} \eprint{1304.5151})

\bibitem{Antognini:2013lpa}
Antognini J~M, Shappee B~J, Thompson T~A and Amaro-Seoane P 2014 {\em Mon. Not.
  Roy. Astron. Soc.\/} {\bf 439} 1079--1091 (\textit{Preprint}
  \eprint{1308.5682})

\bibitem{2013ApJ...773..187N}
{Naoz} S, {Kocsis} B, {Loeb} A and {Yunes} N 2013 {\em {The Astrophysical
  Journal}\/} {\bf 773} 187 (\textit{Preprint} \eprint{1206.4316})

\bibitem{2014ApJ...781...45A}
{Antonini} F, {Murray} N and {Mikkola} S 2014 {\em {The Astrophysical
  Journal}\/} {\bf 781} 45 (\textit{Preprint} \eprint{1308.3674})

\bibitem{Antonini:2015zsa}
Antonini F, Chatterjee S, Rodriguez C~L, Morscher M, Pattabiraman B, Kalogera V
  and Rasio F~A 2016 {\em Astrophys. J.\/} {\bf 816} 65 (\textit{Preprint}
  \eprint{1509.05080})

\bibitem{Loutrel:2014vja}
Loutrel N, Yunes N and Pretorius F 2014 {\em Phys.Rev.\/} {\bf D90} 104010
  (\textit{Preprint} \eprint{1404.0092})

\bibitem{Huerta:2013qb}
Huerta E~A and Brown D~A 2013 {\em Phys. Rev.\/} {\bf D87} 127501
  (\textit{Preprint} \eprint{1301.1895})

\bibitem{Huerta:2014eca}
Huerta E, Kumar P, McWilliams S~T, O'Shaughnessy R and Yunes N 2014 {\em
  Phys.Rev.\/} {\bf D90} 084016 (\textit{Preprint} \eprint{1408.3406})

\bibitem{PhysRevD.80.084001}
Yunes N, Arun K~G, Berti E and Will C~M 2009 {\em Phys. Rev. D\/} {\bf 80}(8)
  084001 \urlprefix\url{http://link.aps.org/doi/10.1103/PhysRevD.80.084001}

\bibitem{Tessmer:2010sh}
Tessmer M and Schaefer G 2010 {\em Phys. Rev.\/} {\bf D82} 124064
  (\textit{Preprint} \eprint{1006.3714})

\bibitem{Tessmer:2010ii}
Tessmer M and Schaefer G 2011 {\em Annalen Phys.\/} {\bf 523} 813--864
  (\textit{Preprint} \eprint{1012.3894})

\bibitem{Moore:2016qxz}
Moore B, Favata M, Arun K~G and Mishra C~K 2016 {\em Phys. Rev.\/} {\bf D93}
  124061 (\textit{Preprint} \eprint{1605.00304})

\bibitem{Tanay:2016zog}
Tanay S, Haney M and Gopakumar A 2016 {\em Phys. Rev.\/} {\bf D93} 064031
  (\textit{Preprint} \eprint{1602.03081})

\bibitem{Forseth:2015oua}
Forseth E, Evans C~R and Hopper S 2016 {\em Phys. Rev.\/} {\bf D93} 064058
  (\textit{Preprint} \eprint{1512.03051})

\bibitem{Tai:2014bfa}
Tai K~S, McWilliams S~T and Pretorius F 2014 {\em Phys. Rev.\/} {\bf D90}
  103001 (\textit{Preprint} \eprint{1403.7754})

\bibitem{Blanchet:2013haa}
Blanchet L 2014 {\em Living Rev.Rel.\/} {\bf 17} 2 (\textit{Preprint}
  \eprint{1310.1528})

\bibitem{Blanchet379}
Blanchet L and Damour T 1986 {\em Philosophical Transactions of the Royal
  Society of London A: Mathematical, Physical and Engineering Sciences\/} {\bf
  320} 379--430 ISSN 0080-4614

\bibitem{Blanchet383}
Blanchet L 1987 {\em Proceedings of the Royal Society of London A:
  Mathematical, Physical and Engineering Sciences\/} {\bf 409} 383--399 ISSN
  0080-4630

\bibitem{Blanchet:1988}
Blanchet L and Damour T 1988 {\em Phys. Rev. D\/} {\bf 37}(6) 1410--1435
  \urlprefix\url{http://link.aps.org/doi/10.1103/PhysRevD.37.1410}

\bibitem{Blanchet:1992br}
Blanchet L and Damour T 1992 {\em Phys.Rev.\/} {\bf D46} 4304--4319

\bibitem{Blanchet:1993ec}
Blanchet L and Schaefer G 1993 {\em Class. Quant. Grav.\/} {\bf 10} 2699--2721

\bibitem{Blanchet:1995fr}
Blanchet L 1995 {\em Phys. Rev.\/} {\bf D51} 2559--2583 (\textit{Preprint}
  \eprint{gr-qc/9501030})

\bibitem{Rieth:1997mk}
Rieth R and Schaefer G 1997 {\em Class. Quant. Grav.\/} {\bf 14} 2357--2380

\bibitem{Arun:2007rg}
Arun K, Blanchet L, Iyer B~R and Qusailah M~S 2008 {\em Phys.Rev.\/} {\bf D77}
  064034 (\textit{Preprint} \eprint{0711.0250})

\bibitem{Blanchet:1997ji}
Blanchet L 1998 {\em Class. Quant. Grav.\/} {\bf 15} 89--111 (\textit{Preprint}
  \eprint{gr-qc/9710037})

\bibitem{Blanchet:1997jj}
Blanchet L 1998 {\em Class. Quant. Grav.\/} {\bf 15} 113--141 [Erratum: Class.
  Quant. Grav.22,3381(2005)] (\textit{Preprint} \eprint{gr-qc/9710038})

\bibitem{Christodoulou:1991}
Christodoulou D 1991 {\em Phys. Rev. Lett.\/} {\bf 67}(12) 1486--1489
  \urlprefix\url{http://link.aps.org/doi/10.1103/PhysRevLett.67.1486}

\bibitem{Wiseman:1991}
Wiseman A~G and Will C~M 1991 {\em Phys. Rev. D\/} {\bf 44}(10) R2945--R2949
  \urlprefix\url{http://link.aps.org/doi/10.1103/PhysRevD.44.R2945}

\bibitem{Thorne:1992}
Thorne K~S 1992 {\em Phys. Rev. D\/} {\bf 45}(2) 520--524
  \urlprefix\url{http://link.aps.org/doi/10.1103/PhysRevD.45.520}

\bibitem{Arun:2004ff}
Arun K~G, Blanchet L, Iyer B~R and Qusailah M~S~S 2004 {\em Class. Quant.
  Grav.\/} {\bf 21} 3771--3802 [Erratum: Class. Quant. Grav.22,3115(2005)]
  (\textit{Preprint} \eprint{gr-qc/0404085})

\bibitem{Favata:2008yd}
Favata M 2009 {\em Phys. Rev.\/} {\bf D80} 024002 (\textit{Preprint}
  \eprint{0812.0069})

\bibitem{Favata:2011qi}
Favata M 2011 {\em Phys. Rev.\/} {\bf D84} 124013 (\textit{Preprint}
  \eprint{1108.3121})

\bibitem{Lasky:2016knh}
Lasky P~D, Thrane E, Levin Y, Blackman J and Chen Y 2016 {\em Phys. Rev.
  Lett.\/} {\bf 117} 061102 (\textit{Preprint} \eprint{1605.01415})

\bibitem{Garfinkle:2016nhe}
Garfinkle D 2016 {\em Class. Quant. Grav.\/} {\bf 33} 177001 (\textit{Preprint}
  \eprint{1605.06687})

\bibitem{Boyd}
Boyd J~P {\em Acta Applicandae Mathematica\/} {\bf 56} 1--98 ISSN 1572-9036
  \urlprefix\url{http://dx.doi.org/10.1023/A:1006145903624}

\bibitem{blanchet-review}
Blanchet L 2006 {\em Living Rev. Rel.\/} {\bf 9} 4

\bibitem{Mino:1997bx}
Mino Y, Sasaki M, Shibata M, Tagoshi H and Tanaka T 1997 {\em Prog. Theor.
  Phys. Suppl.\/} {\bf 128} 1--121 (\textit{Preprint} \eprint{gr-qc/9712057})

\bibitem{Isaacson:1968ra}
{Isaacson} R~A 1968 {\em Phys. Rev.\/} {\bf 166} 1263--1271

\bibitem{Isaacson:1968gw}
{Isaacson} R~A 1968 {\em Phys. Rev.\/} {\bf 166} 1272--1279

\bibitem{PW}
Poisson E~and~Will C~M 2014 {\em Gravity: Newtonian, Post-Newtonian,
  Relativistic\/} (Cambridge: Cambridge University Press)

\bibitem{Thorne:1980rm}
Thorne K~S 1980 {\em Rev. Mod. Phys.\/} {\bf 52} 299--339

\bibitem{Blanchet:1995fg}
Blanchet L, Damour T and Iyer B~R 1995 {\em Phys.Rev.\/} {\bf D51} 5360
  (\textit{Preprint} \eprint{gr-qc/9501029})

\bibitem{2002PhRvD..65f4005B}
{Blanchet} L, {Iyer} B~R and {Joguet} B 2002 {\em {Phys. Rev. D.}\/} {\bf 65}
  064005--+ (\textit{Preprint} \eprint{gr-qc/0105098})

\bibitem{Blanchet:2001ax}
Blanchet L, Faye G, Iyer B~R and Joguet B 2002 {\em Phys. Rev.\/} {\bf D65}
  061501 [Erratum: Phys. Rev.D71,129902(2005)] (\textit{Preprint}
  \eprint{gr-qc/0105099})

\bibitem{Blanchet:2004ek}
Blanchet L, Damour T, Esposito-Farese G and Iyer B~R 2004 {\em Phys. Rev.
  Lett.\/} {\bf 93} 091101 (\textit{Preprint} \eprint{gr-qc/0406012})

\bibitem{Faye:2012we}
Faye G, Marsat S, Blanchet L and Iyer B~R 2012 {\em Class. Quant. Grav.\/} {\bf
  29} 175004 (\textit{Preprint} \eprint{1204.1043})

\bibitem{Arun:2007sg}
Arun K, Blanchet L, Iyer B~R and Qusailah M~S 2008 {\em Phys.Rev.\/} {\bf D77}
  064035 (\textit{Preprint} \eprint{0711.0302})

\bibitem{Arun:2009mc}
Arun K~G, Blanchet L, Iyer B~R and Sinha S 2009 {\em Phys.Rev.\/} {\bf D80}
  124018 (\textit{Preprint} \eprint{0908.3854})

\bibitem{zbMATH03938612}
{Damour} T and {Deruelle} N 1985 {\em {Ann. Inst. Henri Poincar\'e, Phys.
  Th\'eor.}\/} {\bf 43} 107--132 ISSN 0246-0211

\bibitem{zbMATH04001537}
{Damour} T and {Deruelle} N 1986 {\em {Ann. Inst. Henri Poincar\'e, Phys.
  Th\'eor.}\/} {\bf 44} 263--292 ISSN 0246-0211

\bibitem{Damour:1988mr}
Damour T and Schaefer G 1988 {\em Nuovo Cim.\/} {\bf B101} 127

\bibitem{0264-9381-12-4-009}
Wex N 1995 {\em Classical and Quantum Gravity\/} {\bf 12} 983
  \urlprefix\url{http://stacks.iop.org/0264-9381/12/i=4/a=009}

\bibitem{Schaefer1993196}
Schaefer G and Wex N 1993 {\em Physics Letters A\/} {\bf 174} 196 -- 205 ISSN
  0375-9601
  \urlprefix\url{http://www.sciencedirect.com/science/article/pii/037596019390758R}

\bibitem{Memmesheimer:2004cv}
Memmesheimer R~M, Gopakumar A and Schaefer G 2004 {\em Phys.Rev.\/} {\bf D70}
  104011 (\textit{Preprint} \eprint{gr-qc/0407049})

\bibitem{Damour:2004bz}
Damour T, Gopakumar A and Iyer B~R 2004 {\em Phys. Rev.\/} {\bf D70} 064028
  (\textit{Preprint} \eprint{gr-qc/0404128})

\bibitem{Loutrel-avg}
Liebersbach S, Loutrel N, Yunes N and Pretorius F {\em \textit{in
  preparation}\/}

\bibitem{PetersMathews}
Peters P~C and Mathews J 1963 {\em Physical Review\/} {\bf 131} 435--440

\bibitem{Peters:1964zz}
Peters P 1964 {\em Phys.Rev.\/} {\bf 136} B1224--B1232

\bibitem{NIST}
Olver F~W~J, Lozier D~W, Boisvert R~F and Clark C~W (eds) 2010 {\em {NIST
  Handbook of Mathematical Functions}\/} (Cambridge University Press)

\bibitem{Watson}
Watson G~N 1966 {\em {A Treatise on the Theory of Bessel Functions}\/} 2nd ed
  (Cambridge University Press)

\bibitem{Nikishov:2013}
{Nikishov} A~I 2013 {\em ArXiv e-prints\/} (\textit{Preprint}
  \eprint{1302.0978})

\bibitem{Gradshteyn}
Gradshteyn I~S and Ryzhik I~M 2007 {\em {Table of Integrals, Series, and
  Products}\/} (Academic Press)

\bibitem{Turner:1977}
{Turner} M 1977 {\em {The Astrophysical Journal}\/} {\bf 216} 610--619

\bibitem{Klein:2014bua}
Klein A, Cornish N and Yunes N 2014 {\em Phys. Rev.\/} {\bf D90} 124029
  (\textit{Preprint} \eprint{1408.5158})

\end{thebibliography}
\end{document}